\documentclass[a4paper,12pt, epsfig]{article}
\def\letter{0}\def\pr{0}
\pdfoutput=1 
\usepackage{epsfig}
\usepackage{epstopdf}
\usepackage{graphicx}
\usepackage{ifthen}
\usepackage{ulem}

\usepackage{amsmath}
\usepackage[psamsfonts]{amssymb}
\usepackage{euscript}

\usepackage{latexsym}
\usepackage[arrow,matrix,curve]{xy}

\usepackage[hypertexnames=false]{hyperref}
\hypersetup{
    colorlinks=true,
    linkcolor=blue,
    filecolor=magenta,   
    citecolor=black,   
    urlcolor=cyan,
    }

\newskip\humongous \humongous=0pt plus 1000pt minus 1000pt

\newif\ifdtup

\def\,{\hspace{-.1cm}}
\def\hsp{,\hspace{.7cm}}

\def\fc#1#2 {\frac{n}{q}#1\frac{n}{q}#2}

\def\kt{\kappa}
\def\kt{\mathfrak{K}}
\def\ks{|\kt\rangle}

\def\rv{{\rm{vac}}}

\newcommand{\vac}{\ensuremath{|0\rangle}}

\renewcommand{\sin}{\textrm{sin}}

\renewcommand{\sinh}{\textrm{sinh}}

\renewcommand{\tanh}{\textrm{tanh}}
\newcommand{\sech}{\textrm{sech}}
\newcommand{\csch}{\textrm{csch}}

\def\sign#1{{\rm sign}\left(#1\right)}

\renewcommand{\theequation}{\arabic{section}.\arabic{equation}}
\renewcommand{\(}{\begin{equation}}
\renewcommand{\)}{end{equation} \vspace{-.05in}\linebreak}

\newcounter{saveeqn}
\newcounter{savealpheqn}

\newcommand{\alpheqn}{\setcounter{saveeqn}{\value{equation}}%
  \stepcounter{saveeqn}\setcounter{equation}{0}%
  \renewcommand{\theequation}{\mbox{\arabic{section}.\arabic{saveeqn}
\alph{equation}}}
  \renewcommand{\)}{\end{equation}}}
\def\part#1{\frac{\partial}{\partial{#1}}}%
\def\group#1{\refstepcounter{equation}\setcounter{saveeqn}
 {\value{equation}}%
  \label{#1}\setcounter{equation}{0}%
\renewcommand{\theequation}{\mbox{\arabic{section}.\arabic{saveeqn}
\alph{equation}}}
  \renewcommand{\)}{\end{equation}}}
\newcommand{\reseteqn}{\setcounter{equation}{\value{saveeqn}}%
  \renewcommand{\theequation}{\arabic{section}.\arabic{equation}}%
  \renewcommand{\)}{\end{equation}}}

\newcommand{\aalpheqn}{\setcounter{saveeqn}{\value{equation}}%
  \stepcounter{saveeqn}\setcounter{equation}{0}%
  \renewcommand{\theequation}{\mbox{
        \Alph{subsection}.\arabic{saveeqn}\alph{equation}}}
   \renewcommand{\)}{\end{equation}}}
\newcommand{\areseteqn}{\setcounter{equation}{\value{saveeqn}}%
  \renewcommand{\theequation}{\Alph{subsection}.\arabic{equation}}%
  \renewcommand{\)}{\end{equation}}}

\renewcommand{\thefootnote}{\alph{footnote}}
\renewcommand{\(}{\begin{equation}}
\renewcommand{\)}{\end{equation}}
\newcommand{\ba}{\begin{eqnarray}}
\newcommand{\ea}{\end{eqnarray}}

\renewcommand{\b}{\beta}

\renewcommand{\sl}{{\sqrt{\lambda}}}
\newcommand{\slq}{{\sqrt{\lambda Q_0}}}

\newcommand{\cbp}{\mathop{\vtop{\ialign{##\crcr
   $\hfil\displaystyle{}\hfil$\crcr\noalign{\kern-13pt\nointerlineskip}
   \BIG{)}\hskip0pt\crcr\noalign{\kern3pt}}}}}
\newcommand{\pa}{\mathop{\vtop{\ialign{##\crcr

$\hfil\displaystyle{\oplus}\hfil$\crcr\noalign{\kern+1pt\nointerlineskip
}
   \hspace{.08in}$^{\alpha=0}$\hskip6pt\crcr\noalign{\kern3pt}}}}}
\renewcommand{\hsp}{,\hspace{.3in}}
\newcommand{\p}{^\prime}

\catcode`\@=11
\def\vereq#1#2{\lower3pt\vbox{\baselineskip1.5pt \lineskip1.5pt
\ialign{$\m@th#1\hfill##\hfil$\crcr#2\crcr\sim\crcr}}}
\catcode`\@=12

\renewcommand{\(}{\begin{equation}}
\renewcommand{\)}{\end{equation}}

\def\pin#1{\int \frac{d#1}{2\pi}}
\def\ppin#1{\int\hspace{-17pt}\sum \frac{d#1}{2\pi}}
\def\ppink#1{\int\hspace{-17pt}\sum\frac{d^{#1}k}{(2\pi)^{#1}}}
\def\ppinkp#1{\int\hspace{-17pt}\sum\frac{d^{#1}k\p}{(2\pi)^{#1}}}
\def\dint{\int\hspace{-12pt}\sum }
\def\pink#1{\int \frac{d^{#1}k}{(2\pi)^{#1}}}

\def\pinkp#1{\int \frac{d^{#1}k\p}{(2\pi)^{#1}}}

\def\Bd#1{B^\ddag_{k_{#1}}}
\def\Bdp#1{B^\ddag_{k\p_{#1}}}

\def\cc{\mathcal{C}}
\def\df{\mathcal{D}_{f}}

\def\I{\mathcal{I}}

\def\os{\omega_S}

\def\blu#1{\textcolor{blue}{Jarah: #1}}
\def\red#1{\textcolor{red}{Hui: #1}}

\newcommand{\beas}{\begin{eqnarray*}}
\newcommand{\eeas}{\end{eqnarray*}}

\newcommand{\bquo}{\begin{quote}}
\newcommand{\enqu}{\end{quote}}

\def\lim#1{\stackrel{\rm{lim}}{{}_{#1}}}

\newcommand{\R}{{\mathbb R}}

\newcommand{\g}{\mathfrak g}

\def\ok#1{\omega_{k_{#1}}}
\def\okp#1{\omega_{k\p_{#1}}}
\def\okt#1{\omega_{\kt_{#1}}}
\def\V#1{V^{(#1)}(\sqrt{\lambda}f(x))}
\def\v#1{V^{(#1)}[f(x),x]}
\def\ck{\csch\left(\frac{\pi k}{2\b}\right)}

\def\mb{\mathcal{B}}
\def\mc{\mathcal{C}}
\def\md{\mathcal{D}}
\def\me{\mathcal{E}}

\newcommand{\beq}{\begin{equation}}
\newcommand{\eeq}{\end{equation}}
\newcommand{\bea}{\begin{eqnarray}}
\newcommand{\eea}{\end{eqnarray}}

\newskip\humongous \humongous=0pt plus 1000pt minus 1000pt

\newif\ifdtup

\catcode`\@=11

\ifthenelse{\equal{\letter}{0}}{ 

\@addtoreset{equation}{section}
\def\theequation{\arabic{section}.\arabic{equation}}

\def\@normalsize{\@setsize\normalsize{15pt}\xiipt\@xiipt
\abovedisplayskip 14pt plus3pt minus3pt%
\belowdisplayskip \abovedisplayskip
\abovedisplayshortskip \z@ plus3pt%
\belowdisplayshortskip 7pt plus3.5pt minus0pt}

\def\small{\@setsize\small{13.6pt}\xipt\@xipt
\abovedisplayskip 13pt plus3pt minus3pt%
\belowdisplayskip \abovedisplayskip
\abovedisplayshortskip \z@ plus3pt%
\belowdisplayshortskip 7pt plus3.5pt minus0pt
\def\@listi{\parsep 4.5pt plus 2pt minus 1pt
      \itemsep \parsep
      \topsep 9pt plus 3pt minus 3pt}}

\relax

\def\section{\@startsection{section}{1}{\z@}{3.5ex plus 1ex minus  .2ex}{2.3ex plus .2ex}{\large\bf}}

\def\thesection{\arabic{section}}
\def\thesubsection{\arabic{section}.\arabic{subsection}}

\def\appendix{\setcounter{section}{0}
 \def\thesection{Appendix \Alph{section}}
 \def\thesubsection{\Alph{section}.\arabic{subsection}}
 \def\theequation{\Alph{section}.\arabic{equation}}}
\renewcommand{\theequation}{\arabic{section}.\arabic{equation}}

}{
\renewcommand{\theequation}{\arabic{equation}}

} 

\begin{document}
\def\thefootnote{\fnsymbol{footnote}}
\def\thetitle{The Reflection Coefficient of a Reflectionless Kink}
\def\auttwo{Hui Liu}
\def\autone{Jarah Evslin}
\def\affc{Institute of Physics, Polish Academy of Sciences, Aleja Lotników 32/46, 02-668 Warsaw, Poland}
\def\affb{University of the Chinese Academy of Sciences, YuQuanLu 19A, Beijing 100049, China}
\def\affa{Institute of Modern Physics, NanChangLu 509, Lanzhou 730000, China}


\ifthenelse{\equal{\pr}{1}}{
\title{\thetitle}
\author{\autone}
\author{\auttwo}
\author{\autthree}
\affiliation {\affa}
\affiliation {\affb}

}{

\begin{center}
{\large {\bf \thetitle}}

\bigskip

\bigskip


{\large \noindent  \autone{${}^{1,2}$} \footnote{jarah@impcas.ac.cn} 
and \auttwo{${}^{3}$} \footnote{hliu@ifpan.edu.pl}
}


\vskip.7cm

1) \affa\\
2) \affb\\
3) \affc\\

\end{center}

}

\begin{abstract}
\noindent
Classically, reflectionless kinks transmit all incident radiation.  Recently, we have used an analyticity argument together with a solution of the Lippmann-Schwinger equation to write down the leading quantum correction to the reflection probability.  The argument was fast but rather indirect.  In the present paper, we calculate the reflection coefficient and probability by methodically grinding through the Schrodinger picture time evolution.  We find the same answer.  This answer contains contributions not considered in the traditional calculation of meson-kink scattering in 1991. However, as a result of these contributions, our total result is zero in the case of the Sine-Gordon model, and so is consistent with integrability.

\end{abstract}

%
\setcounter{footnote}{0}
\renewcommand{\thefootnote}{\arabic{footnote}}

\ifthenelse{\equal{\pr}{1}}
{
\maketitle
}{}

\section{Introduction}

The understanding of the interactions of solitons with perturbative excitations has many potential applications, from searches for cosmic strings in the cosmic microwave \cite{stringcmb} and gravity wave \cite{stringgw} backgrounds to soliton-soliton scattering, where soliton-bulk interactions play a key role \cite{mech22,multi22,dorey23,nav23,hahne23}.  

At tree-level, these interactions have long been understood \cite{fk78}.  However, there is reason to believe that quantum corrections qualitatively change the situation, as is thought to be the case for the oscillon \cite{hertzberg,tanmay} and Q-ball \cite{qball14} lifetimes and dynamics \cite{qball23}.  This is because, in the quantum theory, the leading quantum corrections appear to make reflectionless kinks reflect perturbative mesons.  The leading quantum corrections to the scattering of kinks with mesons were studied in a series of papers \cite{kaw89,uehara90,hayashi90} culminating in Ref.~\cite{uehara91}.  Recently, in Ref.~\cite{elas1}, we have used the Lippmann-Schwinger equations to provide a quick derivation of the one-loop quantum corrections to the elastic scattering amplitude.  The result did not agree with Ref.~\cite{uehara91}.  At least some of the differences are due to the fact that some terms were explicitly dropped in Ref.~\cite{uehara91} as they were considered to be loop corrections, however we have shown that in the case of the Sine-Gordon theory, these terms in fact cancel other terms of the form of those that were kept and this cancellation is in fact a consequence of the integrability of the model.

Our derivation made several assumptions about analyticity, and ignored final states that did not correspond to elastic scattering.  While the Sine-Gordon theory did provide a valuable check of our results, more general models possess a cubic coupling at the minima which yields interactions far from the kink that are not present in the Sine-Gordon model.  This, together with the fact that our result disagrees with the standard result of Ref.~\cite{uehara91}, motivates an independent and robust recalculation of this scattering amplitude.

The present paper does just this.  We provide a derivation of the amplitude in gory detail by considering an initial meson wave packet incident on a kink and evolving it in time, evaluating every contributing diagram up to second order in the coupling constant.  

This is done using the linearized soliton perturbation theory of Refs.~\cite{mekink,me2loop}, reviewed in Sec.~\ref{revsez}.  It is a Hamiltonian approach, which uses a decomposition of the fields in normal modes following Ref.~\cite{cahill76}.  In particular, no collective coordinate is introduced, removing many of the complications present in traditional approaches \cite{gs74,tom75}.  The transition from a Hamiltonian to a kink Hamiltonian, central to all approaches to quantum solitons since Ref.~\cite{dhn2}, takes the form of a passive unitary transformation on the regularized theory.  This is in contrast with previous approaches, which regularize the vacuum and kink sectors separately and then need to introduce an arbitrary and often inconsistent matching condition for the regulators \cite{rebhan}.

In Sec.~\ref{calcsez} we calculate all contributions to the scattering amplitude not involving zero modes.  The pieces of the final state containing zero modes are fixed by translation-invariance~\cite{me2loop}.  However, there are contributions to the amplitude involving processes in which zero modes are created and then are absorbed by the free evolution of the kink center of mass.  These are the hardest to calculate.  In Secs.~\ref{foursez} and \ref{twosez} we methodically calculate the final states containing four and two zero modes.  These are, as expected, determined by translation-invariance.  However, in Sec.~\ref{zerosez} we show that these calculations can be easily modified to generate the final states that have no zero modes, but arise from intermediate states involving zero modes.  This provides the final contribution to the elastic scattering amplitude.  

The contributions found here agree precisely with those of Ref.~\cite{elas1}.  This suggests that, in the future, long calculations such as that of the present paper may be unnecessary.  One may simply read the amplitudes off of the solution to the Lippmann-Schwinger equations, as was done in Ref.~\cite{elas1}.

\section{Review} \label{revsez}

\subsection{The Theory}

A number of efficient formalisms are available for treating quantum solitons.  At one loop, as reviewed in Refs.~\cite{goldhaber04,gw22}, reliable and efficient spectral methods have long been available.  Recently a classical-quantum correspondence has been introduced in Refs.~\cite{cq1,cq2} that cannot treat nonlinearities, but has been applied even well beyond the perturbative regime \cite{alb23}.  However, elastic scattering occurs at the next order, and so these formalisms will not be suitable.  

We will instead use linearized soliton perturbation theory.  Linearized soliton perturbation theory was developed at one loop in Ref.~\cite{mekink} and beyond in Ref.~\cite{me2loop}.  So far it has only been applied to 1+1 dimensional models of a scalar field $\phi(x)$ and its conjugate $\pi(x)$
\beq
H=\int d x: \mathcal{H}(x):_a, \quad \mathcal{H}(x)=\frac{\pi^2(x)}{2}+\frac{\left(\partial_x \phi(x)\right)^2}{2}+\frac{V(\sqrt{\lambda} \phi(x))}{\lambda}
\eeq
because in these models all ultraviolet divergences are removed by the normal ordering $::_a$.  However, the formalism is also compatible with a cutoff regularization and counterterms \cite{andy} and so we feel that it can be generalized to more interesting models.

The potential $V$ is required to have degenerate minima, so that there will be classical kink solutions $\phi(x,t)=f(x)$.  We specialize to the case of reflectionless kinks, however we have shown in Ref.~\cite{memult} that calculations such as those that follow are effortlessly generalized to reflective kinks.  In the present context, the leading quantum contribution to the reflection probability would arise from cross terms between the amplitude calculated here, adjusted as in Ref.~\cite{memult}, and the leading order amplitude \cite{fk78,refl}.

We will expand perturbatively in the coupling constant $\lambda$.  In Refs.~\cite{memult,stokes} we have seen that meson multiplication and Stokes scattering occur at order $O(\sl)$ in the amplitude.  We will see that elastic scattering amplitudes begin at order $O(\lambda)$.

The normal ordering will be defined at mass $m$, which in turn is defined by
\beq
m^2=V^{(2)}(\sqrt{\lambda} f(\pm \infty))\hsp
V^{(n)}(\sqrt{\lambda} \phi(x))=\frac{\partial^n V(\sqrt{\lambda} \phi(x))}{\partial( \sqrt{\lambda} \phi(x))^n}
\eeq
where the masses $V^{(2)}(\sqrt{\lambda} f(\infty))$ and $V^{(2)}(\sqrt{\lambda} f(-\infty))$ need to agree in order for a stationary kink state to exist \cite{wstabile}.

\subsection{States and Sectors}

The field $\phi(x)$ has perturbative excitations.  As usual, these are created and destroyed by operators $A^\dag$ and $A$ that are in turn constructed by decomposing $\phi(x)$ and $\pi(x)$ into plane waves.  This is to be expected, as plane waves are the solutions of the linearized classical equations of motion.  We refer to such perturbative excitations as mesons.  The Fock space consisting of the vacuum plus some finite number of mesons will be called the vacuum sector.

In the presence of a kink, the linearized equations of motion become the Sturm-Liouville equation
\beq
\V{2}{\g}(x)=\omega^2{\g}(x)+{\g}^{\prime\prime}(x)\hsp \phi(x,t)=f(x)+e^{-i\omega t}\g(x). \label{sl}
\eeq
The solutions to this equation are normal modes $\g(x)$.  Normal modes can be divided into three categories, depending on their frequency $\omega$.  First, there is a single zero mode
\beq
\g_B(x)=-\frac{f\p(x)}{\sqrt{Q_0}}
\eeq
with frequency $\omega_B=0$.  Here $Q_i$ is the order $O(\lambda^{i/2-1})$ quantum correction to the kink mass, so $Q_0$ is just the classical kink mass.  Second, for every real number $k$ there is a continuum mode with $\ok{}=\sqrt{m^2+k^2}$.  Finally, there may be discrete, real shape modes $\g_S(x)$ with $0<\omega_S<m$.  We chose the convention $\g^*_k=\g_{-k}$ and fix the normalizations via 
\beq
\int dx |{\g}_{B}(x)|^2=1,\
\int dx {\g}_{k_1} (x) {\g}^*_{k_2}(x)=2\pi \delta(k_1-k_2),\ 
\int dx {\g}_{S_1}(x){\g}^*_{S_2}(x)=\delta_{S_1S_2}. \label{comp}
\eeq

Following Ref.~\cite{cahill76}, we may use the normal modes to decompose the Schrodinger picture fields
\bea
\phi(x) &=&\phi_0 \mathfrak{g}_B(x)+\ppin{k} \left(B_k^{\ddag}+\frac{B_{-k}}{2 \omega_k}\right) \mathfrak{g}_k(x) \label{dec}\\
\pi(x) &=&\pi_0 \mathfrak{g}_B(x)+i \ppin{k}\left(\omega_k B_k^{\ddag}-\frac{B_{-k}}{2}\right) \mathfrak{g}_k(x) \nonumber
\eea
where we have defined the shorthand
\beq
B_k^{\ddagger}=\frac{B_k^{\dagger}}{2 \omega_k}\hsp
B_{-S}=B_S\hsp \ppin{k}=\pin{k}+\sum_S.
\eeq

The canonical commutation relations satisfied by $\phi(x)$ and $\pi(x)$ imply that $\phi_0,\ \pi_0,\ B$\ and $B^\ddag$ satisfy the algebra
\beq
\left[\phi_0, \pi_0\right]=i, \quad\left[B_{S_1}, B_{S_2}^{\ddagger}\right]=\delta_{S_1 S_2}, \quad\left[B_{k_1}, B_{k_2}^{\ddagger}\right]=2 \pi \delta\left(k_1-k_2\right).
\eeq
The interpretation of these new operators is straightforward.  In states with a kink, the operator $B_k^\ddag$ creates a continuum normal mode, which we also call a meson.  The operator $B^\ddag_S$ excites an internal shape mode.  The operators $\phi_0$ and $\pi_0$ correspond to the position and momentum of the kink's center of mass.

We refer to the kink ground state plus any number of mesons and shape modes with any wave function composed of $\phi_0$ as a kink sector state.  

\subsection{The Kink Sector}

How do we construct a kink sector state?  In classical field theory, vacuum sector states correspond to fields $\phi(x,t)$ which are close to a minimum of the potential, which we take be zero, while kink sector states correspond to $\phi(x,t)$ close to $f(x)$.  Thus one can turn a vacuum sector state into a kink sector state by shifting $\phi(x,t)\rightarrow\phi(x,t)+f(x)$.   

In quantum field theory, one needs to be careful because such a shift may be incompatible with the regularization \cite{lit}.  Instead, we will work directly in the regularized theory and will, as described below, make use of the unitary displacement operator
\beq
\df={{\rm Exp}}\left[-i\int dx f(x)\pi(x)\right].
\eeq
In the absence of a momentum cutoff, this indeed shifts the field.

The key observation is that acting the operator $\df$ on a vacuum sector state yields a kink sector state, and all kink sector states can be constructed in this way.  Indeed, this is just the old coherent state construction of soliton states \cite{taylor78,co23}.  For example, we may write the soliton ground state as $\df\vac$ where $\vac$ is some state in the vacuum sector, and a Hamiltonian eigenstate with one soliton and one meson as $\df|k_1\rangle$ where $|k_1\rangle$ is another vacuum sector state.

The appearance of a $\df$ factor in every state is annoying, and so we will remove it with a passive transformation.  We stress that this passive transformation is a convenience, merely relabeling the coordinates on the Hilbert space.  The passive transformation is defined as follows.  

We define a {\it{frame}} to be an identification of Hilbert space (projective) vectors with states.  The usual identification of Hilbert space vectors with states is called the {\it{defining frame}}.  Then we define the {\it{kink frame}} as follows.  In the kink frame, the Hilbert space vector $|\psi\rangle$ is identified with the state that is identified with the Hilbert space vector $\df|\psi\rangle$ in the defining frame.  In other words, $|\psi\rangle$ in the kink frame is just our old state $\df|\psi\rangle$ without bothering to write the $\df$.  So in the kink frame we write $\vac$ for the kink ground state and $|k_1\rangle$ for a state with one kink and one meson.

Of course, as is always the case with passive transformations, one needs to simultaneously transform the operators that act on the states.  For example, in the kink frame, time evolution and spatial translations are generated by the kink Hamiltonian and momentum
\beq
H\p=\df^\dag H\df\hsp
P\p=\df^\dag P\df
. \label{df}
\eeq
These are easily evaluated.  The kink momentum is
\beq
P\p=\sqrt{Q_0}\pi_0+P\hsp 
P=-\int dx \pi(x)\partial_x \phi(x)
\eeq
where the $\pi_0$ term is the momentum of the kink center of mass while $P$ represents the momentum in the mesons.  The kink Hamiltonian is
\beq
H\p=\sum_{n=0}^\infty H\p_n\hsp
H\p_0=Q_0\hsp H\p_1=0\hsp
H\p_{n>2}=\lambda^{\frac{n}{2}-1}\int dx \frac{V^{(n)}(\sqrt{\lambda} f(x))}{n !}: \phi^n(x):_a\label{hn}
\eeq
where $H\p_n$ is of order $O(\lambda^{n/2-1})$.  We will write $H\p_2$ momentarily.

\subsection{The Perturbation Theory}

What have we gained by decomposing kink sector states into $\df|\psi\rangle$ and then dropping the $\df$?  The main advantage of this formalism is that $|\psi\rangle$ may be found perturbatively using the eigenvalue equation for $H\p$.  This is the main advantage of linearized perturbation theory, the nonperturbative problem of finding the kink states becomes entirely perturbative.  Similarly, Schrodinger picture time evolution may be performed perturbatively using $e^{-iH\p t}$.

The perturbation theory begins with the free part of the kink Hamiltonian
\begin{equation}
H\p_2=Q_1+H_{\text {free }}, \quad H_{\text {free }}=\frac{\pi_0^2}{2}+ \ppin{k} \omega_k B_k^{\ddag} B_k. \label{h2}
\end{equation}
Recall that $Q_1$ is a scalar, it is just the one-loop correction to the kink mass.  The $\pi_0^2/2$ term is the kinetic energy of the kink center of mass while the other terms are quantum harmonic oscillators for the shape and continuum modes.  We will always work in the center of mass frame.  The ground state $\vac_0$ of the free Hamiltonian is the quantum field theory state which is the ground state of all of these quantum mechanical models, in other words it is the unique state that satisfies
\beq
\pi_0\vac_0=B_k\vac_0=B_S\vac_0=0. \label{v0}
\eeq

We can write any state in the kink sector by applying creation operators $B^\ddag$ and zero modes $\phi_0$ to this state.  $B^\ddag$ converts $H\p_2$ eigenstates into other $H\p_2$ eigenstates, which we will denote with a subscript $0$
\beq
\Bd{1}\cdots\Bd{n}\vac_0=|k_1\cdots k_n\rangle_0. \label{i0}
\eeq

We are interested not in eigenstates $|k_1\cdots k_n\rangle_0$ of the free Hamiltonian $H\p_2$, but rather in eigenstates $|k_1\cdots k_n\rangle$ of the full Hamiltonian $H\p$.  To find these, perturbatively, we decompose them in powers of the coupling
\beq
|k_1\cdots k_n\rangle=\sum_{i=0}^\infty |k_1\cdots k_n\rangle_i
\eeq
where $|k_1\cdots k_n\rangle_i$ is of order $O(\lambda^{i/2})$ when expanded in the basis that we will describe shortly.  The perturbative expansion starts with the approximation $i=0$ given in Eq.~(\ref{i0}).

As the Hamiltonian is translation-invariant, we may specialize to states that are translation-invariant.  In other words, we are only interested in states annihilated by $P\p$.  Now the states are described by a wave function in the kink center of mass position $\phi_0$, but translation-invariance means that if we find the part of a state near\footnote{This crude notation means that we decompose the state into eigenvalues of $\phi_0$ and then consider components with eigenvalues close to zero.  It is explained more precisely in Ref.~\cite{menorm}.} $\phi_0=0$, then we can use translation-invariance to reconstruct it elsewhere.  Thus we expand about $\phi_0=0$.  In terms of operators, this means that we consider a polynomial expansion in $\phi_0$, which is a good approximation for the part of the state near the zero eigenvalue of $\phi_0$.  In summary, a basis of states is given by
\beq
\phi_0^m\Bd{1}\cdots\Bd{n}\vac_0.
\eeq

We refer to the part of a state\footnote{We use the letter $m$ as both a nonnegative integer index counting zero modes and as a real, positive number describing the meson mass.} with $m=0$ as the primary part and the $m>0$ part as the descendants.  In Ref.~\cite{me2loop} we showed that all of the descendants are determined by translation invariance $P\p|\psi\rangle=0$.  Therefore we only use perturbation theory to determine the primaries.  

The last ingredient that we will need for our perturbative treatment is Wick's theorem~\cite{mewick}, which relates the normal ordering $::_a$ to a normal ordering $::_b$ in which $\pi_0$ and $B_k$ appear at the end
\bea
:\phi^j(x):_a&=&\sum_{m=0}^{\lfloor{\frac{j}{2}}\rfloor}\frac{j!}{2^m m!(j-2m)!}\I^m(x):\phi^{j-2m}(x):_b\label{wick}\\
\I(x)&=&\pin{k}\frac{\left|\g_{k}(x)\right|^2-1}{2\omega_k}+\sum_S\frac{\left|{\g}_{S}(x)\right|^2}{2\omega_S}.\nonumber
\eea
The contraction factor $\I(x)$ will be represented pictorially below as a loop that begins and ends at the same vertex.  This theorem lets us convert the formula (\ref{hn}) for the interactions in the kink Hamiltonian into the formulas that will appear in the text.

\section{Contributions with No Zero Modes} \label{calcsez}

We are interested in the following process.  Meson 1 strikes the kink from the left.  An interaction occurs at order $O(\lambda)$ and meson 2 leaves the kink, again to the left.  The initial and final states both contain a single unexcited kink and a single meson.

\subsection{Generalities}
\subsubsection{Initial Condition}

More precisely, our system begins in the state
\beq
|t=0\rangle=\pin{k_1} e^{-\sigma^2(k_1-k_0)^2-i(k_1-k_0)x_0}|k_1\rangle \label{wp}
\eeq
where meson 1 is centered at a position $x_0<0$ relative to the kink in a wave packet of width $\sigma$ and average momentum $k_0>0$.  Recall that $|k_1\rangle$ is the translation-invariant $H\p$ eigenstate consisting of a single kink and a single meson with momentum $k_1$.  It is invariant under simultaneous translations of the kink and the meson, preserving their separation.  The state was constructed explicitly up to order $O(\lambda)$ in Ref.~\cite{menorm}.

We will be interested in the limits
\beq
\frac{x_0}{\sigma}\rightarrow -\infty\hsp
m\sigma\rightarrow\infty.
\eeq
The first limit states that the initial meson wave packet does not overlap with the kink while the second limit states that the wave packet is nearly monochromatic.  

As $|k_1\rangle$ is a translation-invariant Hamiltonian eigenstate,  $|t=0\rangle$ is also translation-invariant.  However it is not a Hamiltonian eigenstate, as each $|k_1\rangle$ has a different eigenvalue.  The Hamiltonian and momentum commute $[H,P]=[H\p,P\p]=0$ and so, evolving in time, the state will remain translation-invariant.



\subsubsection{Evolution Operator}

The Schrodinger picture evolution operator is
\beq
U(t)=e^{-iH\p t}=\sum_{n=0}^\infty  U_n(t).
\eeq
Here we have decomposed it into the order $O(\lambda^{n/2})$ contributions $U_n$.  Up to order $O(\lambda)$ these are
\bea
U_0(t)&=&e^{-iH\p_2 t}\hsp U_1(t)=-i \int_{0}^t d\tau_1 e^{-iH\p_2(t-\tau_1)}H\p_3 e^{-iH\p_2\tau_1 }\\
U_2(t)&=&-i \int_{0}^t d\tau_1 e^{-iH\p_2(t-\tau_1)}H\p_4 e^{-iH\p_2\tau_1 }- \int_{0}^t d\tau_1\int_{0}^t d\tau_2 e^{-iH\p_2(t-\tau_2)}H\p_3 e^{-iH\p_2(\tau_2-\tau_1) }H\p_3 e^{-iH\p_2\tau_1 }.
\nonumber
\eea

We will define $x_t$, which, before the collision, is the meson's position at time $t$, and also $t_c$, the collision time, by
\beq
x_t=x_0+\frac{k_0}{\ok 0} t\hsp t_c=-\frac{\ok 0}{k_0}x_0. \label{xtdef}
\eeq
We will be interested in the limit $(t-t_c)/\sigma\rightarrow\infty$, so that by the end of the experiment meson 2 is far from the kink.

As $m\sigma\rightarrow\infty$, in the support of the Gaussian $e^{-\sigma^2(k_1-k_0)^2}$ we may approximate $k_1\sim k_0$ and so linearly expand
\beq
\ok 1=\ok 0+\frac{k_0}{\ok 0}(k_1-k_0). \label{ok}
\eeq

\subsection{One Interaction}

\begin{figure}[htbp]
\centering
\includegraphics[width = 0.6\textwidth]{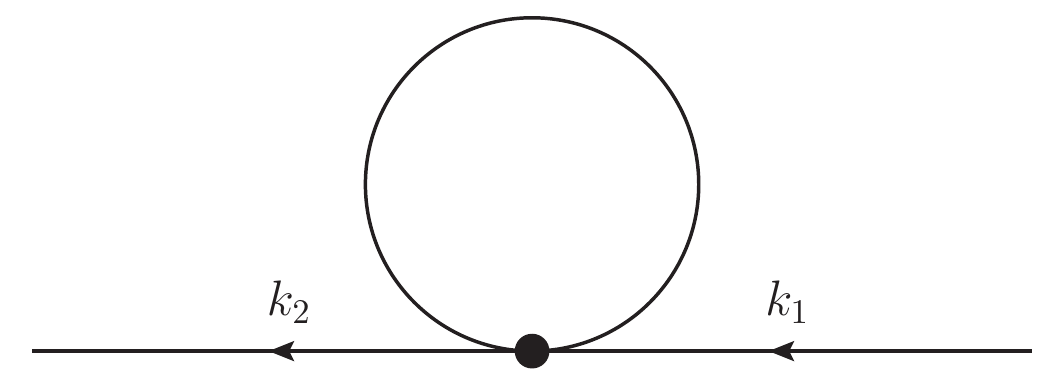}
\caption{Time runs to the left.  This is a schematic drawing of the following process.  Meson 1 travels.  Then it emits and absorbs the same virtual particle and in the process becomes meson 2.  This interaction is proportional to $\I(x)$ which falls exponentially in $mx$ far from the kink, thus the interaction necessarily happens close to the kink.  However the kink is not drawn.} \label{figb}
\end{figure}

The simplest process that leads to elastic scattering is drawn in Fig.~\ref{figb}.  Meson 1, with momentum $k_1$, interacts via the interaction
\beq
H_4^{(1)\prime}=\frac{\lambda}{2}\pink{2}V_{\I-k_1k_2}\Bd 2\frac{B_{k_1}}{2\ok 1}\label{h41}
\eeq
at time $\tau_1$.  Here $H_4^{(1)\prime}$ is a term in $H\p_4$. This interaction involves a virtual meson which it both creates and annihilates, and it leaves meson 2, with momentum $k_2$.  Each loop at the same vertex gives a factor of the function $\I(x)$.

Here we have used the shorthand $V$ to denote an $n$-point function defined as follows
\bea
V_{k_1\cdots k_n}&=&\int d x \V{n} \g_{k_1}(x)\cdots \g_{k_n}(x)\\
V_{\I k_1\cdots k_{n-2}}&=&\int d x \V{n} \I(x) \g_{k_1}(x)\cdots \g_{k_{n-2}}(x)\nonumber
\eea
where we remind the reader that the loop factor $\I(x)$ was defined in Eq.~(\ref{wick}).

This interaction is proportional to $\lambda$ already, and so a final state proportional to $\lambda$ may only arise if one acts it on a state of order $O(\lambda^0)$.  In other words, we must act it on the leading order term of the Hamiltonian eigenstate $|k_1\rangle$
\beq
|k_1\rangle_0=\Bd 1 \vac_0.
\eeq
This is not a Hamiltonian eigenstate, but it is an eigenstate of the free Hamiltonian $H\p_2$.

Acting the interaction (\ref{h41}) on $|k_1\rangle_0$ one finds
\beq
H_4^{(1)\prime}|k_1\rangle_0=\frac{\lambda}{4\ok 1}\pin{k_2}V_{\I -k_1 k_2}|k_2\rangle_0. \label{h4ac}
\eeq
Our goal is to obtain $U(t)|t=0\rangle$.  Now we are ready to calculate one term, the contribution from the interaction $H^{(1)\prime}_4$. 

Let us write the corresponding part of the evolution operator as
\beq
U_2(t)=-i \int_{0}^t d\tau_1 e^{-iH\p_2(t-\tau_1)}H^{(1)\prime}_4 e^{-iH\p_2\tau_1 }. \label{u2aa}
\eeq
This is an abuse of our notation, as we have already defined $U_2(t)$ to be the complete evolution operator at order $O(\lambda)$ and (\ref{u2aa}) is just one term in $U_2(t)$, however it would be cumbersome to give separate names to every term in the evolution operator.

Evolving the initial state we find
\bea
U_2(t)|t=0\rangle&=&-i \int_{0}^t d\tau_1 e^{-iH\p_2(t-\tau_1)}H^{(1)\prime}_4 e^{-iH\p_2\tau_1 }\pin{k_1} e^{-\sigma^2(k_1-k_0)^2-i(k_1-k_0)x_0}|k_1\rangle_0\\
&=&-i \int_{0}^t d\tau_1 e^{-iH\p_2(t-\tau_1)}H^{(1)\prime}_4 \pin{k_1} e^{-\sigma^2(k_1-k_0)^2-i(k_1-k_0)x_0-i\ok 1 \tau_1 }|k_1\rangle_0\nonumber\\
&=&-i \int_{0}^t d\tau_1 e^{-iH\p_2(t-\tau_1)}H^{(1)\prime}_4 e^{-i\ok 0 \tau_1 }\pin{k_1} e^{-\sigma^2(k_1-k_0)^2-i(k_1-k_0)x_{\tau_1}}|k_1\rangle_0.\nonumber
\eea
Using (\ref{h4ac}) one finds
\bea
U_2(t)|t=0\rangle&=&-i\frac{\lambda}{4}\pink{2} \frac{V_{\I -k_1 k_2}}{\ok 1}\int_{0}^t d\tau_1 e^{-i\ok 2 (t-\tau_1)} e^{-i \ok 0 \tau_1}e^{-\sigma^2(k_1-k_0)^2-i(k_1-k_0)x_{\tau_1}}|k_2\rangle_0.\nonumber
\eea

Recall that, as $m\sigma\rightarrow\infty$, $k_1$ is very close to $k_0$ in the support of the Gaussian weight.  This means that $\ok 1$ is very close to $\ok 0$, and so we replace the $\ok 1$ in the denominator with $\ok 0$.  However we cannot do the same with phase factors of the form $k_1 x_0$, for example, because $|x_0|\gg \sigma$ and so this would create an error in the phase of order $x_0/\sigma$ which is very large.  In summary, we will make the approximations
\beq
\ok 1=\ok 0\hsp \g_{-k_1}(x)=\g_{-k_0}(x)e^{i(k_1-k_0)x}
\eeq
but we will not drop the $(k_1-k_0)x$ terms.  The second approximation comes from the fact that, for a reflectionless kink, $\g_k(x)$ consists of $e^{-ikx}$ times various terms that vary with respect to $k$ with a characteristic scale of order $O(m)$, which is much greater than $1/\sigma$ and so these terms may be considered to be constant over the width of the Gaussian $e^{-\sigma^2(k_1-k_0)^2}$.

This leaves
\bea
U_2(t)|t=0\rangle&=&-i\frac{\lambda}{4\ok 0}\pin{k_2} \int_{0}^t d\tau_1 e^{-i\ok 2(t-\tau_1)-i \ok 0\tau_1}\int dx \V4 \I(x) \g_{-k_0}(x) \g_{k_2}(x) \nonumber\\
&&\times \pin{k_1}e^{-\sigma^2(k_1-k_0)^2-i(k_1-k_0)(x_{\tau_1}-x)}|k_2\rangle_0\nonumber\\
&=&-i\frac{\lambda}{4\ok 0}\frac{\sqrt{\pi}}{2\pi\sigma}\pin{k_2} e^{-i\ok 2 t}\int_{0}^t d\tau_1 e^{-i(\ok 0-\ok 2)\tau_1}\nonumber\\
&&\times\int dx \V4 \I(x) \g_{-k_0}(x) \g_{k_2}(x) e^{-(x_{\tau_1}-x)^2/(4\sigma^2)}|k_2\rangle_0.
\eea
Now $\I(x)$ has its support at $x\sim O(1/m)$ and so $x/\sigma$ tends to $0$ in our limit.  So can we drop the $x/\sigma$ term in the Gaussian factor?  A shift in $x$ of order $O(1/m)$ would shift the dummy variable $x_{\tau_1}$ and so $\tau_1$ by of order $O(1/m)$ for relativistic mesons.  This would in turn shift the phase factor $e^{-i(\ok 0-\ok 2)\tau_1}$ by a phase of order $(\ok 0-\ok 2)/m$.  However, as we will see momentarily and is anyway clear from momentum conservation, $\ok 0$ and $\ok 2$ are quite close, differing by of order $O(1/\sigma)$, and so the corresponding phase shift would be of order $O(1/(m\sigma))$ which vanishes in our limit.

In conclusion, we may safely drop the $x$ from the Gaussian term, and so pull it out of the $x$ integral, leaving
\bea
U_2(t)|t=0\rangle&=&-i\frac{\lambda}{4\ok 0}\frac{\sqrt{\pi}}{2\pi\sigma}\pin{k_2} V_{\I k_0-k_2} e^{-i\ok 2 t}\int_{0}^t d\tau_1 e^{-x_{\tau_1}^2/(4\sigma^2)-i(\ok 0-\ok 2)\tau_1}|k_2\rangle_0\nonumber\\
&=&-i\frac{\lambda}{4 k_0}\pin{k_2} V_{\I k_0-k_2} e^{-i\ok 2 t}  e^{-\sigma^2(\ok 0-\ok 2)^2\ok{0}^2/k_0^2-i(\ok 0-\ok 2)t_c}|k_2\rangle_0.
\eea
The expression $(\ok 0-\ok 2)$ vanishes at $k_2=\pm k_0$ and so the Gaussian factor has two peaks.  The $k_2=k_0$ peak corresponds to forward scattering.  We are not interested in it, so we will drop it.  About the other peak we may use (\ref{ok}) to rewrite the $(\ok 0-\ok 2)$ terms as $(k_0+k_2)k_0/\ok 0$ and so
\bea
U_2(t)|t=0\rangle&=&-i\frac{\lambda}{4 k_0}\pin{k_2} e^{-i\ok 2 t}V_{\I k_0-k_2}  e^{-\sigma^2(k_0+k_2)^2+i(k_0+k_2)x_0}|k_2\rangle_0. \label{a1}
\eea

\subsection{A Tadpole} \label{tadsez}

\begin{figure}[htbp]
\centering
\includegraphics[width = 0.8\textwidth]{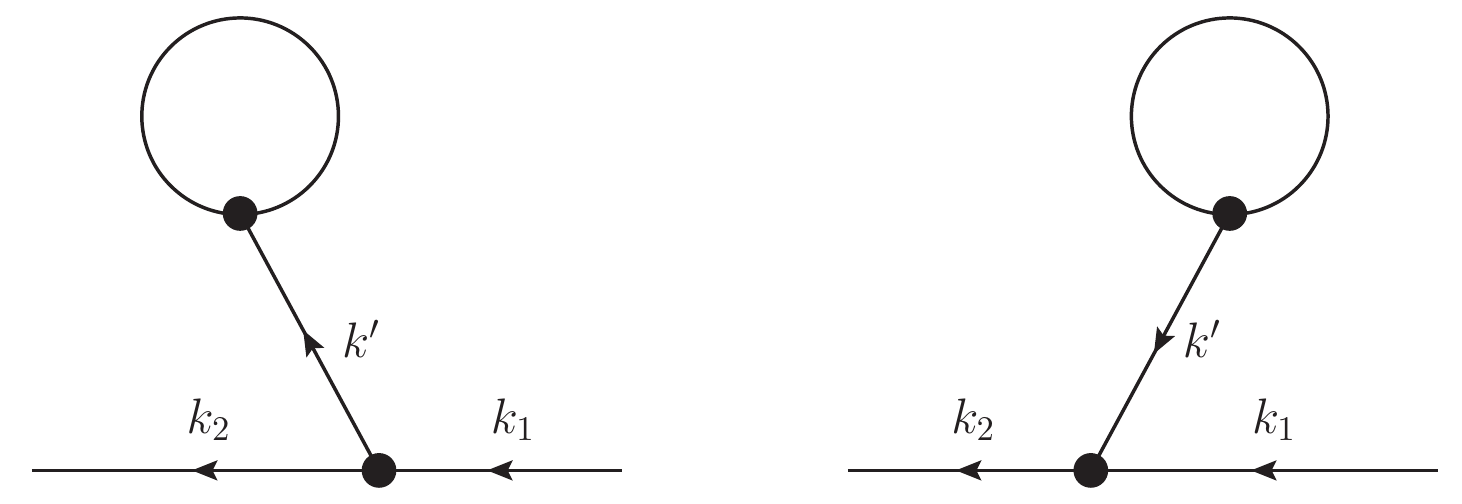}
\caption{In the right panel, while meson 1 approaches, a virtual particle pair comes in and out of existence, leaving behind it a virtual particle.  We will say that the virtual particle is created by a tadpole, although one might rightly note that it is emitted by the kink which is never drawn.  This virtual particle merges with meson 1, leaving meson 2.  In the left panel, meson 1 emits a virtual particle, becoming meson 2.  This emitted virtual particle decays via a virtual particle pair tadpole process.} \label{figc}
\end{figure}

All other contributions to elastic scattering involve two $H\p_3$ interactions.  In this subsection we will consider the interactions
\bea
H^{(1)\prime}_3&=&\frac{\sl}{2}\ppin{k_1}\ppin{k_2}\ppin{k\p} V_{-k_1k_2k\p}\Bd 2\left(\Bdp{}+\frac{B_{-k\p}}{2\okp{}}\right)\frac{B_{k_1}}{2\ok 1}\label{vkkk}\\
H^{(2)\prime}_3&=&\frac{\sl}{2}\ppin{k\p}V_{\I k\p} \left(\Bdp{}+\frac{B_{-k\p}}{2\okp{}}\right). \nonumber
\eea
In the interaction $H^{(1)\prime}_3$, at time $\tau_1$ the meson $k_1$ changes to $k_2$ and a virtual meson of momentum $k\p$ is emitted or absorbed.  At this point we allow both $k\p$ and also $k_2$ to be a continuum or a shape mode, since we do not yet know which will be the virtual meson.    In the tadpole interaction $H^{(2)\prime}_3$, at time $\tau_2$ the virtual meson is absorbed or emitted and another virtual meson travels in a loop to the same vertex.   Finally, we will restrict our attention to final states in which meson 2 is a continuum excitation.  This restriction is not really necessary, as it is not hard to show that if the final state consists, instead, of a kink and an excited shape mode that, since this cannot be on-shell, the amplitude vanishes.

As drawn in Fig.~\ref{figc}, the interactions may occur in either order.  The evolution operator is
\beq
U_2^A(t)=-\int_{0}^t d\tau_1\int_{\tau_1}^t d\tau_2e^{-iH\p_2(t-\tau_2)}H^{(2)\prime}_3e^{-iH\p_2(\tau_2-\tau_1)}H^{(1)\prime}_3e^{-iH\p_2\tau_1} \label{ev1}
\eeq
if $\tau_1<\tau_2$ and otherwise it is
\beq
U_2^B(t)=-\int_0^t d\tau_1 \int_0^{\tau_1}d\tau_2
e^{-iH\p_2(t-\tau_1)}H^{(1)\prime}_3e^{-iH\p_2(\tau_1-\tau_2)}H^{(2)\prime}_3e^{-iH\p_2\tau_2}. \label{ev2}
\eeq

\subsubsection{The Case $\tau_1<\tau_2$}

In this case, projecting out the three-meson sector and remembering the factor of two from the choice of contractions of $B_{k\p}$, the interaction terms act as
\bea
H^{(1)\prime}_3|k_1\rangle_0&=&\frac{\sl}{2}\ppin{k_2}\ppin{k\p} \frac{V_{-k_1k_2k\p}}{2\ok 1}|k_2k\p\rangle_0\label{vkkka}\\
H^{(2)\prime}_3|k_2k\p\rangle_0&=&\frac{\sl}{2} \frac{V_{\I -k\p}}{2\okp{}}|k_2\rangle_0+\frac{\sl}{2} \frac{V_{\I -k_2}}{2\ok 2}|k\p\rangle_0.\nonumber
\eea

As always when considering the leading contribution to the initial state, one begins at time $\tau_1$ with
\beq
e^{-iH\p_2\tau_1}|t=0\rangle_0=e^{-i\ok 0\tau_1}\pin{k_1}e^{-\sigma^2(k_1-k_0)^2-i(k_1-k_0)x_{\tau_1}}|k_1\rangle_0 \label{init}
\eeq
where $x_{\tau_1}$ is defined in the first expression in Eq.~(\ref{xtdef}).  At time $t$ this evolves to
\bea
U_2^A(t)|t=0\rangle_0&=&-\frac{\lambda}{8}\pin{k_2}\int_{0}^t d\tau_1\int_{\tau_1}^t d\tau_2 
\ppin{k\p}e^{-i\ok 0\tau_1-i\okp{}(\tau_2-\tau_1)-i\ok 2(t-\tau_1)}\label{tad}\\
&&\hspace{-2cm}\ \ \times\pin{k_1}e^{-\sigma^2(k_1-k_0)^2-i(k_1-k_0)x_{\tau_1}}\frac{V_{-k_1k_2k\p} V_{\I-k\p}}{\ok 1 \okp{}}|k_2\rangle_0\nonumber\\
&&\hspace{-2cm}=-\frac{\lambda}{8\ok 0}\pin{k_2}\int_{0}^t d\tau_1\int_{\tau_1}^t d\tau_2 
\ppin{k\p}\frac{V_{\I-k\p}}{ \okp{}}e^{-i\ok 0\tau_1-i\okp{}(\tau_2-\tau_1)-i\ok 2(t-\tau_1)}\nonumber\\
&&\hspace{-2cm}\ \ \times\int dx \V3 \g_{-k_0}(x)\g_{k_2}(x)\g_{k\p}(x)\pin{k_1}e^{-\sigma^2(k_1-k_0)^2-i(k_1-k_0)(x_{\tau_1}-x)}|k_2\rangle_0\nonumber\\
&&\hspace{-2cm}=-\frac{\lambda}{8\ok 0}\frac{\sqrt{\pi}}{2\pi\sigma} \pin{k_2}\int_{0}^t d\tau_1\int_{\tau_1}^t d\tau_2 
\ppin{k\p}\frac{V_{\I-k\p}}{ \okp{}}e^{-i\ok 0\tau_1-i\okp{}(\tau_2-\tau_1)-i\ok 2(t-\tau_1)}\nonumber\\
&&\hspace{-2cm}\ \ \times\int dx \V3 \g_{-k_0}(x)\g_{k_2}(x)\g_{k\p}(x)e^{-(x_{\tau_1}-x)^2/(4\sigma^2)}|k_2\rangle_0.\nonumber
\eea

\subsubsection{Showing that the First Interaction Occurs Near the Kink}

Unlike the previous process, the $x$ integrand no longer obviously has compact support unless $k\p$ is a shape mode.  To see that it in fact does have compact support, even if $k\p$ is not a shape mode, when integrated over $k\p$ and $\tau_2$, let us first multiply the integrand by a normalized bump function $e^{-(x-\hat x)^2/(4\hat\sigma^2)}/(2\sqrt{\pi}\hat\sigma)$
\bea
U_2^A(\hat x,t)|t=0\rangle_0&=&-\frac{\lambda}{8\ok 0}\frac{\sqrt{\pi}}{2\pi\sigma}\frac{\sqrt{\pi}}{2\pi\hat\sigma}\\
&&\hspace{-2cm}\ \ \times\pin{k_2}\int_{0}^t d\tau_1\int_{\tau_1}^t d\tau_2 
\ppin{k\p}\frac{V_{\I-k\p}}{ \okp{}}e^{-i\ok 0\tau_1-i\okp{}(\tau_2-\tau_1)-i\ok 2(t-\tau_1)}\nonumber\\
&&\hspace{-2cm}\ \ \times\int dx \V3 \g_{-k_0}(x)\g_{k_2}(x)\g_{k\p}(x)e^{-(x_{\tau_1}-x)^2/(4\sigma^2)-(x-\hat x)^2/(4\hat\sigma^2)}|k_2\rangle_0\nonumber
\eea
where $|\hat{x}|\gg \hat\sigma\gg 1/m$ and $\hat\sigma\ll\sigma$.  This will allow us to determine the contribution to the integral arising from $x\sim\hat{x}$.  We will now show that it vanishes for all $\hat{x}$ satisfying $|\hat{x}|\gg \hat\sigma\gg 1/m$.

As the $x$ integral now has support at $|x|\sim|\hat x|\gg 1/m$, we may replace
\beq
\V3 \g_{-k_0}(x)\g_{k_2}(x)\g_{k\p}(x)
\eeq
by its asymptotic value at $|x|\gg 1/m$.  In the case of classically reflectionless kinks, this is
\bea
\g_k(x)&=&\left\{\begin{tabular}{lll}
$\mb_ke^{-ikx}$&\rm{if} & $x\ll  -1/m$\\
$\md_ke^{-ikx}$&\rm{if} & $x\gg 1/m$\\
\end{tabular}
\right. \label{gk}\\
|\mb_k|^2&=&|\md_k|^2=1\hsp
\mb^*_k=\mb_{-k}\hsp
\md^*_k=\md_{-k}\nonumber
\eea
where the phases $\mb_k$ and $\md_k$ vary slowly with respect to $k$.  

For concreteness, choose $\hat x<0$, as the following argument proceeds identically with the other sign choice.  Then we replace  $\V3 \g_{-k_0}(x)\g_{k_2}(x)\g_{k\p}(x)$ with $V^{(3)L}_{-k_0k_2k\p}e^{-i(-k_0+k_2+k\p)x}$ where
\beq
V^{(3)L}_{-k_0k_2k\p}=V^{(3)}(\sl f(-\infty))\mb_{-k_0}\mb_{k_2}\mb_{k\p}.
\eeq
The support of the state near $\hat x$ is then
\bea
U_2^A(\hat x,t)|t=0\rangle_0&=&-\frac{\lambda}{8\ok 0}\frac{\sqrt{\pi}}{2\pi\sigma}\frac{\sqrt{\pi}}{2\pi\hat\sigma}\pin{k_2}\int_{0}^t d\tau_1\int_{\tau_1}^t d\tau_2 
\pin{k\p}e^{-i\ok 0\tau_1-i\okp{}(\tau_2-\tau_1)-i\ok 2(t-\tau_1)}\nonumber\\
&&\hspace{-2cm}\ \ \times\frac{V_{\I-k\p}}{ \okp{}}V^{(3)L}_{-k_0k_2k\p} \int dx  e^{-i(-k_0+k_2+k\p)x}e^{-(x_{\tau_1}-x)^2/(4\sigma^2)}e^{-(x-\hat x)^2/(4\hat\sigma^2)}|k_2\rangle_0.\nonumber
\eea
As $\hat\sigma\ll\sigma$, in the support of the bump function, we may replace $e^{-(x_{\tau_1}-x)^2/(4\sigma^2)}$ with $e^{-(x_{\tau_1}-\hat x)^2/(4\sigma^2)}$ and pull it out of the $x$ integral.  Then
\bea
U_2^A(\hat x,t)|t=0\rangle_0&=&-\frac{\lambda}{8\ok 0}\frac{\sqrt{\pi}}{2\pi\sigma}\pin{k_2}\int_{0}^t d\tau_1\int_{\tau_1}^t d\tau_2 
\pin{k\p}e^{-i\ok 0\tau_1-i\okp{}(\tau_2-\tau_1)-i\ok 2(t-\tau_1)}\nonumber\\
&&\hspace{-2cm}\ \ \times\frac{V_{\I-k\p}}{ \okp{}}V^{(3)L}_{-k_0k_2k\p} e^{-(x_{\tau_1}-\hat x)^2/(4\sigma^2)-\hat\sigma^2(-k_0+k_2+k\p)^2} e^{-i(-k_0+k_2+k\p)\hat x}|k_2\rangle_0.\nonumber
\eea
Now $k\p$ is close to $k_0-k_2$ as a result of the Gaussian $e^{-\hat\sigma^2(-k_0+k_2+k\p)^2}.$  Physically this is because the virtual meson is created at $\hat x$, which is far from the kink where mesons cannot transfer momentum to the kink.  This means that we may expand about $k\p\sim k_0-k_2$
\beq
\okp{}=\omega_{k_0-k_2}+\frac{k_0-k_2}{\omega_{k_0-k_2}}(-k_0+k_2+k\p).
\eeq

We then find
\bea
U_2^A(\hat x,t)|t=0\rangle_0&=&-\frac{\lambda }{8\ok 0}\frac{\sqrt{\pi}}{2\pi\sigma}\pin{k_2}V^{(3)L}_{-k_0,k_2,k_0-k_2}\int_{0}^t d\tau_1\int_{\tau_1}^t d\tau_2 
e^{-i\ok 0\tau_1-i\omega_{k_0-k_2}(\tau_2-\tau_1)-i\ok 2(t-\tau_1)}\nonumber\\
&&\hspace{-2cm}\ \ \times e^{-(x_{\tau_1}-\hat x)^2/(4\sigma^2)}\pin{k\p}\frac{V_{\I-k\p}}{ \omega_{k_2-k_0}} e^{-\hat\sigma^2(-k_0+k_2+k\p)^2} e^{-i(-k_0+k_2+k\p)\left(\hat x+\frac{k_0-k_2}{\omega_{k_0-k_2}}(\tau_2-\tau_1)\right)}|k_2\rangle_0\nonumber\\
&&\hspace{-2cm}=-\frac{\lambda}{8\ok 0}\frac{\sqrt{\pi}}{2\pi\sigma}\pin{k_2}V^{(3)L}_{-k_0,k_2,k_0-k_2}\int_{0}^t d\tau_1\int_{\tau_1}^t d\tau_2 
e^{-i\ok 0\tau_1-i\omega_{k_0-k_2}(\tau_2-\tau_1)-i\ok 2(t-\tau_1)}\nonumber\\
&&\hspace{-2cm}\ \ \times e^{-(x_{\tau_1}-\hat x)^2/(4\sigma^2)}\frac{1}{ \omega_{k_2-k_0}}\int dy V^{(3)}(\sl f(y)) \I(y)\g_{k_2-k_0}(y)\nonumber\\
&&\hspace{-2cm}\ \ \times\pin{k\p} e^{-\hat\sigma^2(-k_0+k_2+k\p)^2} e^{-i(-k_0+k_2+k\p)\left(\hat x-y+\frac{k_0-k_2}{\omega_{k_0-k_2}}(\tau_2-\tau_1)\right)}|k_2\rangle_0\nonumber\\
&&\hspace{-2cm}=-\frac{\lambda}{8\ok 0}\frac{\sqrt{\pi}}{2\pi\sigma}\frac{\sqrt{\pi}}{2\pi\hat\sigma}\pin{k_2}V^{(3)L}_{-k_0,k_2,k_0-k_2}\int_{0}^t d\tau_1\int_{\tau_1}^t d\tau_2 
e^{-i\ok 0\tau_1-i\omega_{k_0-k_2}(\tau_2-\tau_1)-i\ok 2(t-\tau_1)}\nonumber\\
&&\hspace{-4cm}\ \ \times e^{-(x_{\tau_1}-\hat x)^2/(4\sigma^2)}\frac{1}{ \omega_{k_2-k_0}}\int dy V^{(3)}(\sl f(y)) \I(y)\g_{k_2-k_0}(y) e^{-\left(\hat x-y+\frac{k_0-k_2}{\omega_{k_0-k_2}}(\tau_2-\tau_1)\right)^2/(4\hat\sigma^2)}|k_2\rangle_0.\nonumber
\eea

Now unlike $x$, which was the location of the first interaction, $y$, the location of the second interaction, must be close to the kink.  This is mandated by the $\I(y)$ term which has support at $y\sim O(1/m)$.  Therefore $y/\hat\sigma$ can be set to zero, implying that the corresponding Gaussian factor is $y$-independent and can be pulled out of the $y$ integral
\bea
U_2^A(\hat x,t)|t=0\rangle_0&=&-\frac{\lambda }{8\ok 0}\frac{\sqrt{\pi}}{2\pi\sigma}\frac{\sqrt{\pi}}{2\pi\hat\sigma}\pin{k_2}V^{(3)L}_{-k_0,k_2,k_0-k_2}\int_{0}^t d\tau_1\int_{\tau_1}^t d\tau_2 
e^{-i\ok 0\tau_1-i\omega_{k_0-k_2}(\tau_2-\tau_1)-i\ok 2(t-\tau_1)}\nonumber\\
&&\hspace{-2cm}\ \ \times e^{-(x_{\tau_1}-\hat x)^2/(4\sigma^2)} e^{-\left(\hat x+\frac{k_0-k_2}{\omega_{k_0-k_2}}(\tau_2-\tau_1)\right)^2/(4\hat\sigma^2)}\frac{V_{\I,k_2-k_0}}{ \omega_{k_2-k_0}}|k_2\rangle_0.\nonumber
\eea

Finally, consider the $\tau_2$ Gaussian integration.  Depending on the values of $\tau_1$ and $\hat{x}$, the range of integration may or may not overlap with the support of the second Gaussian factor.  If it does not overlap, this integral trivially vanishes.  If it does overlap, then it overlaps for a range of $\hat\sigma\omega_{k_0-k_2}/(k_0-k_2)>\hat\sigma$.  During this time, the phase $e^{-i\omega_{k_0-k_2}\tau_2}$ decreases by more than $\omega_{k_0-k_2}\hat\sigma>m\hat\sigma$ units.  Thus the integral yields a factor of less than $e^{-m^2\hat\sigma^2}$, which vanishes in our limit $m\hat\sigma\rightarrow \infty$.  We thus conclude that, including a bump function near $x=\hat x$, 
\beq
U_2^A(\hat x,t)|t=0\rangle_0=0
\eeq
for $|\hat x|\gg 1/m$.  In other words, there is no contribution to $U_2^A(t)|t=0\rangle_0$ from $x$ near $\hat{x}$.  As a result, the position $x$ of the first interaction is necessarily inside the kink $x\sim O(1/m)$, where the mesons and kink may exchange momentum.

To make this statement more quantitative, assume for a moment that the limit $|\hat{x}|/\sigma$ is nonzero.  As the limit $m\sigma$ tends to $\infty$, in this case $m\hat{x}$ also tends to $\infty$.  One therefore can choose $\hat\sigma$ so that $|\hat{x}|\gg \hat\sigma\gg 1/m$.  Now the results of this subsubsection imply that such a $\hat{x}$ does not contribute to the integral.  Thus, contributions to the integral can only arise when the limit of $|\hat{x}|/\sigma$ tends to zero.  In other words, the support of our original integral is at the limit $|x|/\sigma\rightarrow 0$, where we may drop the $x/\sigma$ term in the Gaussian exponential.

\subsubsection{Continuing with the Computation}

This long argument has been made to justify dropping the $x/\sigma$ term in Eq.~(\ref{tad}), as the
$x$ integral has support at $|x|\ll\sigma$
\bea
U_2^A(t)|t=0\rangle_0&=&-\frac{\lambda}{8\ok 0}\frac{\sqrt{\pi}}{2\pi\sigma}\pin{k_2}\int_{0}^t d\tau_1\int_{\tau_1}^t d\tau_2 
\ppin{k\p}\frac{V_{\I-k\p}}{ \okp{}}e^{-i\ok 0\tau_1-i\okp{}(\tau_2-\tau_1)-i\ok 2(t-\tau_1)}\nonumber\\
&&\times e^{-x_{\tau_1}^2/(4\sigma^2)}V_{-k_0k_2k\p}|k_2\rangle_0\nonumber\\
&=&-i\frac{\lambda}{8\ok 0}\frac{\sqrt{\pi}}{2\pi\sigma}\pin{k_2}e^{-i\ok 2 t}
\ppin{k\p}\frac{V_{-k_0k_2k\p}V_{\I-k\p}}{ \okp{}^2}\nonumber\\
&&\times \int_{0}^t d\tau_1 e^{-x_{\tau_1}^2/(4\sigma^2)}e^{-i(\ok 0-\ok 2)\tau_1}\left(e^{-i\okp{}(t-\tau_1)}-1
\right)|k_2\rangle_0\nonumber\\
&=&-i\frac{\lambda}{8k_0}\pin{k_2}e^{-i\ok 2 t}
\ppin{k\p}\frac{V_{-k_0k_2k\p}V_{\I-k\p}}{ \okp{}^2}
e^{-i(\ok 0-\ok 2)t_c}\nonumber\\
&&\times  \left(e^{-i\okp{}(t-t_c)}
e^{-\sigma^2\frac{\ok{0}^2}{k_0^2}(\ok 0-\ok 2-\okp{})^2}-e^{-\sigma^2\frac{\ok{0}^2}{k_0^2}(\ok 0-\ok 2)^2}
\right)|k_2\rangle_0.\label{guscio}
\eea

Consider the first term in the parenthesis.  This has support at $\ok 0\sim \ok 2+\okp{}$, where the virtual meson is on-shell.  In fact, it is unrelated to elastic scattering, instead it represents a quantum correction to meson multiplication.  

Now consider the $k\p$ integral of that term.  In the support of the Gaussian, $\okp{}$ may be expanded to  linear order in $k\p$ as in Eq.~(\ref{ok}).  Recall that the linear coefficient is the group velocity.    Then, the size of the support of the Gaussian factor is equal to $1/\sigma$ times the ratio of the $k_0$ to the $k\p$ velocities, which is of order unity.   Over this range, the phase $e^{-i\okp{}(t-t_c)}$ changes by of order $(t-t_c)/\sigma$.  This leads to a suppression factor of less than $e^{-(t-t_c)^2/\sigma^2}$ after $k\p$ integration, and so this term vanishes.  This argument of course does not apply if $k\p$ is a shape mode, in which case it is discrete.  We will turn to that case in \ref{stokesez}.

What about the second term in the parenthesis?  This has two peaks, at $k_2=\pm k_0$.  The positive sign corresponds to forward scattering, which we are not interested in here.  Therefore we keep the negative sign
\bea
U_2^A(t)|t=0\rangle_0&=&i\frac{\lambda}{8k_0}\pin{k_2}e^{-i\ok 2 t}
\ppin{k\p}\frac{V_{-k_0k_2k\p}V_{\I-k\p}}{ \okp{}^2}
e^{-\sigma^2(k_0+k_2)^2+i(k_0+k_2)x_0}
|k_2\rangle_0.\nonumber
\eea

\subsubsection{The Case $\tau_1>\tau_2$}

If the tadpole creates the virtual meson which is then absorbed by the incoming meson, then the interaction terms act as follows
\bea
H^{(2)\prime}_3|k_1\rangle_0&=&\frac{\sl}{2}\ppin{k\p}V_{\I k\p} |k_1k\p\rangle_0\\
H^{(1)\prime}_3|k_1k\p\rangle_0&=&\frac{\sl}{4\ok 1\okp{}}\ppin{k_2} V_{-k_1k_2-k\p}|k_2\rangle_0\nonumber
\eea
leading to the final state
\bea
U_2^B(t)|t=0\rangle_0&=&-\frac{\lambda}{8\ok 0}\frac{\sqrt{\pi}}{2\pi\sigma}\pin{k_2}\int_{0}^t d\tau_1\int_{0}^{\tau_1} d\tau_2 
\ppin{k\p}\frac{V_{\I-k\p}}{ \okp{}}e^{-i\ok 0\tau_1-i\okp{}(\tau_1-\tau_2)-i\ok 2(t-\tau_1)}\nonumber\\
&&\hspace{-2cm}\ \ \times\int dx \V3 \g_{-k_0}(x)\g_{k_2}(x)\g_{k\p}(x)e^{-(x_{\tau_1}-x)^2/(4\sigma^2)}|k_2\rangle_0.\label{initcorr}
\eea
The argument proceeds identically to the previous case, with a caveat at $\tau_2=0$ that we will return to, leading to the same result.  Summing them, yields a factor of two
\bea
\left(U_2^A(t)+U_2^B(t)\right)|t=0\rangle_0&=&i\frac{\lambda}{4k_0}\pin{k_2}e^{-i\ok 2 t}e^{-\sigma^2(k_0+k_2)^2+i(k_0+k_2)x_0}
\label{a2}\\
&&\times
\ppin{k\p}\frac{V_{-k_0k_2k\p}V_{\I-k\p}}{ \okp{}^2}
|k_2\rangle_0.\nonumber
\eea

\subsubsection{Initial State Corrections}

There are also two initial state corrections, corresponding intuitively to the case in which either of these interactions has occurred in the distant past.  More precisely, these correspond to the evolution of the $|k_1\rangle_1$ subleading term in the $|k_1\rangle$ in Eq.~(\ref{wp}).   The corresponding amplitudes are again of order $O(\lambda)$, but now the initial state is suppressed by a factor of $\sl$ while the evolution operator is order $O(\sl)$.  

We will not draw these, but given any diagram in this paper, one may arrive at the corresponding diagram for initial state corrections as follows.  First choose a time $\tau$.  Then remove the part of the diagram at earlier times $\tau\p<\tau$, corresponding to everything that appears to the right of the time $\tau$. 

In the first case, one considers a virtual meson in the meson cloud about the kink.  After a time $t_c$, the incoming meson strikes the virtual meson and creates the final meson.  The virtual meson contributes a phase factor of $e^{-i(\okp{}+\ok 0-\ok 2) t_c}$ which oscillates rapidly with respect to $k\p$ unless the $\ok 2=\ok 0+\okp{}$, corresponding to the limit in which the virtual meson is on-shell.  Like the first term in the parenthesis in Eq.~(\ref{guscio}), the $k\p$ integration over a domain of order $O(1/\sigma)$ leads to interference in the $e^{-i\okp{}t_c}$ phase which annihilates this correction.

The second initial state contribution arises from a quantum correction to the incoming meson, which consists of two mesons of momenta $k_2$ and $k_1-k_2$, one of which interacts with a virtual meson created by the kink once they arrive at the kink, after a time $\tau_2\sim t_c$.  One needs to integrate over $\tau_2$, and each value is weighted by a phase $e^{-i\omega_{k_0-k_2}\tau_2}$.  As $\omega_{k_0-k_2}>m$, one finds of order $mt_c$ oscillations, and so after integrating over $k\p$ this contribution is hopelessly suppressed. 

What if the virtual meson is a shape mode?  Then $k\p$ is discrete and cannot be integrated, so this argument fails.  The shape mode contribution to the meson cloud falls exponentially with the distance from the kink, so one can ignore the second initial state contribution.

What about the first initial state contribution?  What if the meson interacts with a virtual shape mode in the kink cloud?  In fact, whether the virtual meson is a shape mode or not, this initial state correction exactly cancels with the $\tau_2=0$ boundary term from the $\tau_2$ integration in Eq.~(\ref{initcorr}).

\subsection{A Bubble}

\begin{figure}[htbp]
\centering
\includegraphics[width = 0.9\textwidth]{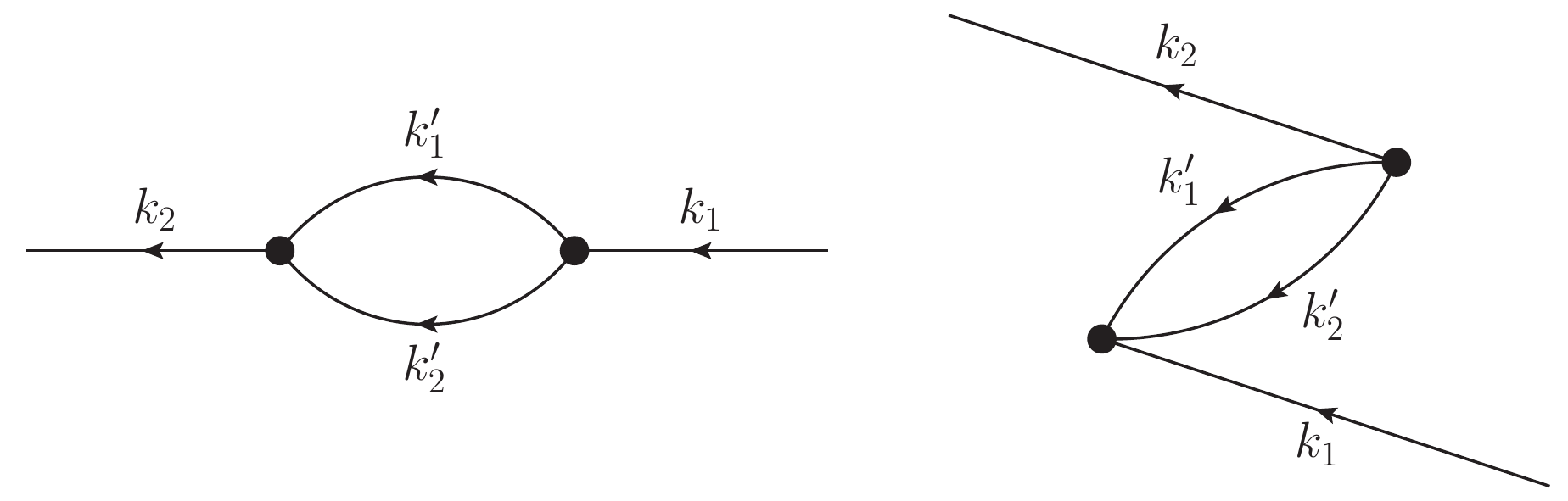}
\caption{In the left panel we see meson 1 splitting into two virtual mesons which recombine into meson 2.  In the right panel, two virtual mesons and meson 2 are created together, and later the two virtual mesons annihilate themselves together with meson 1.} \label{figd}
\end{figure}

The contribution that motivates our project is drawn in Fig.~\ref{figd}.  There are again two interactions.  At time $\tau_1$, the interaction
\beq
H^{(1)\prime}_3=\frac{\sl}{4}\ppin{k_1}\ppinkp{2} V_{-k_1k\p_1k\p_2}\left(\Bdp{1}\Bdp{2}+\frac{B_{-k\p_1}B_{-k\p_2}}{12\okp{1}
\okp{2}}\right)\frac{B_{k_1}}{\ok 1}
\eeq
connects the incoming meson 1 with two virtual mesons $1\p$ and $2\p$.  These might lie in the continuum, but they may also be shape modes, or perhaps one of each.  In particular, if both are shape modes, this corresponds to an unstable resonance.  Next, at time $\tau_2$, the interaction
\beq
H^{(2)\prime}_3=\frac{\sl}{6}\ppin{k_2}\ppinkp{2} V_{k_2k\p_1k\p_2}\Bd 2\left(\Bdp{1}\Bdp{2}+\frac{3B_{-k\p_1}B_{-k\p_2}}{4\okp{1}
\okp{2}}\right)
\eeq
connects the two virtual mesons to the outgoing meson 2.  We expect the amplitude to have a peak at the energy of the twice-excited shape mode.

\subsubsection{The Case $\tau_1<\tau_2$}

In this case, projecting out five-meson final states and remembering a factor of two from the choice of which annihilation operator annihilates which virtual meson, the interactions act as
\beq
H^{(1)\prime}_3|k_1\rangle_0=\frac{\sl}{4}\ppinkp{2} \frac{V_{-k_1k\p_1k\p_2}}{\ok 1}|k\p_1k\p_2\rangle_0\hsp
H^{(2)\prime}_3|k\p_1k\p_2\rangle_0=\frac{\sl}{4}\ppin{k_2}\frac{V_{k_2-k\p_1-k\p_2}}{\okp{1}
\okp{2}}|k_2\rangle_0.
\eeq

The evolution operator (\ref{ev1}) and the early state (\ref{init}) then yield
\bea
U_2^A(t)|t=0\rangle_0&=&-\frac{\lambda}{16}\pin{k_2}\int_{0}^t d\tau_1\int_{\tau_1}^t d\tau_2 
\ppinkp{2}e^{-i\ok 0\tau_1-i(\okp 1+\okp 2)(\tau_2-\tau_1)-i\ok 2(t-\tau_2)}\nonumber\\
&&\hspace{-3cm}\ \ \times\pin{k_1}e^{-\sigma^2(k_1-k_0)^2-i(k_1-k_0)x_{\tau_1}}\frac{V_{-k_1k\p_1k\p_2} V_{k_2-k\p_1-k\p_2}}{\ok 1 \okp 1\okp 2}|k_2\rangle_0\nonumber\\
&&\hspace{-3cm}=-\frac{\lambda}{16\ok 0}\pin{k_2}\int_{0}^t d\tau_1\int_{\tau_1}^t d\tau_2 
\ppinkp{2}\frac{V_{k_2-k\p_1-k\p_2}}{ \okp 1\okp 2}e^{-i\ok 0\tau_1-i(\okp 1+\okp 2)(\tau_2-\tau_1)-i\ok 2(t-\tau_2)}\nonumber\\
&&\hspace{-3cm}\ \ \times\int dx \V3 \g_{-k_0}(x)\g_{k\p_1}(x)\g_{k\p_2}(x)\pin{k_1}e^{-\sigma^2(k_1-k_0)^2-i(k_1-k_0)(x_{\tau_1}-x)}|k_2\rangle_0\nonumber\\
&&\hspace{-3cm}=-\frac{\lambda}{16\ok 0}\frac{\sqrt{\pi}}{2\pi\sigma}\pin{k_2}\int_{0}^t d\tau_1\int_{\tau_1}^t d\tau_2 
\ppinkp{2}\frac{V_{k_2-k\p_1-k\p_2}}{ \okp 1\okp 2}e^{-i\ok 0\tau_1-i(\okp 1+\okp 2)(\tau_2-\tau_1)-i\ok 2(t-\tau_2)}\nonumber\\
&&\hspace{-3cm}\ \ \times\int dx \V3 \g_{-k_0}(x)\g_{k\p_1}(x)\g_{k\p_2}(x)e^{-(x_{\tau_1}-x)^2/(4\sigma^2)}|k_2\rangle_0. \label{xpic}
\eea

\subsubsection{Showing that the First Interaction Occurs Near the Kink}

Again, we would like to drop the $x/\sigma$ term when $k\p_1$ and $k\p_2$ are both continuum modes so that the $x$ integrand does not have compact support.  In this subsubsection we will try to argue that, even when $k\p_1$ and $k\p_2$ are continuum modes, after performing the other integrals, the $x$ integral vanishes except when $x/\sigma$ tends to zero.  The argument will be similar to the tadpole case, but not quite the same.

As in the previous case, to see that it has compact support if integrated over $k\p$ and $\tau_2$, we multiply the integrand by the normalized bump function $e^{-(x-\hat x)^2/(4\hat\sigma^2)}/(2\sqrt{\pi}\hat\sigma)$ where $|\hat{x}|\gg \hat\sigma\gg 1/m$ and $\hat\sigma\ll\sigma$.  We also choose $\hat{x}<0$, promising the reader that the manipulations are identical in the case $\hat{x}>0$.

Again, this allows us to replace $\V3 \g_{-k_0}(x)\g_{k\p_1}(x)\g_{k\p_2}(x)$ with $V^{(3)L}_{-k_0k\p_1k\p_2}e^{-i(-k_0+k\p_1+k\p_2)x}$.  The support of the state near $\hat x$ is then
\bea
U_2^A(\hat x,t)|t=0\rangle_0&=&-\frac{\lambda }{16\ok 0}\frac{\sqrt{\pi}}{2\pi\sigma}\frac{\sqrt{\pi}}{2\pi\hat\sigma}\pin{k_2}\int_{0}^t d\tau_1\int_{\tau_1}^t d\tau_2 
\pinkp{2}\frac{V_{k_2-k\p_1-k\p_2}V^{(3)L}_{-k_0k\p_1k\p_2}}{ \okp 1\okp 2}\nonumber\\
&&\hspace{-3.8cm}\ \ \times e^{-i\ok 0\tau_1-i(\okp 1+\okp 2)(\tau_2-\tau_1)-i\ok 2(t-\tau_2)} \int dx  e^{-i(-k_0+k\p_1+k\p_2)x}e^{-(x_{\tau_1}-x)^2/(4\sigma^2)}e^{-(x-\hat x)^2/(4\hat\sigma^2)}|k_2\rangle_0\nonumber\\
&=&-\frac{\lambda}{16\ok 0}\frac{\sqrt{\pi}}{2\pi \sigma}\pin{k_2}\int_{0}^t d\tau_1\int_{\tau_1}^t d\tau_2 
\pinkp{2}\frac{V_{k_2-k\p_1-k\p_2}V^{(3)L}_{-k_0k\p_1k\p_2}}{ \okp 1\okp 2}\nonumber\\
&&\hspace{-3.8cm}\ \ \times e^{-i\ok 0\tau_1-i(\okp 1+\okp 2)(\tau_2-\tau_1)-i\ok 2(t-\tau_2)}  e^{-(x_{\tau_1}-\hat x)^2/(4\sigma^2)-\hat\sigma^2(-k_0+k\p_1+k\p_2)^2} e^{-i(-k_0+k\p_1+k\p_2)\hat x}|k_2\rangle_0.
\eea

Now $k\p_2$ is close to $k_0-k\p_1$ as a result of the Gaussian $e^{-\hat\sigma^2(-k_0+k\p_1+k\p_2)^2}.$  Again, this is because the virtual mesons are created at $\hat x$, which is far from the kink where mesons cannot transfer momentum to the kink.  Expanding $k\p_2$ about $k\p_2\sim k_0-k\p_1$
\beq
\okp{2}=\omega_{k_0-k\p_1}+\frac{k_0-k\p_1}{\omega_{k_0-k\p_1}}(-k_0+k\p_1+k\p_2).
\eeq

We then find
\bea \label{ybig}
U_2^A(\hat x,t)|t=0\rangle_0&=&-\frac{\lambda }{16\ok 0}\frac{\sqrt{\pi}}{2\pi \sigma}\pin{k_2}\int_{0}^t d\tau_1\int_{\tau_1}^t d\tau_2 e^{-(x_{\tau_1}-\hat x)^2/(4\sigma^2)}\\
&&\hspace{-3cm}\ \ \times
\pin{k\p_1}e^{-i\ok 0\tau_1-i(\okp 1+\omega_{k_0-k\p_1})(\tau_2-\tau_1)-i\ok 2(t-\tau_2)} \nonumber\\
&&\hspace{-3cm}\ \ \times \pin{k\p_2} \frac{V_{k_2-k\p_1-k\p_2}V^{(3)L}_{-k_0k\p_1k\p_2}}{ \okp 1\okp 2}e^{-\hat\sigma^2(-k_0+k\p_1+k\p_2)^2-i(-k_0+k\p_1+k\p_2)\left(\hat x+\frac{k_0-k\p_1}{\omega_{k_0-k\p_1}}(\tau_2-\tau_1)\right)}|k_2\rangle_0\nonumber\\
&&\hspace{-3cm}=-\frac{\lambda }{16\ok 0}\frac{\sqrt{\pi}}{2\pi \sigma}\pin{k_2}e^{-i\ok 2 t}\int_{0}^t d\tau_1\int_{\tau_1}^t d\tau_2 e^{-(x_{\tau_1}-\hat x)^2/(4\sigma^2)}\nonumber\\
&&\hspace{-3cm}\ \ \times
\pin{k\p_1}V^{(3)L}_{-k_0,k\p_1,k_0-k\p_1}\frac{e^{-i\ok 0\tau_1-i(\okp 1+\omega_{k_0-k\p_1})(\tau_2-\tau_1)+i\ok 2\tau_2}}{ \okp 1\omega_{k_0-k\p_1}}\nonumber\\
&&\hspace{-3cm}\ \ \times\int dy  V^{(3)}(\sl f(y)) \g_{k_2}(y)\g_{-k\p_1}(y)\g_{k\p_1-k_0}(y) \nonumber\\
&&\hspace{-3cm}\ \ \times 
\pin{k\p_2} e^{-\hat\sigma^2(-k_0+k\p_1+k\p_2)^2-i(-k_0+k\p_1+k\p_2)\left(\hat x-y+\frac{k_0-k\p_1}{\omega_{k_0-k\p_1}}(\tau_2-\tau_1)\right)}|k_2\rangle_0\nonumber\\
&&\hspace{-3cm}=-\frac{\lambda}{16\ok 0}\frac{\sqrt{\pi}}{2\pi \sigma}\frac{\sqrt{\pi}}{2\pi\hat\sigma}\pin{k_2}e^{-i\ok 2 t}\int_{0}^t d\tau_1\int_{\tau_1}^t d\tau_2 e^{-(x_{\tau_1}-\hat x)^2/(4\sigma^2)}\nonumber\\
&&\hspace{-3cm}\ \ \times
\pin{k\p_1}V^{(3)L}_{-k_0,k\p_1,k_0-k\p_1}\frac{e^{-i\ok 0\tau_1-i(\okp 1+\omega_{k_0-k\p_1})(\tau_2-\tau_1)+i\ok 2\tau_2}}{ \okp 1\omega_{k_0-k\p_1}} \nonumber\\
&&\hspace{-3cm}\ \ \times 
\int dy  V^{(3)}(\sl f(y)) \g_{k_2}(y)\g_{-k\p_1}(y)\g_{-k_0+k\p_1}(y)e^{-\left(\hat x-y+\frac{k_0-k\p_1}{\omega_{k_0-k\p_1}}(\tau_2-\tau_1)\right)^2/(4\hat\sigma^2)}|k_2\rangle_0.\nonumber 
\eea

First we studied the one vertex interaction, in which we found that $x$ must be close to the kink because of the $\I(x)$ loop factor.  Then we turned to a tadpole interaction, in which $x$ was not obviously close, but $y$ was close because of an $\I(y)$ term, which allowed us to show that $x$ is close.  However in the case of the present interaction, even $y$ is not obviously small.

To show that the $y$ integral has support at small $y$, after integration over $\tau_2$, we will insert another normalized bump function $e^{-(y-\hat y)^2/(4\hat\sigma^2)}/(2\hat\sigma\sqrt{\pi})$ into the $y$ integral, which satisfies the same limits as the $x$ bump function, in particular $m|\hat y|\gg 1$.  Again, for concreteness we will make the irrelevant choice $\hat{y}<0$.  Then we may replace $V^{(3)}(\sl f(y)) \g_{k_2}(y)\g_{-k\p_1}(y)\g_{-k_0+k\p_1}(y)$ with $V^{(3)L}_{k_2,-k\p_1,-k_0+k\p_1} e^{-i(k_2-k_0)y}$ and the localized final state is
\bea
U_2^A(\hat x,\hat y,t)|t=0\rangle_0&=&-\frac{\lambda}{16\ok 0}\frac{\sqrt{\pi}}{2\pi\hat\sigma}\pin{k_2}e^{-i\ok 2 t}\int_{0}^t d\tau_1\int_{\tau_1}^t d\tau_2 e^{-(x_{\tau_1}-\hat x)^2/(4\sigma^2)}\\
&&\hspace{-3cm}\ \ \times
\pin{k\p_1}V^{(3)L}_{-k_0,k\p_1,k_0-k\p_1}V^{(3)L}_{k_2,-k\p_1,-k_0+k\p_1}\frac{e^{-i\ok 0\tau_1-i(\okp 1+\omega_{k_0-k\p_1})(\tau_2-\tau_1)+i\ok 2\tau_2}}{ \okp 1\omega_{k_0-k\p_1}} 
\nonumber\\&&\hspace{-3cm}\ \ \times 
\int dy e^{-i(k_2-k_0)y}e^{-\left(\hat x-y+\frac{k_0-k\p_1}{\omega_{k_0-k\p_1}}(\tau_2-\tau_1)\right)^2/(4\hat\sigma^2)}|k_2\rangle_0\nonumber\\
&&\hspace{-3cm}=-\frac{\lambda}{16\ok 0}\pin{k_2}e^{-i\ok 2 t}\int_{0}^t d\tau_1\int_{\tau_1}^t d\tau_2 e^{-(x_{\tau_1}-\hat x)^2/(4\sigma^2)}\nonumber\\
&&\hspace{-3cm}\ \ \times
\pin{k\p_1}V^{(3)L}_{-k_0,k\p_1,k_0-k\p_1}V^{(3)L}_{k_2,-k\p_1,-k_0+k\p_1}\frac{e^{-i\ok 0\tau_1-i(\okp 1+\omega_{k_0-k\p_1})(\tau_2-\tau_1)+i\ok 2\tau_2}}{ \okp 1\omega_{k_0-k\p_1}} 
\nonumber\\&&\hspace{-3cm}\ \ \times 
e^{-\hat\sigma^2(k_2-k_0)^2-i(k_2-k_0)\left(\hat x+\frac{k_0-k\p_1}{\omega_{k_0-k\p_1}}(\tau_2-\tau_1)\right)}|k_2\rangle_0.\nonumber
\eea
The term $e^{-\hat\sigma^2(k_2-k_0)^2}$ ensures that the outgoing meson 2 has the same momentum as the incoming meson 1.  Thus this process describes forward scattering, which we are not interested in.  The reason, of course, is that we chose both $|x|$ and $|y|$ to be greater than $O(1/m)$, so that both interactions occurred far from the kink.  Thus no momentum could be exchanged between the kink and the mesons.

We therefore conclude that only $y\sim O(1/m)$ can contribute to elastic scattering if $|x|\gg O(1/m)$. In particular, $|y/\hat\sigma|$ limits to zero and so may be dropped in Eq.~(\ref{ybig}), leading to
\bea
U_2^A(\hat x,t)|t=0\rangle_0&=&-\frac{\lambda}{16\ok 0}\frac{\sqrt{\pi}}{2\pi\sigma}\frac{\sqrt{\pi}}{2\pi\hat\sigma}\pin{k_2}e^{-i\ok 2 t}\int_{0}^t d\tau_1\int_{\tau_1}^t d\tau_2 e^{-(x_{\tau_1}-\hat x)^2/(4\sigma^2)}\\
&&\times
\pin{k\p_1}\frac{e^{-i\ok 0\tau_1-i(\okp 1+\omega_{k_0-k\p_1})(\tau_2-\tau_1)+i\ok 2\tau_2}}{ \okp 1\omega_{k_0-k\p_1}} 
e^{-\left(\hat x+\frac{k_0-k\p_1}{\omega_{k_0-k\p_1}}(\tau_2-\tau_1)\right)^2/(4\hat\sigma^2)} \nonumber\\
&&\times V^{(3)L}_{-k_0,k\p_1,k_0-k\p_1}V_{k_2,-k\p_1,-k_0+k\p_1}|k_2\rangle_0\nonumber . 
\eea
Finally we turn to the integrals of the interaction times.  The $\tau_2$ integral yields a Gaussian whose exponential is equal to $-\hat\sigma^2(\okp 1+\omega_{k_0-k\p_1}-\ok 2)^2$ divided by a velocity squared, while the $\tau_1$ integral yields a Gaussian whose exponential is $-\sigma^2(k_0+k_2)^2$, where we have chosen the sign of $k_2$ to yield elastic scattering and not forward scattering.  In the support of this later Gaussian, we may replace $\ok 2$ by $\ok 0$ in the former Gaussian, so that its exponential is $-\hat\sigma^2(\okp 1+\omega_{k_0-k\p_1}-\ok 0)^2$.  This is of order $-\hat\sigma^2m^2$ for all values of $k_0$ and $k\p_1$, as two-body decay to two particles of the same mass as the original particle cannot simultaneously conserve momentum and energy.  Therefore the first exponential vanishes, and we find that $U_2^A(\hat x,t)|t=0\rangle_0$ vanishes when the first interaction is localized near any $\hat x$ that is not of order $O(1/m)$, as was the case for the previous two interactions.

\subsubsection{Continuing with the Computation}

Finally we are justified in dropping the $x/\sigma$ factor in Eq.~(\ref{xpic}), which leaves
\bea
U_2^A(t)|t=0\rangle_0&=&-\frac{\lambda}{16\ok 0}\frac{\sqrt{\pi}}{2\pi\sigma}\pin{k_2}e^{-i\ok 2 t}\int_{0}^t d\tau_1\int_{\tau_1}^t d\tau_2 
\ppinkp{2}\frac{V_{k_2-k\p_1-k\p_2}V_{-k_0k\p_1k\p_2}}{ \okp 1\okp 2}\nonumber\\
&&\hspace{-2cm}\ \ \times e^{-x_{\tau_1}^2/(4\sigma^2)-i(\ok 0-\okp 1-\okp 2)\tau_1-i(\okp 1+\okp 2-\ok 2)\tau_2} |k_2\rangle_0\nonumber\\
&&\hspace{-2cm}=i\frac{\lambda}{16\ok 0}\frac{\sqrt{\pi}}{2\pi\sigma}\pin{k_2}e^{-i\ok 2 t}\int_{0}^t d\tau_1
\ppinkp{2}\frac{V_{k_2-k\p_1-k\p_2}V_{-k_0k\p_1k\p_2}}{ \okp 1\okp 2(\ok 2-\okp 1-\okp 2)}\nonumber\\
&&\hspace{-2cm}\ \ \times e^{-x_{\tau_1}^2/(4\sigma^2)-i(\ok 0-\okp 1-\okp 2)\tau_1}\left(e^{-i(\okp 1+\okp 2-\ok 2)t} -e^{-i(\okp 1+\okp 2-\ok 2)\tau_1}\right) |k_2\rangle_0\nonumber\\
&&\hspace{-2cm}=i\frac{\lambda}{16 k_0}e^{-i\ok 0 t_c}\pin{k_2}e^{-i\ok 2 (t-t_c)}
\ppinkp{2}\frac{V_{k_2-k\p_1-k\p_2}V_{-k_0k\p_1k\p_2}}{ \okp 1\okp 2(\ok 2-\okp 1-\okp 2)}\nonumber\\
&&\hspace{-2cm}\ \ \times \left(e^{-\sigma^2\frac{\ok{0}^2}{k_0^2}(\ok 0-\okp 1-\okp 2)^2-i(\okp 1+\okp 2-\ok 2)(t-t_c)} -e^{-\sigma^2\frac{\ok{0}^2}{k_0^2}(\ok 0-\ok 2)^2  } \right) |k_2\rangle_0.
\eea
Note that there is no pole at $\ok 2=\okp 1+\okp 2$, as the sum of the two terms in the parenthesis has a simple zero there, leaving a term proportional to $t-t_c$.  Of course this does nonetheless diverge if one naively takes a $t\rightarrow\infty$ limit before integrating over the meson momenta.

As in the tadpole case, the first term in the parentheses corresponds not to elastic scattering, but rather to meson multiplication.  One may again note that over the support of the Gaussian its phase varies many times, and so it should not contribute once the virtual meson momenta have been integrated.  This argument applies here as it did there away from $\ok 2=\okp 1+\okp 2$.  What about at $\ok 2=\okp 1+\okp 2$, where the momenta cannot be freely varied as the surface is constrained?

Since the integrand is in fact everywhere finite, there is a vanishingly small contribution from any vanishingly small neighborhood of $\ok 2=\okp 1+\okp 2$.  One may therefore remove such a neighborhood from the domain of integration, in other words one may evaluate the integral close to $\ok 2=\okp 1+\okp 2$ using a principal value prescription without changing the value of the integral
\bea
U_2^A(t)|t=0\rangle_0&=&
i\frac{\lambda}{16 k_0}e^{-i\ok 0 t_c}\pin{k_2}e^{-i\ok 2 (t-t_c)}
\ppinkp{2}\frac{V_{k_2-k\p_1-k\p_2}V_{-k_0k\p_1k\p_2}}{ \okp 1\okp 2}\\
&&\hspace{-3cm}\ \ \times e^{-\sigma^2\frac{\ok{0}^2}{k_0^2}(\ok 0-\ok 2)^2}{\rm PV}\left[\frac{ e^{-\sigma^2\frac{\ok{0}^2}{k_0^2}\left[(\ok 0-\okp 1-\okp 2)^2-(\ok 0-\ok 2)^2\right]-i(\okp 1+\okp 2-\ok 2)(t-t_c)} 
-1   }{\ok 2-\okp 1-\okp 2}\right] |k_2\rangle_0.\nonumber
\eea
The principal value is additive, so the two terms in the numerator may be separated, yielding the sum of two principal values.  

In the support of the overall Gaussian, we may replace $V_{k_2-k\p_1-k\p_2}$ with $V_{k_0-k\p_1-k\p_2}$.  We do not replace the $k_2$ in the phase, as it is multiplied by a group velocity factor times $t$, which is the scale at which the naive divergence is cut off.  

Now consider the $k\p_2$ integral of the first term
\beq
\frac{e^{-\sigma^2 (\ok{0}^2/k_0^2) (\ok 0-\okp 1-\okp 2)^2} e^{-i(\okp 1+\okp 2-\ok 2) (t-t_c)}}{\ok 2-\okp 1-\okp 2}.
\eeq
In the limit $m(t-t_c)\rightarrow\infty$, the phase rotates so quickly that the $k\p_2$ integral is exponentially suppressed, being roughly of order exp$(-(t-t_c)^2/\sigma^2)$.  This vanishes as we take $(t-t_c)/\sigma\rightarrow\infty$ so that the final wave packet has no overlap with the kink.  However, when the denominator is less than this exponentially factor, as occurs near the poles, this argument fails.  The poles lie at
\beq
k\p_2=\pm\sqrt{(\ok 2-\okp 1)^2-m^2}= \pm k_I
\eeq
where we have introduced the positive momentum notation $k_I$.  Therefore we must evaluate the contribution from a neighborhood of order $O(1/(t-t_c))$ of the poles.

Near each of these poles, the contribution to the principal value is nonzero as a result of the phase factor.   Near each pole, the phase decreases as $\okp 2$ increases, and so as $|k\p_2|$ increases.  As a result, near the $k\p_2=-k_I$ pole, the phase increases with $k\p_2$ and near the $k\p_2=k_I$ pole it decreases.  This implies that the principal value is $\mp i\pi$ times the residue at the $k\p_2=\pm k_I$ pole.  The residue is $-\okp{2}/k\p_2$, times the various coefficients of the square brackets evaluated at the pole, at both poles.  Summing the contributions at the two poles one finds
\beq
i\pi\frac{\okp{2}}{|k\p_2|} \left(\delta(k\p_2- k_I)+\delta(k\p_2+ k_I)\right)= i\pi\delta(\okp 2+\okp 1-\ok 2).
\eeq

We have argued that we may replace the first term in square brackets with $i\pi \delta(-\ok 2+\okp 1+\okp 2)$.  This may in turn be absorbed into the other principal value term using the Sokhitski-Plemelj theorem
\beq
i\pi \delta(-\ok 2+\okp 1+\okp 2)+{\rm PV}\left[\frac{ 
1   }{-\ok 2+\okp 1+\okp 2}\right]=\frac{ 
1   }{-\ok 2+\okp 1+\okp 2-i\epsilon}.
\eeq

In conclusion, we may replace the first term in the parenthesis with an $\epsilon$ shift.  Now, we are interested in elastic, not forward scattering, so we will choose the sign of $k_2$ in the Gaussian peak considered, removing the forward scattering part, yielding
\bea
U_2^A(t)|t=0\rangle_0&=&
-i\frac{\lambda}{16 k_0}e^{-i\ok 0 t_c}\pin{k_2}e^{-i\ok 2 (t-t_c)}e^{-\sigma^2(k_0+k_2)^2}
\\
&&\ \ \times \ppinkp{2}\frac{V_{k_0-k\p_1-k\p_2}V_{-k_0k\p_1k\p_2}}{ \okp 1\okp 2(\ok 0-\okp 1-\okp 2+i\epsilon)}|k_2\rangle_0.\nonumber
\eea
In the denominator we have replaced $\ok{2}$ with $\ok{0}$, using the fact that they are equal in the support of the Gaussian in our $m\sigma\rightarrow\infty$ limit.  We recognize the $+i\epsilon$ in the final state as the usual one appearing in the in states in the Lippmann-Schwinger equation.

\subsubsection{The Case $\tau_1>\tau_2$}

This case is identical, except that the virtual mesons exchange their creation and annihilation operators.  This leads to the final state
\bea
U_2^B(t)|t=0\rangle_0&=&-\frac{\lambda}{16\ok 0}\frac{\sqrt{\pi}}{2\pi\sigma}\pin{k_2}\int_{0}^t d\tau_1\int_{0}^{\tau_1} d\tau_2 
\ppinkp{2}\frac{V_{k_2-k\p_1-k\p_2}}{ \okp 1\okp 2}e^{-i\ok 0\tau_1-i(\okp 1+\okp 2)(\tau_1-\tau_2)}\nonumber\\
&&\hspace{-1cm}\ \ \times e^{-i\ok 2(t-\tau_2)}\int dx \V3 \g_{-k_0}(x)\g_{k\p_1}(x)\g_{k\p_2}(x)e^{-(x_{\tau_1}-x)^2/(4\sigma^2)}|k_2\rangle_0. 
\eea

Therefore an identical derivation to the one above follows.  The $\tau_2$ integral leads to a $(\ok 2+\okp 1+\okp 2)$ in the denominator so there is not even superficially a pole, and no $i\epsilon$ is required.  The $\tau_1$ integral again gives two terms, and this time it is the second term that corresponds to an on-shell $k\p$ and vanishes upon integration.  As these two terms differ by a sign, and as it is the first and not the second term that remains, one obtains an overall sign flip with respect to the $\tau_1<\tau_2$ case, yielding
\bea
U_2^B(t)|t=0\rangle_0&=&
i\frac{\lambda}{16 k_0}e^{-i\ok 0 t_c}\pin{k_2}e^{-i\ok 2 (t-t_c)}e^{-\sigma^2(k_0+k_2)^2}\\
&&\ \ \times \ppinkp{2}\frac{V_{k_0-k\p_1-k\p_2}V_{-k_0k\p_1k\p_2}}{ \okp 1\okp 2(\ok 0+\okp 1+\okp 2)}|k_2\rangle_0.\nonumber
\eea

Adding these two contributions we find
\bea
\left(U_2^A(t)+U_2^B(t)\right)|t=0\rangle_0&=&-i\frac{\lambda}{8 k_0}\pin{k_2}e^{-i\ok 2 t}e^{-\sigma^2(k_0+k_2)^2+i(k_0+k_2)x_0} \label{a3}
\\
&&\ \ \times \ppinkp{2}\frac{(\okp 1+\okp 2) V_{k_0-k\p_1-k\p_2}V_{-k_0k\p_1k\p_2}}{ \okp 1\okp 2\left(\ok {0}^2-\left(\okp 1+\okp 2\right)^2+i\epsilon\right)} |k_2\rangle_0.\nonumber
\eea

\section{$\phi_0^4$ Terms} \label{foursez}

Recall that translation invariance dictates all terms with zero modes \cite{me2loop}.  These terms have two contributions.  First, there is the cloud of mesons around the incoming or outcoming meson.  Next, there is the cloud of mesons around the kink.  In both cases, the quantum corrections contain more mesons than the leading order kets, or more precisely more $B^\ddag$ operators, except when the incoming or outgoing meson is close to the kink, in which case the incoming or outgoing meson may be absorbed by the kink \cite{measy}.  In particular, in the asymptotic past and future, when the incoming and outgoing meson are far from the kink, these quantum corrections to components with zero modes $\phi_0$ will all have at least two mesons.  

This argument implies that there should not be any terms with zero modes and only one meson, or more precisely terms of the form $\phi_0^m B^\ddag\vac_0$ with $m>0$, at times $t$ late enough that the meson has traveled far from the kink.  In the current section, we will verify that this is indeed the case for terms with $\phi_0^4$ in the final state $U(t)|t=0\rangle$ at order $O(\lambda)$, which is the leading order at which $\phi_0^4$ may arise.  

\subsection{The Main Contribution}

\begin{figure}[htbp]
\centering
\includegraphics[width = 0.6\textwidth]{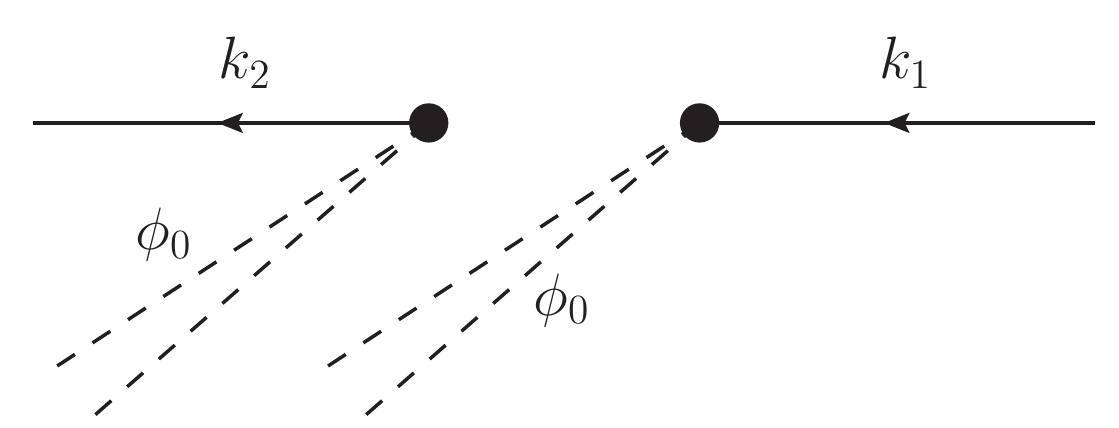}
\caption{Meson 1 is absorbed by the kink, leaving two zero modes.  The kink also emits meson 2, together with two more zero modes.} \label{fig4}
\end{figure}

Let us begin with the case in which $e^{-iHt}$ is evaluated at order $O(\lambda)$ and $|t=0\rangle$ at order $O(1)$.

We will consider the interactions 
\beq \label{h32}
H^{(1)\prime}_3=\frac{\sl}{2}\pin{k_1} V_{BB-k_1}\frac{B_{k_1}}{2\ok{1}}\phi_0^2 \hsp H^{(2)\prime}_3=\frac{\sl}{2}\pin{k_2} V_{BBk_2}\Bd {2}\phi_0^2.
\eeq
In this case, meson 1 is annihilated by $H^{(1)\prime}_3$ at time $\tau_1$ while meson $2$ is created by $H^{(2)\prime}_3$ at time $\tau_2$.  This is drawn in Fig.~\ref{fig4}. 

There are two cases to consider, corresponding to the sign of $\tau_1-\tau_2$. 

\subsubsection{$\tau_1<\tau_2$}

First consider the case $\tau_1<\tau_2$, in which meson 1 is absorbed by the kink before meson 2 is emitted.  Now the interactions act as
\beq
H^{(1)\prime}_3|k_1\rangle_0={\sl}{}\frac{V_{BB-k_1}}{4\ok{1}}\phi_0^2\vac_0\hsp
H^{(2)\prime}_3\phi_0^2\vac_0=\frac{\sl}{2}\pin{k_2} V_{BBk_2}
\phi_0^4|k_2\rangle_0.
\eeq
The corresponding contribution to the final state is
\bea
U_2^A(t)|t=0\rangle_0&=&-\frac{\lambda}{8}\pink{2}\int_{0}^t d\tau_1\int_{\tau_1}^t d\tau_2
\frac{V_{BB-k_1}V_{BB k_2}}{\ok 1}\label{f4a}\\
&&\times
e^{-i\ok 2(t-\tau_2)-i\ok 1\tau_1}
e^{-\sigma^2(k_1-k_0)^2-i(k_1-k_0)x_0}
\phi_0^4|k_2\rangle_0.\nonumber
\eea

\subsubsection{$\tau_1>\tau_2$}

Next we turn to the case in which meson 2 is emitted before meson 1 is absorbed.  Now the interactions act as
\bea
H^{(2)\prime}_3|k_1\rangle_0&=&\frac{\sl}{2}\pin{k_2} V_{BBk_2}
\phi_0^2|k_1k_2\rangle_0\\
H^{(1)\prime}_3\phi_0^2|k_1 k_2\rangle_0&=& {\sl}{}\frac{V_{BB-k_2}}{4\ok{2}}\phi_0^4|k_1\rangle_0 + {\sl}{}\frac{V_{BB-k_1}}{4\ok{1}}\phi_0^4|k_2\rangle_0 \nonumber
\eea
leading to the contribution
\bea
U_2^B(t)|t=0\rangle_0&=&-\frac{\lambda}{8}\pink{2}\int_{0}^t d\tau_1\int_{0}^{\tau_1} d\tau_2
\frac{V_{BB-k_1}V_{BB k_2}}{\ok 1}\label{f4b}\\
&&\times
e^{-i\ok 2(t-\tau_2)-i\ok 1\tau_1}
e^{-\sigma^2(k_1-k_0)^2-i(k_1-k_0)x_0}
\phi_0^4|k_2\rangle_0\nonumber
\eea
where we have removed the forward scattering part, proportional to $|k_1\rangle_0$.

The integrand is equal to the previous case, and so these contributions are easily added
\bea
\left(U_2^A(t)+U_2^B(t)\right)|t=0\rangle_0&=&-\frac{\lambda}{8}\pink{2}
\frac{V_{BB-k_1}V_{BB k_2}}{\ok 1}e^{-\sigma^2(k_1-k_0)^2-i(k_1-k_0)x_0}e^{-i\ok 2 t}\nonumber\\
&&\hspace{-5cm}\ \ \times
\int_{0}^t d\tau_1\int_{0}^{t} d\tau_2e^{i\ok 2\tau_2-i\ok 1\tau_1}
\phi_0^4|k_2\rangle_0\nonumber\\
&&\hspace{-5cm}=\frac{\lambda}{8}\pink{2}
\frac{V_{BB-k_1}V_{BB k_2}}{\ok {1}^2\ok 2}e^{-\sigma^2(k_1-k_0)^2-i(k_1-k_0)x_0}\left(1-e^{-i\ok 2 t}\right)
\left(1-e^{-i\ok 1 t}\right)
\phi_0^4|k_2\rangle_0\nonumber\\
&&\hspace{-5cm}=\frac{\lambda}{8}\pink{2}
\frac{V_{BB-k_1}V_{BB k_2}}{\ok {1}^2\ok 2}e^{-\sigma^2(k_1-k_0)^2}\left(1-e^{-i\ok 2 t}\right)
\nonumber\\&&\hspace{-3cm}\ \ \times
\left(e^{-i(k_1-k_0)x_0}-e^{-i\ok 0 t}e^{-i(k_1-k_0)x_t}\right)
\phi_0^4|k_2\rangle_0. \label{f4}
\eea

The Gaussian factor implies that $k_1$ has its support in a domain of width of order $O(1/\sigma)$.  The phase changes rapidly in this domain, $x_0/\sigma$ times and $x_t/\sigma$ times in the first and the second terms of the last parenthesis.  This leads to an exponential suppression, after integrating over $k_1$, of order $O(e^{-x_0^2/(4\sigma^2)})$ and $O(e^{-x_t^2/(4\sigma^2)})$ respectively.  These both converge rapidly to $0$ in our limit in which $\sigma/t$ and $\sigma/x_0$ tend to zero.  We thus conclude that there is no $\phi_0^4$ contribution.

\subsection{Initial State Contributions}

Contributions may also arise from subleading terms in the initial state $|t=0\rangle$.  Were $|t=0\rangle$ an eigenstate of the full Hamiltonian $H\p$, there would be three contributions, arising from terms of form $\phi_0^2|k_1k_2\rangle_0$, $\phi_0^2\vac_0$ and $\phi_0^4|k_2\rangle_0$, with $k_2\neq k_1$, in the initial state.  However $|t=0\rangle$ is not a Hamiltonian eigenstate, it is an asymptotic state.  As shown in Ref.~\cite{measy}, where the asymptotic states are evaluated explicitly, the second and third terms are therefore not present.  This fact can be derived directly by considering the Hamiltonian eigenstate and integrating over the wave packet (\ref{wp}).  Terms in which the $k_1$ meson has been annihilated contain an integral over $k_1$ that vanishes similarly.

This leaves terms of the first form.  There is only one such quantum correction \cite{measy}
\beq
|k_1\rangle_1\Big|_{\phi_0^2}=-\frac{\sl}{2}\ppin {k_2} \frac{V_{BBk_2}}{\ok 2}\phi_0^2|k_1k_2\rangle_0. \label{k11}
\eeq
This yields a quantum correction to the initial wave packet $|t=0\rangle$
\bea
|t=0\rangle_1&=&\pin{k_1} e^{-\sigma^2(k_1-k_0)^2-i(k_1-k_0)x_0}|k_1\rangle_1\\
&=&-\frac{\sl}{2}\pin{k_1}\ppin{k_2} e^{-\sigma^2(k_1-k_0)^2-i(k_1-k_0)x_0} \frac{V_{BB k_2}}{\ok 2} \phi_0^2|k_1 k_2\rangle_0.\nonumber
\eea
We evolve this with 
\beq
U_1(t)=-i\int_0^t d\tau_1 e^{-iH\p_2(t-\tau_1)}H^{(1)\prime}_3e^{-iH\p_2 \tau_1}
\eeq
to produce the contribution
\bea
U_1(t)|t=0\rangle_1&=&\frac{i\lambda}{8}\int_0^t d\tau_1\pink{2}  \frac{V_{BB-k_1}V_{BB k_2}}{\ok 1\ok 2} e^{-\sigma^2(k_1-k_0)^2-i(k_1-k_0)x_0-i\ok 1 \tau_1-i\ok 2 t}\phi_0^4|k_2\rangle_0\nonumber\\
&&\hspace{-2cm}=\frac{\lambda}{8}\pink{2}  \frac{V_{BB-k_1}V_{BB k_2}}{\ok {1}^2\ok 2} \left(1-e^{-i\ok 1 t} \right)e^{-\sigma^2(k_1-k_0)^2-i(k_1-k_0)x_0-i\ok 2 t}\phi_0^4|k_2\rangle_0\label{u1con}
\eea
to the final state, where we removed the forward scattering part  in the first line.  We also removed the contribution from final states in which there is an excited shape mode and no continuum mesons, as these terms do not correspond to elastic scattering and anyway vanish as they can never conserve energy on shell.  

The contributions arising from the continuum $k_2$ integral cancels the second term in the first parentheses in the last expressions in Eq.~(\ref{f4}).  We have already argued that these terms each vanish at large $t$, but for completeness if we add the present contribution to (\ref{f4}) we obtain
\beq
U(t)|t=0\rangle=\frac{\lambda}{8}\pink{2}
\frac{V_{BB-k_1}V_{BB k_2}}{\ok {1}^2\ok 2}e^{-\sigma^2(k_1-k_0)^2}
\left(e^{-i(k_1-k_0)x_0}-e^{-i\ok 0 t}e^{-i(k_1-k_0)x_t}\right)
\phi_0^4|k_2\rangle_0.
\eeq
As argued above, this vanishes upon performing the $k_1$ integration.  It would not vanish were $x_t$ close to zero, reflecting the fact that during the meson-kink collision, there are indeed nonvanishing $\phi_0^4$ terms with a single meson.  We will see below that these terms are important, as they lead to $\phi_0^2$ terms that are necessary to maintain translation invariance.

Eq.~(\ref{u1con}) also includes contributions in which $k_2$ is a shape mode.  In this case, the final state is not a kink and a meson, but instead an excited kink.   It therefore does not correspond to elastic scattering.  In the case of this process, the final  energy is necessarily less than that of the initial state and so this can never be on-shell, and so one can show that after $k_1$ integration the amplitude vanishes exponentially in $t-t_c$.

\subsection{A Generalization}

\begin{figure}[htbp]
\centering
\includegraphics[width = 0.8\textwidth]{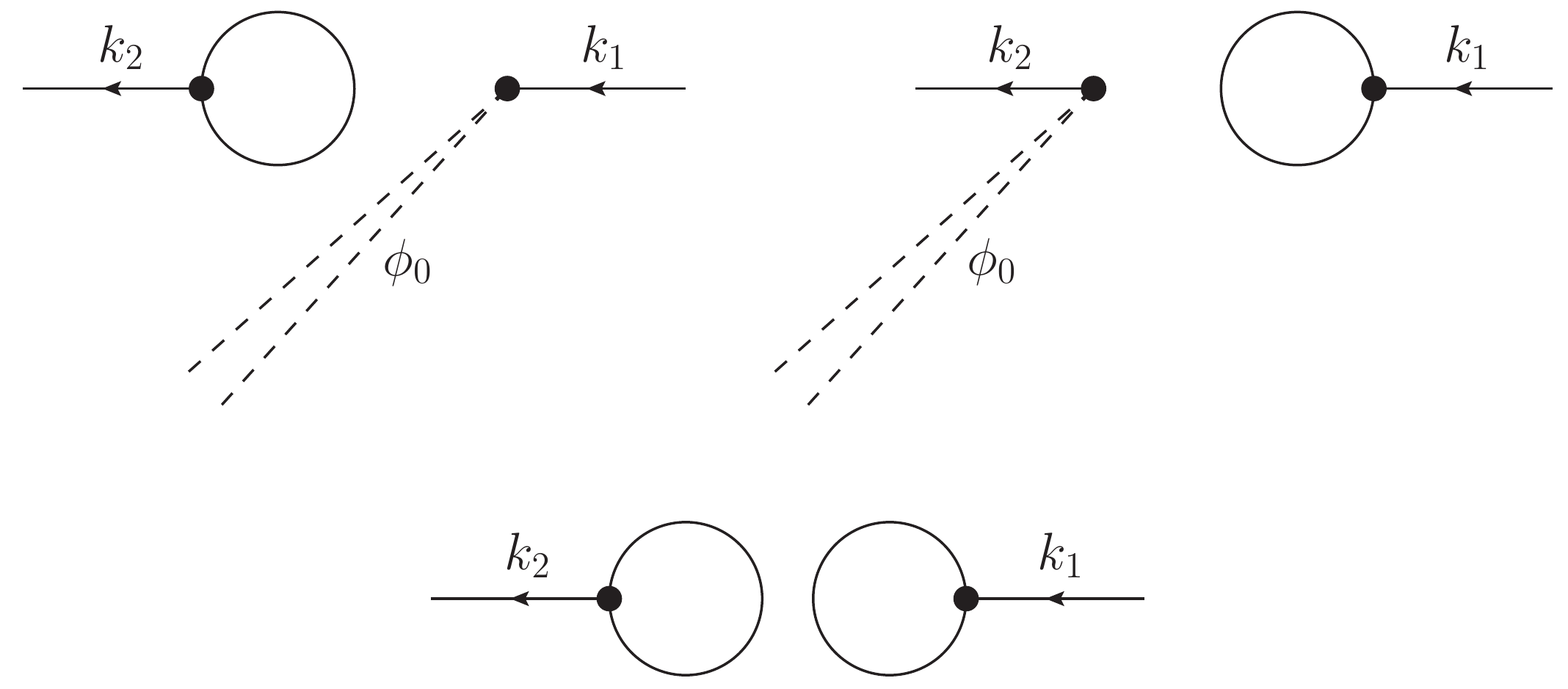}
\caption{Meson 1 is destroyed by a tadpole (right and bottom) or converted into two zero modes (left) and meson 2 is created by a tadpole (left and bottom) or together with two zero modes (right).} \label{fig4i}
\end{figure}

We have just shown that the interaction terms (\ref{h32}) in $H\p_3$, those that are proportional to $\phi_0^2$, do not lead to any contribution proportional to $\phi_0^4$ at any time $t$ except within of order $O(\sigma)$ of $t_c$.  In particular such contributions vanish at large times, when the experiment ends.  The argument relied on the fact that this term is proportional to $\g_B^2(x)$, which is localized at $|x|\sim 1/m\ll \sigma$, which let us drop $x/\sigma$ terms.

The interaction $H\p_3\Big|_{\I}$  possesses a similar term
\beq
H\p_3\Big|_{\I}=\frac{\sl}{2}\ppin{k\p} V_{\I k\p}\left(\Bdp{}
+\frac{B_{-k\p}}{2\okp{}}\right)\label{h3i}.
\eeq
The same arguments may then be applied to calculate the final state of the process shown in the bottom panel of Fig.~\ref{fig4i} to show that there is no contribution to the state $U(t)|t=0\rangle$ proportional to $\I^2(x)$.  

What about the initial state contribution?  Again from Ref.~\cite{measy} the leading correction to the $|k_1\rangle$ asymptotic state is
\beq
|k_1\rangle_1\Big|_{\I}=-\frac{\sl}{2}\ \ppin{k_2}\frac{V_{\I k_2}}{\ok 2}|k_1k_2\rangle_0
\eeq
which is identical to (\ref{k11}) except the $\phi_0^2$ is missing and the $\g_B^2(x)$ has been replaced by $\I(x)$, which again is supported at $|x|\sim 1/m$.  Thus even this contribution can be calculated identically.

In fact, one can do better.  One can repeat the argument with the sum of these two contributions
\beq
H\p_3\Big|_{\I,\phi_0^2}=\frac{\sl}{2}\ \ppin{k\p} \left(V_{\I k\p}+V_{BBk\p}\phi_0^2\right)\left(\Bdp{}
+\frac{B_{-k\p}}{2\okp{}}\right).
\eeq
The argument again proceeds identically, but now one can see that even terms with one $\I$ and one $\phi_0^2$, seen in the top of Fig.~\ref{fig4i},  vanish at all times $t$.  

\section{$\phi_0^2$ Terms} \label{twosez}

In this section we systematically study the components of the state at a time $t$ that have two zero modes, or a more precisely a factor of $\phi_0^2|k_2\rangle_0$.  Contributions to such states can be decomposed into four categories, to each of which we dedicate a subsection.  First we consider contributions with a single, four-point interaction.  The other three categories each contain two three-point interactions.  Of these, in the first, both zero modes arise from the same interaction.  In the second, one zero mode arises from each interaction.  In the last, each interaction generates two zero modes, as in Sec.~\ref{foursez}, but two of these zero modes are eliminated by the kinetic term for the kink center of mass.

\subsection{A Single Interaction}

\begin{figure}[htbp]
\centering
\includegraphics[width = 0.6\textwidth]{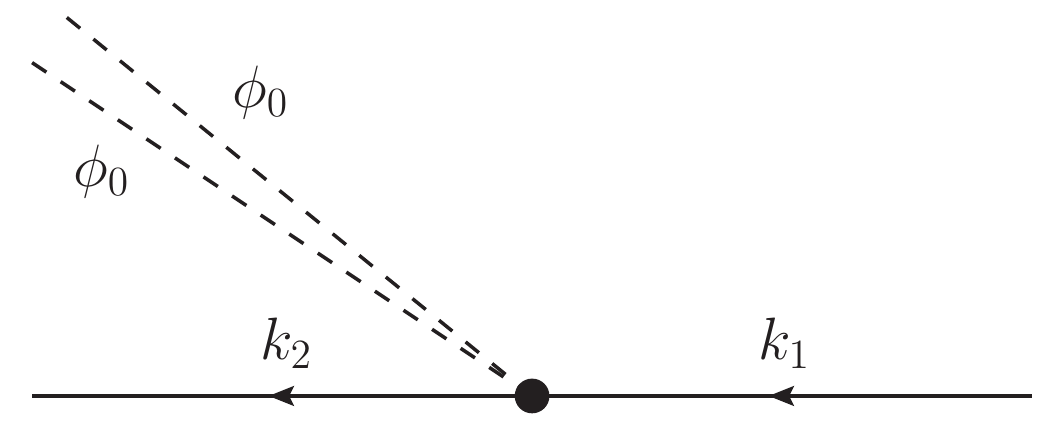}
\caption{Meson 1 converts into meson 2, emitting two zero-modes in the process.} \label{figb2}
\end{figure}

The simplest contribution to final states of the form $\phi_0^2|k_2\rangle_0$ arises from a single interaction
\beq
H^{(1)\prime}_4=\frac{\lambda}{2}\pink{2}V_{BB-k_1 k_2} \Bd 2 \frac{B_{k_1}}{2\ok 1}\phi_0^2. \label{h42}
\eeq
Acting on an initial meson $|k_1\rangle_0$ it yields
\beq
H^{(1)\prime}_4|k_1\rangle_0=\frac{\lambda}{4\ok 1}\pin{k_2}V_{BB-k_1 k_2} \phi_0^2|k_2\rangle_0.
\eeq
This leads to the final state
\beq
U_2(t)|t=0\rangle_0=-i\frac{\lambda}{4}\pink{2} \frac{V_{BB-k_1k_2}}{\ok 1} e^{-i\ok 2 t} \int_0^t d\tau_1e^{-i(\ok 0-\ok 2)\tau_1}e^{-\sigma^2(k_1-k_0)^2-i(k_1-k_0)x_{\tau_1}}\phi_0^2|k_2\rangle_0. \label{f2a}
\eeq
The corresponding process is drawn in Fig.~\ref{figb2}.

\subsection{A Virtual Meson that Decays to Two Zero Modes} \label{tad2sez}

\begin{figure}[htbp]
\centering
\includegraphics[width = 0.6\textwidth]{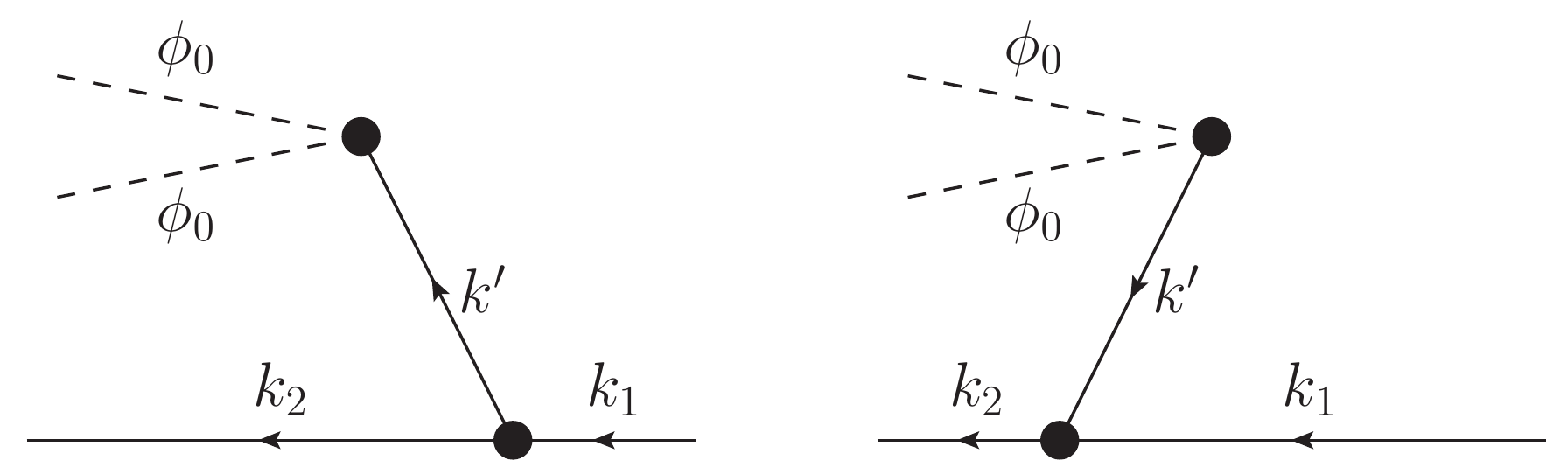}
\caption{In the right panel, a virtual meson is created together with two zero modes.  The virtual meson strikes meson 1 and turns it into meson 2.  In the left panel, meson 1 nucleates a virtual meson which decays into two zero modes.} \label{figc2}
\end{figure}

Next let us consider the contribution with two $H\p_3$ interactions drawn in Fig.~\ref{figc2}.  In the first, at time $\tau_1$ meson $1$ changes to meson $2$ and a virtual meson of momentum $k\p$ is emitted or absorbed.  In the second, at time $\tau_2$ the virtual meson is absorbed or emitted and two zero modes are created.

The two relevant interactions are
\bea\label{tad2}
H^{(1)\prime}_3&=&\frac{\sl}{2}\ppink{2}\ppin{k\p} V_{-k_1k_2k\p}\Bd 2\left(\Bdp{}+\frac{B_{-k\p}}{2\okp{}}\right)\frac{B_{k_1}}{2\ok 1}\\
H^{(2)\prime}_3&=&\frac{\sl}{2}\ppin{k\p}V_{BBk\p} \left(\Bdp{}+\frac{B_{-k\p}}{2\okp{}}\right)\phi_0^2. \nonumber
\eea

\subsubsection{The Case $\tau_1<\tau_2$}
In this case, the virtual meson is emitted by meson 1
\beq
H^{(1)\prime}_3|k_1\rangle_0=\frac{\sl}{2}\ppin{k_2}\ppin{k\p} \frac{V_{-k_1k_2k\p}}{2\ok 1}|k_2k\p\rangle_0
\eeq
and it is then absorbed by the kink
\beq
H^{(2)\prime}_3|k_2k\p\rangle_0=\frac{\sl}{2} \frac{V_{BB-k\p}}{2\okp{}}\phi_0^2|k_2\rangle_0 + \frac{\sl}{2} \frac{V_{BB-k_2}}{2\ok{2}}\phi_0^2|k\p\rangle_0.
\eeq

The resulting final state is
\bea
U_2^A(t)|t=0\rangle_0&=&-\frac{\lambda}{8}\pink{2}\ppin{k\p}\int_{0}^t d\tau_1\int_{\tau_1}^t d\tau_2
\frac{V_{-k_1k_2k\p}V_{BB-k\p}}{\ok 1\okp{}}\label{f2bi}\\
&&\times
e^{-i\ok 2(t-\tau_1)-i\okp{}(\tau_2-\tau_1)-i\ok 0\tau_1}
e^{-\sigma^2(k_1-k_0)^2-i(k_1-k_0)x_{\tau_1}}
\phi_0^2|k_2\rangle_0\nonumber\\
&=&i\frac{\lambda}{8}\pink{2}\ppin{k\p}
\frac{V_{-k_1k_2k\p}V_{BB-k\p}}{\ok 1\okp{}^2}e^{-i\ok 2 t}\nonumber\\
&&\times\int_{0}^t d\tau_1
e^{-i(\ok 0-\ok 2)\tau_1}
e^{-\sigma^2(k_1-k_0)^2-i(k_1-k_0)x_{\tau_1}}
\phi_0^2|k_2\rangle_0\nonumber
\eea
where, in the $\tau_2$ integration, we have dropped the boundary term at $\tau_2=t$ as it corresponds to the limit in which the virtual meson goes on-shell.  Like the two-process cases above, this term vanishes after $k\p$ is integrated as its phase oscillates rapidly.

\subsubsection{The Case $\tau_1>\tau_2$}

In this case the virtual meson is first emitted by the kink
\beq
H^{(2)\prime}_3|k_1\rangle_0=\frac{\sl}{2}\ppin{k\p}V_{BBk\p}\phi_0^2|k_1 k\p\rangle_0
\eeq
and then it is absorbed by meson 1
\beq
H^{(1)\prime}_3 |k_1k\p\rangle_0=\frac{\sl}{4} \ppin{k_2} \frac{V_{-k_1k_2-k\p}}{\ok 1\okp{}} |k_2\rangle_0
\eeq
leading to the final state
\bea
U_2^B(t)|t=0\rangle_0&=&-\frac{\lambda}{8}\pink{2}\ppin{k\p}\int_{0}^t d\tau_1\int_{0}^{\tau_1} d\tau_2
\frac{V_{-k_1k_2k\p}V_{BB-k\p}}{\ok 1\okp{}}\label{f2bi2}\\
&&\times
e^{-i\ok 2(t-\tau_1)-i\okp{}(\tau_1-\tau_2)-i\ok 0\tau_1}
e^{-\sigma^2(k_1-k_0)^2-i(k_1-k_0)x_{\tau_1}}
\phi_0^2|k_2\rangle_0\nonumber\\
&=&i\frac{\lambda}{8}\pink{2}\ppin{k\p}
\frac{V_{-k_1k_2k\p}V_{BB-k\p}}{\ok 1\okp{}^2}e^{-i\ok 2 t}\nonumber\\
&&\times\int_{0}^t d\tau_1
e^{-i(\ok 0-\ok 2)\tau_1}
e^{-\sigma^2(k_1-k_0)^2-i(k_1-k_0)x_{\tau_1}}
\phi_0^2|k_2\rangle_0.\nonumber
\eea
This time, when performing the $\tau_2$ integral, we have dropped the contribution from $\tau_2=0$.  This term is in fact exactly canceled by an initial state contribution, but anyway corresponds to the on-shell limit of our virtual meson in which the $k\p$ integration yields zero.

This contribution to the final state is equal to that of Eq.~(\ref{f2bi}) with the other ordering.  Adding them then yields a factor of two.  Using the Ward Identity (\ref{warda2}), this can be summarized
\bea
\left( U_2^A(t)+U_2^B(t)
\right)|t=0\rangle_0&=&i\frac{\lambda}{4\sqrt{\lambda Q_0}} \pink{2}\ppin{k\p}
\frac{V_{-k_1k_2k\p}\Delta_{-k\p B}}{\ok 1}e^{-i\ok 2 t}\label{f2b}\\
&&\times\int_{0}^t d\tau_1
e^{-i(\ok 0-\ok 2)\tau_1}
e^{-\sigma^2(k_1-k_0)^2-i(k_1-k_0)x_{\tau_1}}
\phi_0^2|k_2\rangle_0.\nonumber
\eea
Here we have used the shorthand
\beq
\Delta_{ij}=\int dx \g_i(x)\g\p_j(x) \label{ddef}
\eeq
where $i$ and $j$ run over the normal mode indices $B$, $S$ and $k$.  Intuitively, the matrix $\Delta$ represents the momentum operator acting on the mesons.

\subsection{One Zero-Mode at Each Vertex}

\begin{figure}[htbp]
\centering
\includegraphics[width = 0.6\textwidth]{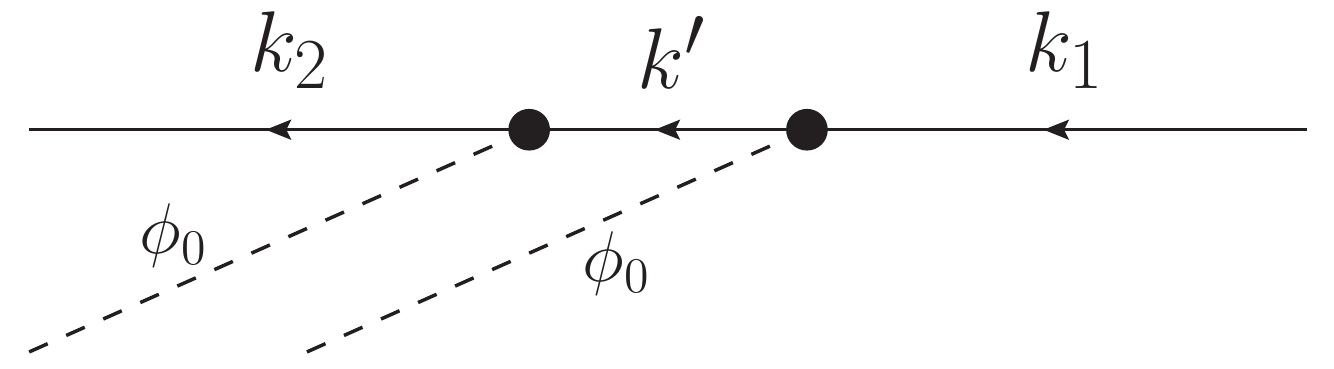}
\caption{Meson 1 turns into a virtual meson, emitting a zero mode.  The virtual meson emits yet another zero mode, converting into meson 2.} \label{figa2}
\end{figure}

Next we turn to the case in which there is a single zero mode created at each interaction of the form $(\sl/2)\int dx \g_B(x)\phi_0 :\phi^2(x):_b$.  At times $\tau_1$ and $\tau_2$ we place the interactions
\bea
H^{(1)\prime}_3&=&\frac{\sl}{2}\ppin{k_1}\ppin{k\p} V_{B-k_1k\p}\left(2\Bdp{}+\frac{B_{-k\p}}{2\okp{}}\right)\frac{B_{k_1}}{2\ok 1}\phi_0\\
H^{(2)\prime}_3&=&\frac{\sl}{2}\ppin{k_2}\ppin{k\p} V_{Bk\p k_2}\Bd 2\left(\Bdp{}+\frac{B_{-k\p}}{\okp{}}\right)\phi_0\nonumber
\eea
respectively, bearing in mind that we are interested in the components of the final state with a single meson.  This is drawn in Fig.~\ref{figa2}.

\subsubsection{The Case $\tau_1<\tau_2$}

At each interaction, the meson interacts with the kink, exciting a single zero mode
\beq
H^{(1)\prime}_3|k_1\rangle_0=\sl \ \ppin{k\p} \frac{V_{B-k_1k\p}}{2\ok 1}\phi_0|k\p\rangle_0\hsp
H^{(2)\prime}_3\phi_0|k\p\rangle_0=\sl \ \ppin{k_2} \frac{V_{B-k\p k_2}}{2\okp{}}\phi_0^2|k_2\rangle_0.
\eeq
The corresponding contribution to the final state is
\bea
U_2^A(t)|t=0\rangle_0&=&-\frac{\lambda}{4}\pink{2}\ppin{k\p}\int_{0}^t d\tau_1\int_{\tau_1}^t d\tau_2
\frac{V_{B-k_1k\p}V_{B-k\p k_2}}{\ok 1\okp{}}\label{f2c}\\
&&\times
e^{-i\ok 2(t-\tau_2)-i\okp{}(\tau_2-\tau_1)-i\ok 1\tau_1}
e^{-\sigma^2(k_1-k_0)^2-i(k_1-k_0)x_0}
\phi_0^2|k_2\rangle_0.\nonumber
\eea
If we first integrate $\tau_1$ from $0$ to $\tau_2$, dropping the vanishing contribution from $\tau_1=0$, we obtain
\bea
U_2^A(t)|t=0\rangle_0&=&-i\frac{\lambda}{4}\pink{2}e^{-i\ok 2 t} \ppin{k\p}\int_{0}^t d\tau_2
\frac{V_{B-k_1k\p}V_{B-k\p k_2}}{\ok 1\okp{}(\ok 1-\okp{})}\label{int1}\\
&&\times
e^{-i(\ok 1-\ok 2)\tau_2}
e^{-\sigma^2(k_1-k_0)^2-i(k_1-k_0)x_0}
\phi_0^2|k_2\rangle_0\nonumber\\
&=&-i\frac{\lambda}{4}\pink{2}e^{-i\ok 2 t} \ppin{k\p}\int_{0}^t d\tau_2
\frac{V_{B-k_1k\p}V_{B-k\p k_2}}{\ok 1\okp{}(\ok 1-\okp{})}\nonumber\\
&&\times
e^{-i(\ok 0-\ok 2)\tau_2}
e^{-\sigma^2(k_1-k_0)^2-i(k_1-k_0)x_{\tau_2}}
\phi_0^2|k_2\rangle_0.\nonumber
\eea
If instead we first integrate $\tau_2$ from $\tau_1$ to $t$, and drop the vanishing contribution at $\tau_2=t$, then we obtain
\bea
U_2^A(t)|t=0\rangle_0&=&-i\frac{\lambda}{4}\pink{2}e^{-i\ok 2 t} \ppin{k\p}\int_{0}^t d\tau_1
\frac{V_{B-k_1k\p}V_{B-k\p k_2}}{\ok 1\okp{}(\ok 2-\okp{})}\\
&&\times
e^{-i(\ok 1-\ok 2)\tau_1}
e^{-\sigma^2(k_1-k_0)^2-i(k_1-k_0)x_0}
\phi_0^2|k_2\rangle_0\nonumber\\
&=&-i\frac{\lambda}{4}\pink{2}e^{-i\ok 2 t} \ppin{k\p}\int_{0}^t d\tau_2
\frac{V_{B-k_1k\p}V_{B-k\p k_2}}{\ok 1\okp{}(\ok 2-\okp{})}\nonumber\\
&&\times
e^{-i(\ok 0-\ok 2)\tau_2}
e^{-\sigma^2(k_1-k_0)^2-i(k_1-k_0)x_{\tau_2}}
\phi_0^2|k_2\rangle_0.\nonumber
\eea
Of course this must equal (\ref{int1}), as the finite $\tau_i$ integrals commute.  In particular, both must equal their average, which will be more convenient below
\bea
U_2^A(t)|t=0\rangle_0&=&i\frac{\lambda}{8}\pink{2}e^{-i\ok 2 t} \ppin{k\p}\int_{0}^t d\tau_2
\frac{V_{B-k_1k\p}V_{B-k\p k_2}}{\ok 1\okp{}} \label{f2ca}\\
&&\times
\left[ \frac{1}{\okp{}-\ok 2}+\frac{1}{\okp{}-\ok 1}
\right] e^{-i(\ok 0-\ok 2)\tau_2}
e^{-\sigma^2(k_1-k_0)^2-i(k_1-k_0)x_{\tau_2}}
\phi_0^2|k_2\rangle_0.\nonumber
\eea

\subsubsection{The Case $\tau_1>\tau_2$}

Now the first interaction creates two new mesons
\beq
H^{(2)\prime}_3|k_1\rangle_0=\frac{\sl}{2}\ \ppin{k_2}\ppin{k\p} V_{Bk\p k_2}\phi_0|k_1k_2k\p\rangle_0
\eeq
while the second destroys one of these together with meson 1
\beq
H^{(1)\prime}_3\phi_0|k_1k_2k\p\rangle_0=\frac{\sl}{4} \frac{V_{B-k_1-k\p}}{\ok 1\okp{}}\phi_0^2|k_2\rangle_0 + \frac{\sl}{4} \frac{V_{B-k_1-k_2}}{\ok 1\ok 2}\phi_0^2|k\p\rangle_0 + \frac{\sl}{4} \frac{V_{B-k_2-k\p}}{\ok 2\okp{}}\phi_0^2|k_1\rangle_0
\eeq
where the last term will correspond to forward scattering and we will remove it when calculating the final state.  As $k\p$ and $k_2$ are both dummy variables, in the case of the $|k\p\rangle$ term, we can and will exchange their names, so that the final state is proportional to $|k_2\rangle$ and the first two terms on the right hand side are equal.

Evolving to time $t$ we find the state
\bea
U_2^B(t)|t=0\rangle_0&=&-\frac{\lambda}{4}\pink{2}\ppin{k\p}\int_{0}^t d\tau_1\int_{0}^{\tau_1} d\tau_2
\frac{V_{B-k_1k\p}V_{B-k\p k_2}}{\ok 1\okp{}}\label{f2ca2}\\
&&\times
e^{-i\ok 2(t-\tau_2)-i\okp{}(\tau_1-\tau_2)-i\ok 1\tau_1}
e^{-\sigma^2(k_1-k_0)^2-i(k_1-k_0)x_0}
\phi_0^2|k_2\rangle_0.\nonumber
\eea

Integration over $\tau_1$ from $\tau_2$ to $t$, dropping $\tau_1=t$, yields
\bea
U_2^B(t)|t=0\rangle_0
&=&i\frac{\lambda}{4}\pink{2}e^{-i\ok 2 t} \ppin{k\p}\int_{0}^t d\tau_2
\frac{V_{B-k_1k\p}V_{B-k\p k_2}}{\ok 1\okp{}(\ok 1+\okp{})}\nonumber\\
&&\times
e^{-i(\ok 0-\ok 2)\tau_2}
e^{-\sigma^2(k_1-k_0)^2-i(k_1-k_0)x_{\tau_2}}
\phi_0^2|k_2\rangle_0\nonumber
\eea
whereas integration over $\tau_2$, dropping $\tau_2=0$, would instead yield
\bea
U_2^B(t)|t=0\rangle_0&=&i\frac{\lambda}{4}\pink{2}e^{-i\ok 2 t} \ppin{k\p}\int_{0}^t d\tau_2
\frac{V_{B-k_1k\p}V_{B-k\p k_2}}{\ok 1\okp{}(\ok 2+\okp{})}\nonumber\\
&&\times
e^{-i(\ok 0-\ok 2)\tau_2}
e^{-\sigma^2(k_1-k_0)^2-i(k_1-k_0)x_{\tau_2}}
\phi_0^2|k_2\rangle_0.\nonumber
\eea
Averaging one finds
\bea
U_2^B(t)|t=0\rangle_0&=&i\frac{\lambda}{8}\pink{2}e^{-i\ok 2 t} \ppin{k\p}\int_{0}^t d\tau_2
\frac{V_{B-k_1k\p}V_{B-k\p k_2}}{\ok 1\okp{}}\\
&&\times
\left[ \frac{1}{\ok 2+\okp{}}+\frac{1}{\ok 1+\okp{}}
\right]e^{-i(\ok 0-\ok 2)\tau_2}
e^{-\sigma^2(k_1-k_0)^2-i(k_1-k_0)x_{\tau_2}}
\phi_0^2|k_2\rangle_0\nonumber.
\eea

\subsubsection{Conclusions}

Finally, we add the contribution (\ref{f2ca}) from the case $\tau_1<\tau_2$ to obtain
\bea
\left( U_2^A(t)+U_2^B(t)
\right)|t=0\rangle_0&=&i\frac{\lambda}{4}\pink{2}e^{-i\ok 2 t} \ppin{k\p}\int_{0}^t d\tau_2
\frac{V_{B-k_1k\p}V_{B-k\p k_2}}{\ok 1}\\
&&\hspace{-2cm}\times
\left[ \frac{1}{\okp{}^2-\ok {2}^2}+\frac{1}{\okp{}^2-\ok {1}^2}
\right]e^{-i(\ok 0-\ok 2)\tau_2}
e^{-\sigma^2(k_1-k_0)^2-i(k_1-k_0)x_{\tau_2}}
\phi_0^2|k_2\rangle_0\nonumber.
\eea
Using the Ward Identity (\ref{warda}) this can be simplified somewhat
\bea
\left( U_2^A(t)+U_2^B(t)
\right)|t=0\rangle_0&=&i\frac{\lambda}{4\sqrt{\lambda Q_0}}\pink{2}e^{-i\ok 2 t} \ppin{k\p} \frac{\left(V_{B-k_1k\p}\Delta_{-k\p k_2}+\Delta_{k\p-k_1}V_{B-k\p k_2}\right)}{\ok 1}\nonumber
\\
&&\times
\int_{0}^t d\tau_2e^{-i(\ok 0-\ok 2)\tau_2}
e^{-\sigma^2(k_1-k_0)^2-i(k_1-k_0)x_{\tau_2}}
\phi_0^2|k_2\rangle_0. \label{f2cb}
\eea

Now we can see the reason that we chose the complicated prescription of averaging over the two orders of time integration.  Although of course these integrals commute, we see that the average prescription used here leads to the combination $V\Delta+\Delta V$ in round brackets in (\ref{f2cb}) which is the same as that in the Ward identity (\ref{ward}), even without setting $k_0=-k_2$.  

Could we have simply set $k_0=-k_2$ and just chose one ordering for the time integrals?  Well, the uncertainty principle says that $k_0+k_2$ will be of order $O(1/t)$, which indeed tends to zero at large $t$ although it is dimensionful and so one needs to be more careful.  The problem, as we will see below, is that the $e^{-i\pi_0^2 t/2}$ term in the evolution operator contains, at first order, $-i\pi_0^2 t/2$ which leads to a zero-mode-free term proportional to $t$.  In all, this contribution would be proportional to $t(k_0+k_2)$, which is indeed dimensionless and does not tend to zero at large $t$.  Therefore, in terms with zero modes we need to be careful about factors of $k_0+k_2$ or equivalently $\ok 2-\ok 0$ or, even worse, $\ok 2-\ok 1$. 

We note that there are neither initial nor final state corrections, as they would consist of a single meson and a $\Delta_{kB}$ term which vanishes when folded into the initial or final wave packet, which is far from the kink, or more precisely the support of $\g_B(x)$.

\subsection{Two Zero-Modes from Four Zero-Modes} \label{da4a2}

\begin{figure}[htbp]
\centering
\includegraphics[width = 0.9\textwidth]{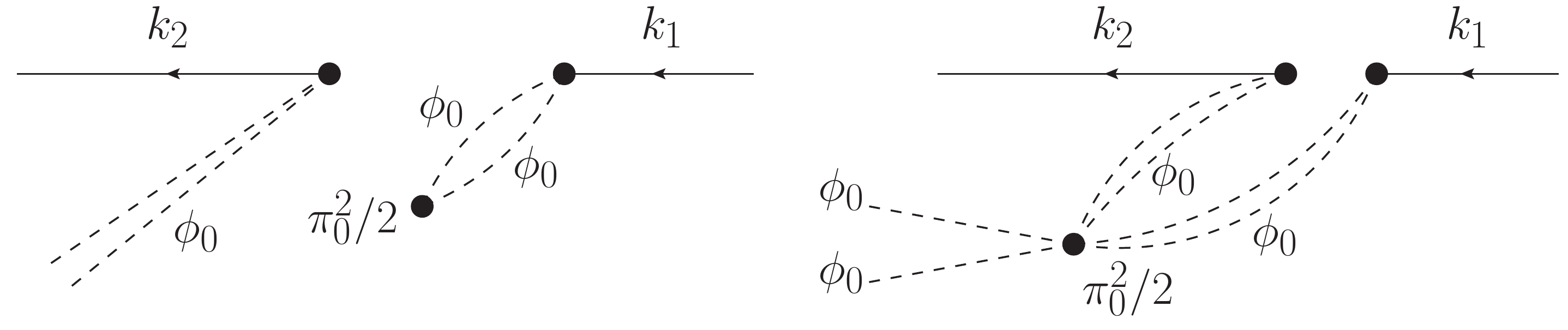}
\caption{This process is as in Fig.~\ref{fig4}.  However, two of the four zero modes are annihilated by the $\pi_0^2/2$ term in the free Hamiltonian $H\p_2$.}\label{fig42}
\end{figure}

The final contribution to the two zero-mode sector of the final state arises from interactions in which four zero modes are created, two by each of two $H\p_3$ terms in (\ref{h32}), and then two of these four zero-modes are destroyed by the $\pi_0^2/2$ in the free Hamiltonian $H\p_2$.  This process is depicted in Fig.~\ref{fig42}.

The free propagator $H\p_2$ consists of a $\pi_0^2/2$ term, as well as harmonic oscillator terms for the normal modes.  These all commute, and so the respective parts of the free propagator may be factorized.  Concretely, consider a basis element of the kink sector $\phi_0^m|k_1\cdots k_n\rangle_0$.  Then the free propagator acts as
\beq
e^{-iH\p_2 T}\phi_0^m|k_1\cdots k_n\rangle_0=e^{-i\omega T}e^{-i\pi_0^2 T/2}\phi_0^m|k_1\cdots k_n\rangle_0
\hsp
\omega=\sum_{i=1}^n\ok{n}.
\eeq
The contribution of interest in this subsection uses a single $\pi_0^2$, to reduce the number of zero modes from $4$ to $2$, and so corresponds to the term
\beq
e^{-i\omega T}\left(-i\frac{\pi_0^2}{2}T\right)\phi_0^m|k_1\cdots k_n\rangle_0=i\frac{m(m-1)T}{2}e^{-i\omega T}\phi_0^{m-2}|k_1\cdots k_n\rangle_0.
\eeq

Now observe that $e^{-i\omega T}\phi_0^m|k_1\cdots k_n\rangle_0$ is the result of the free evolution in which no zero modes are annihilated.  And so, once one has calculated the $m$ zero-mode sector at an arbitrary time $\tau$ as an integral over the various interaction times, one need only include a factor of $im(m-1)T/2$ in the integrand to obtain the contribution to the $m-2$ zero-mode sector.  This needs to be done during the free evolution between each pair of interactions, as two zero modes may in principle be annihilated between any pair of interactions.  Here $T$ is the time that passes between the pair of interactions.

In kink-meson elastic scattering at order $O(\lambda)$, the only pair of interactions that creates four zero modes is written as an integral of interaction times in Eqs.~(\ref{f4a}) and (\ref{f4b}).  Consider first the case $\tau_1<\tau_2$.  Then, including the factors of $im(m-1)T/2$ where $m=2$ between the interactions and $m=4$ after both, one obtains the final state contribution
\beq
U_2^A(t)|t=0\rangle_0=-\frac{\lambda}{8}\pink{2}
\frac{V_{BB-k_1}V_{BB k_2}}{\ok 1} e^{-i\ok 2 t} I_A e^{-\sigma^2(k_1-k_0)^2-i(k_1-k_0)x_0}
\phi_0^2|k_2\rangle_0
\eeq
where
\bea
I_A=i\int_{0}^t d\tau_1\int_{\tau_1}^t d\tau_2e^{i\ok 2\tau_2-i\ok 1\tau_1}\left((\tau_2-\tau_1)+6(t-\tau_2)\right).
\eea
Despite the linear growth in $t$, the arguments above show that the $\tau_2=t$ contribution vanishes exponentially and so we may drop it
\bea
I_A&=&i\int_0^t d\tau_1 e^{-i\ok 1\tau_1} \left(6t-\tau_1+5i\frac{\partial}{\partial\ok 2}\right) \int_{\tau_1}^t d\tau_2 e^{i\ok 2\tau_2}\\
&=&-\int_0^t d\tau_1 e^{-i\ok 1\tau_1} \left(6t-\tau_1+5i\frac{\partial}{\partial\ok 2}\right) \frac{e^{i\ok 2\tau_1}}{\ok 2}\nonumber\\
&=&\int_0^t d\tau_1 \frac{e^{-i(\ok 1-\ok 2)\tau_1}}{\ok 2} \left(-6t+6\tau_1+\frac{5i}{\ok 2}\right) .\nonumber
\eea
Integrating $\tau_1$ first, and dropping $\tau_1=0$ would instead yield
\bea
I_A&=&i \int_{0}^t d\tau_2 e^{i\ok 2\tau_2}\left(6t-5\tau_2-i\frac{\partial}{\partial\ok 1}\right) \int_0^{\tau_2} d\tau_1 e^{-i\ok 1\tau_1}\\
&=&\int_0^t d\tau_1 \frac{e^{-i(\ok 1-\ok 2)\tau_1}}{\ok 1} \left(-6t+6\tau_1-\frac{i}{\ok 1}\right) .\nonumber
\eea

In the case $\tau_1>\tau_2$ one finds
\beq
U_2^B(t)|t=0\rangle_0=-\frac{\lambda}{8}\pink{2}
\frac{V_{BB-k_1}V_{BB k_2}}{\ok 1} e^{-i\ok 2 t} I_B e^{-\sigma^2(k_1-k_0)^2-i(k_1-k_0)x_0}
\phi_0^2|k_2\rangle_0
\eeq
where
\bea
I_B=i\int_{0}^t d\tau_1\int_{0}^{\tau_1} d\tau_2e^{i\ok 2\tau_2-i\ok 1\tau_1}\left((\tau_1-\tau_2)+6(t-\tau_1)\right).
\eea
Now we drop the vanishing $\tau_2=0$ contribution to arrive at
\bea
I_B&=&i\int_0^t d\tau_1 e^{-i\ok 1\tau_1} \left(6t-5\tau_1+i\frac{\partial}{\partial\ok 2}\right) \int_{0}^{\tau_1} d\tau_2 e^{i\ok 2\tau_2}\\
&=&\int_0^t d\tau_1 \frac{e^{-i(\ok 1-\ok 2)\tau_1}}{\ok 2} \left(6t-6\tau_1-\frac{i}{\ok 2}\right) \nonumber
\eea
while integrating $\tau_1$ first and then renaming $\tau_2$ would give
\bea
I_B
&=&\int_0^t d\tau_1 \frac{e^{-i(\ok 1-\ok 2)\tau_1}}{\ok 1} \left(6t-6\tau_1+\frac{5i}{\ok 1}\right).
\eea

We see that the naively divergent $(t-\tau_1)$ terms cancel in $I_A+I_B$.   This linear divergence would be caused by the fact that the constant $\phi_0^4$ term, created at time $\tau_1$ or $\tau_2$, would create $\phi_0^2$ at a constant rate as a result of the $\pi_0^2/2$ in $H\p_2$.  The cancellation occurs because, as we have shown, the $\phi_0^4$ term itself vanishes at late times.

Summing the two cases, and again replacing $I_A$ and $I_B$ by the average of the expressions obtained from the two integration orders, one finds the contribution to the final state to be
\bea
\left( U_2^A(t)+U_2^B(t)
\right)|t=0\rangle_0&=&-i\frac{\lambda}{4}\pink{2}e^{-i\ok 2 t}  \frac{V_{BB-k_1}V_{BBk_2}}{\ok {1}}\left(\frac{1}{\ok {1}^2}+\frac{1}{\ok{2}^2}
\right)\\
&&\times
\int_{0}^t d\tau_2e^{-i(\ok 0-\ok 2)\tau_2}
e^{-\sigma^2(k_1-k_0)^2-i(k_1-k_0)x_{\tau_2}}
\phi_0^2|k_2\rangle_0.\nonumber
\eea
Again it will be convenient to rewrite this using a Ward identity
\bea
\left( U_2^A(t)+U_2^B(t)
\right)|t=0\rangle_0&=&-i\frac{\lambda}{4\sqrt{\lambda Q_0}}\pink{2}e^{-i\ok 2 t}  \frac{\left(V_{BB-k_1}\Delta_{k_2B}+V_{BBk_2}\Delta_{-k_1 B}\right)}{\ok {1}}\nonumber
\\
&&\times
\int_{0}^t d\tau_2e^{-i(\ok 0-\ok 2)\tau_2}
e^{-\sigma^2(k_1-k_0)^2-i(k_1-k_0)x_{\tau_2}}
\phi_0^2|k_2\rangle_0.\label{f2d}
\eea

\subsection{The Total}

Finally we are ready to add the 2 zero-mode, 1 meson contributions to the elastic scattering of the final state given in Eqs.~(\ref{f2a}), Eq.~(\ref{f2b}), Eq.~(\ref{f2cb}) and Eq.~(\ref{f2d})
\bea
U(t)|t=0\rangle&=&i\frac{\lambda}{4}\pink{2}e^{-i\ok 2 t}  \frac{S_2}{\ok {1}}
\int_{0}^t d\tau_2e^{-i(\ok 0-\ok 2)\tau_2}
e^{-\sigma^2(k_1-k_0)^2-i(k_1-k_0)x_{\tau_2}}
\phi_0^2|k_2\rangle_0\nonumber
\eea
where
\bea
S_2&=&-V_{BB-k_1k_2}+\frac{1}{\slq}\Bigg[-V_{BB-k_1}\Delta_{k_2B}-V_{BBk_2}\Delta_{-k_1 B}\nonumber\\
&&+\ppin{k\p}\left(V_{-k_1k_2k\p}\Delta_{-k\p B}+V_{B-k_1k\p}\Delta_{-k\p k_2}+\Delta_{k\p-k_1}V_{B-k\p k_2}\right)\Bigg]=0. \label{s2}
\eea
The last equality is a result of the Ward identity (\ref{ward}) for translation invariance.  This implies that no $\phi_0^2$ terms appear at first order in the one-meson sector, as is demanded by translation-invariance. 

\section{From Zero Modes to No Zero Modes} \label{zerosez}

Recall that any translation-invariant state in the kink sector is entirely determined by its primary components, those with no zero modes.  Furthermore, the reduced inner product of Ref.~\cite{menorm} allows one to compute amplitudes using only the no zero-mode sector of the final state.  Therefore, the computation of any initial value problem reduces to the computation of the no zero mode sector of the final state.  

So then why have we wasted so much space calculating the sector of the final state with zero modes?  Because, following the strategy of Subsec.~\ref{da4a2}, we can easily modify those computations to yield the zero-mode free parts of the final state resulting from interactions that create zero modes, which are later destroyed by the free $H\p_2$ evolution.

\subsection{A Single Interaction}

\begin{figure}[htbp]
\centering
\includegraphics[width = 0.6\textwidth]{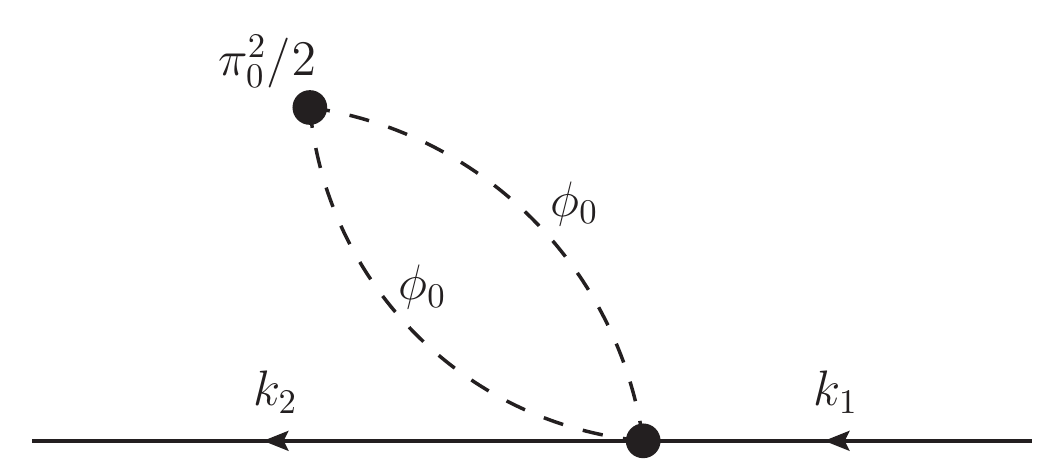}
\caption{This interaction is as in Fig.~\ref{figb2}.  However the two zero modes are absorbed by the $\pi_0^2/2$ kinetic term for the kink center of mass.}\label{figb20}
\end{figure} 

As always, the simplest case is that with a single interaction, in this case that of Eq.~(\ref{h42}).  This creates $m=2$ zero modes, and so we must insert a factor of
\beq
im(m-1)T/2=i(t-\tau_1)
\eeq
where $T=t-\tau_1$ is the time after the creation of the zero modes.  This changes the $\phi_0^2$ part of the final state, given in Eq.~(\ref{f2a}), into the $\phi_0^0$ part
\bea
U_2(t)|t=0\rangle_0&=&\frac{\lambda}{4}\pink{2} \frac{V_{BB-k_1k_2}}{\ok 1} e^{-i\ok 2 t} \label{f0a}\\
&&\times \int_0^td\tau_1(t-\tau_1)e^{-i(\ok 0-\ok 2)\tau_1}e^{-\sigma^2(k_1-k_0)^2-i(k_1-k_0)x_{\tau_1}}|k_2\rangle_0. \nonumber
\eea
This process in drawn in Fig.~\ref{figb20}.

\subsection{A Virtual Meson that Decays to Two Zero Modes}

\begin{figure}[htbp]
\centering
\includegraphics[width = 0.9\textwidth]{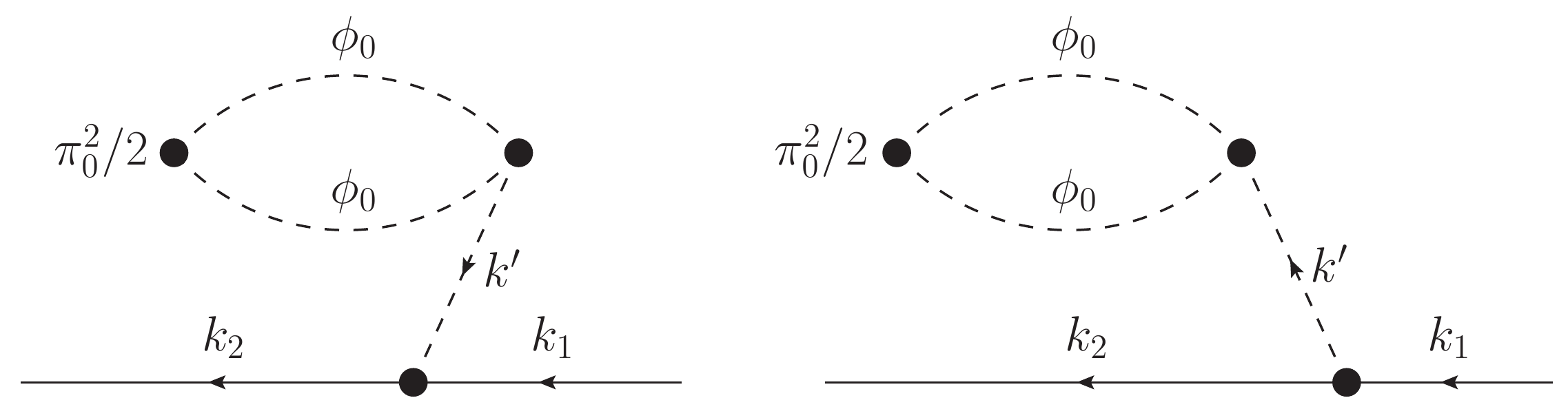}
\caption{This interaction is as in Fig.~\ref{figc2}.  However the two zero modes are annihilated by the $\pi^2_0/2$ term in the free evolution.} \label{figc20}
\end{figure}

Next we turn to the interactions (\ref{tad2}) in which a virtual meson is emitted by meson 1 at time $\tau_1$ and it is absorbed by the kink, creating two zero modes, at time $\tau_2$.  The process in which these two zero modes are removed by the free evolution, drawn in Fig.~\ref{figc20}, contains, a factor of
\beq
im(m-1)T/2=i(t-\tau_2)
\eeq
with respect to the $\phi_0^2$ contributions calculated in Subsec.~\ref{tad2sez}.  

Including this factor in Eq.~(\ref{f2bi}), one finds that the contribution from the case $\tau_1<\tau_2$ is
\bea
U_2^A(t)|t=0\rangle_0&=&-i\frac{\lambda}{8}\pink{2}\ppin{k\p}\int_{0}^t d\tau_1\int_{\tau_1}^t d\tau_2
\frac{V_{-k_1k_2k\p}V_{BB-k\p}}{\ok 1\okp{}}(t-\tau_2)\\
&&\times
e^{-i\ok 2(t-\tau_1)-i\okp{}(\tau_2-\tau_1)-i\ok 0\tau_1}
e^{-\sigma^2(k_1-k_0)^2-i(k_1-k_0)x_{\tau_1}}
|k_2\rangle_0\nonumber\\
&=&-\frac{\lambda}{8}\pink{2}\ppin{k\p}
\frac{V_{-k_1k_2k\p}V_{BB-k\p}}{\ok 1\okp{}^2}e^{-i\ok 2 t}\nonumber\\
&&\times\int_{0}^t d\tau_1\left(t-\tau_1+\frac{i}{\okp{}}\right)
e^{-i(\ok 0-\ok 2)\tau_1}
e^{-\sigma^2(k_1-k_0)^2-i(k_1-k_0)x_{\tau_1}}
|k_2\rangle_0\nonumber
\eea
where again we have dropped the contribution at $\tau_2=t$
\bea
&&i\frac{\lambda}{8}\pink{2}\ppin{k\p}
\frac{V_{-k_1k_2k\p}V_{BB-k\p}}{\ok 1\okp{}^3}e^{-i(\ok 2+\okp{}) t}\label{finale}\\
&&\times\int_{0}^t d\tau_1
e^{-i(\ok 0-\ok 2-\okp{})\tau_1}
e^{-\sigma^2(k_1-k_0)^2-i(k_1-k_0)x_{\tau_1}}
|k_2\rangle_0\nonumber\\
&=&i\frac{\lambda}{8}\frac{\sqrt{\pi}}{2\pi\sigma}\pin{k_2}\ppin{k\p}
\frac{V_{-k_0k_2k\p}V_{BB-k\p}}{\ok 0\okp{}^3}e^{-i(\ok 2+\okp{}) t}
\int_{0}^t d\tau_1
e^{-i(\ok 0-\ok 2-\okp{})\tau_1}
e^{-x_{\tau_1}^2/(4\sigma^2)}
|k_2\rangle_0\nonumber\\
&=&i\frac{\lambda}{8\sqrt{\lambda Q_0}}e^{-i\ok 0t_c}\pin{k_2}\ppin{k\p}
\frac{V_{-k_0k_2k\p}\Delta_{-k\p B}}{k_0 \okp{}}
e^{-\sigma^2(\ok 0/k_0)^2(\ok 0-\ok 2-\okp{})^2-i(\ok 2+\okp{}) (t-t_c)}
|k_2\rangle_0.\nonumber
\eea
The $k\p$ integration causes this term to vanish, as the integrand oscillates quickly.  This argument fails if $k\p$ is a discrete shape mode, and so we will handle this case separately in \ref{stokesez}.

Similarly, in the case $\tau_1>\tau_2$, we include the factor in Eq.~(\ref{f2bi2})
\bea
U_2^B(t)|t=0\rangle_0&=&-i\frac{\lambda}{8}\pink{2}\ppin{k\p}\int_{0}^t d\tau_1\int_{0}^{\tau_1} d\tau_2
\frac{V_{-k_1k_2k\p}V_{BB-k\p}}{\ok 1\okp{}}(t-\tau_2)\\
&&\times
e^{-i\ok 2(t-\tau_1)-i\okp{}(\tau_1-\tau_2)-i\ok 0\tau_1}
e^{-\sigma^2(k_1-k_0)^2-i(k_1-k_0)x_{\tau_1}}
|k_2\rangle_0\nonumber\\
&=&-\frac{\lambda}{8}\pink{2}\ppin{k\p}
\frac{V_{-k_1k_2k\p}V_{BB-k\p}}{\ok 1\okp{}^2}e^{-i\ok 2 t}\nonumber\\
&&\times\int_{0}^t d\tau_1\left(t-\tau_1-\frac{i}{\okp{}}\right)
e^{-i(\ok 0-\ok 2)\tau_1}
e^{-\sigma^2(k_1-k_0)^2-i(k_1-k_0)x_{\tau_1}}
|k_2\rangle_0.\nonumber
\eea
Adding the two contributions one finds
\bea
\left( U_2^A(t)+U_2^B(t)
\right)|t=0\rangle_0&=&-\frac{\lambda}{4\sqrt{\lambda Q_0}}\pink{2}\ppin{k\p}
\frac{V_{-k_1k_2k\p}\Delta_{-k\p B}}{\ok 1}e^{-i\ok 2 t}\label{f0b}\\
&&\times\int_{0}^t d\tau_1 (t-\tau_1)
e^{-i(\ok 0-\ok 2)\tau_1}
e^{-\sigma^2(k_1-k_0)^2-i(k_1-k_0)x_{\tau_1}}
|k_2\rangle_0.\nonumber
\eea

\subsection{One Zero-Mode at Each Vertex}
\begin{figure}[htbp]
\centering
\includegraphics[width = 0.8\textwidth]{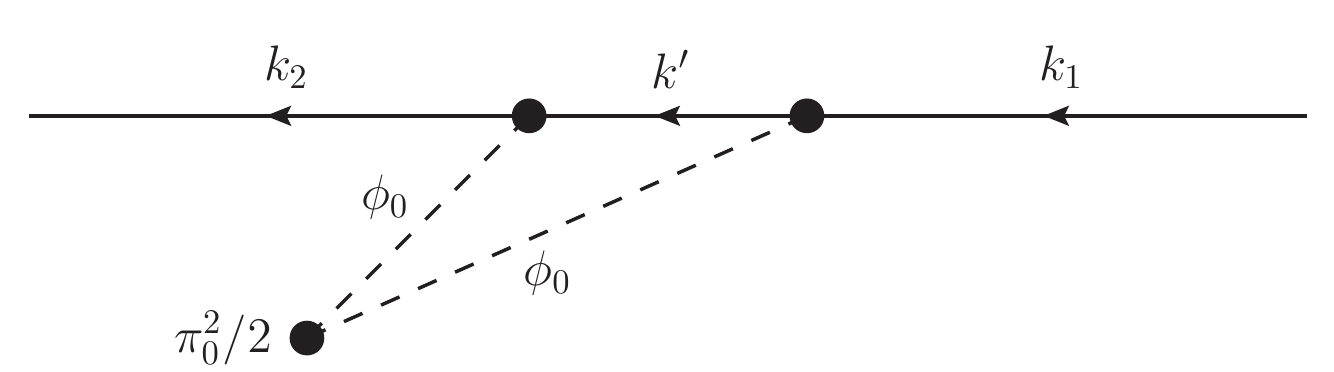}
\caption{Most of this paper is about the computation of this term, which is the only contribution to the scattering amplitude resulting from zero modes.  The process is as in Fig.~\ref{figa2}, except that the two zero modes are annihilated by the $\pi_0^2/2$ in the free evolution operator.} \label{figa}
\end{figure}

Now consider the case in which each vertex creates a single zero mode $\phi_0$.  Since the only operator in the free Hamiltonian that annihilates zero-modes is $\pi_0^2/2$, no zero modes can be annihilated until both are created.  The time $T$ will therefore be equal to $t$ minus whichever of $\tau_1$ and $\tau_2$ is greater.  This process is drawn in Fig.~\ref{figa}.

If $\tau_1<\tau_2$ then Eq.~(\ref{f2c}) is modified to
\bea
U_2^A(t)|t=0\rangle_0&=&-i\frac{\lambda}{4}\pink{2}e^{-i\ok 2t} \ppin{k\p}I_A
\frac{V_{B-k_1k\p}V_{B-k\p k_2}}{\ok 1\okp{}}\\
&&\times
e^{-\sigma^2(k_1-k_0)^2-i(k_1-k_0)x_0}
|k_2\rangle_0\nonumber\\
I_A&=&\int_{0}^t d\tau_1 \int_{\tau_1}^{t} d\tau_2 e^{-i(\okp{}-\ok 2)\tau_2-i(\ok 1-\okp{})\tau_1}(t-\tau_2).\nonumber
\eea
Integrating $\tau_1$ first yields a factor of
\bea
I_A&=&\int_{0}^t d\tau_2 e^{-i(\okp{}-\ok 2)\tau_2}(t-\tau_2)\int_{0}^{\tau_2} d\tau_1 e^{-i(\ok 1-\okp{})\tau_1}\\
&=&\frac{i}{\ok 1-\okp{}}\int_{0}^t d\tau_2 e^{-i(\ok 1-\ok 2)\tau_2} (t-\tau_2)\nonumber
\eea
whereas integrating $\tau_2$ first would yield
\bea
I_A&=&\int_{0}^t d\tau_1 e^{-i(\ok 1-\okp{})\tau_1} \left(t-i\frac{\partial}{\partial \okp{}}\right)\int_{\tau_1}^{t} d\tau_2 e^{-i(\okp{}-\ok 2)\tau_2}\\
&=&\frac{i}{\ok 2-\okp{}}\int_{0}^t d\tau_1 e^{-i(\ok 1-\ok 2)\tau_1} \left(t-\tau_1-\frac{i}{\ok 2-\okp{} }\right).\nonumber
\eea
Again, the integrals commute and so these expressions are equal.  It will be convenient to use the average.

If $\tau_1>\tau_2$ then Eq.~(\ref{f2ca2}) is modified to
\bea
U_2^B(t)|t=0\rangle_0&=&-i\frac{\lambda}{4}\pink{2}e^{-i\ok2 t}\ppin{k\p} I_B
\frac{V_{B-k_1k\p}V_{B-k\p k_2}}{\ok 1\okp{}}\\
&&\times
e^{-\sigma^2(k_1-k_0)^2-i(k_1-k_0)x_0}
|k_2\rangle_0\nonumber\\
I_B&=&\int_{0}^t d\tau_1 \int_{0}^{\tau_1} d\tau_2 e^{i(\okp{}+\ok 2)\tau_2-i(\ok 1+\okp{})\tau_1}(t-\tau_1).\nonumber
\eea
Integrating $\tau_1$ first
\bea
I_B&=& \int_{0}^{t} d\tau_2  e^{i(\okp{}+\ok 2)\tau_2}\left(t-i\frac{\partial}{\partial\okp{}}\right)\int_{\tau_2}^t d\tau_1 e^{-i(\ok 1+\okp{})\tau_1}\\
&=&-\frac{i}{\okp{}+\ok 1}\int_{0}^{t} d\tau_2  e^{-i(\ok 1-\ok 2)\tau_2}\left(t-\tau_2+\frac{i}{\okp{}+\ok 1}\right)\nonumber
\eea
while integrating $\tau_2$ first
\bea
I_B&=&-\frac{i}{\okp{}+\ok 2}\int_{0}^t d\tau_1  e^{-i(\ok 1-\ok 2)\tau_1}(t-\tau_1).
\eea

Now, replacing all dummy variables $\tau_2$ with $\tau_1$ and averaging over the integral orderings one finds
\bea
I_A+I_B&=&\int_{0}^t d\tau_1  e^{-i(\ok 1-\ok 2)\tau_1}\left[i \left(\frac{\okp{}}{\ok{1}^2-\okp{}^2}+\frac{\okp{}}{\ok{2}^2-\okp{}^2}\right)(t-\tau_1)\right.\\
&&\left.+\frac{1}{2(\ok 1+\okp{})^2}+\frac{1}{2(\ok 2-\okp{})^2}
\right].\nonumber
\eea
Reinserting these integrals in the equations for the final states one finds
\bea
\left( U_2^A(t)+U_2^B(t)
\right)|t=0\rangle_0&=&\frac{\lambda}{4}\pink{2}\ppin{k\p}
\frac{V_{B-k_1k\p}V_{B-k\p k_2}}{\ok 1}e^{-i\ok 2 t}\nonumber\\
&&\hspace{-4cm}\ \ \times\int_{0}^t d\tau_1 \left[(t-\tau_1)\left(\frac{1}{\ok{1}^2-\okp{}^2}+\frac{1}{\ok{2}^2-\okp{}^2}\right)
-\frac{i}{2\okp{}(\ok 1+\okp{})^2}-\frac{i}{2\okp{}(\ok 2-\okp{})^2}
\right]\nonumber\\
&&\hspace{-4cm}\ \ \times
e^{-i(\ok 0-\ok 2)\tau_1}
e^{-\sigma^2(k_1-k_0)^2-i(k_1-k_0)x_{\tau_1}}
|k_2\rangle_0=A+B\nonumber
\eea
where
\bea\label{f0c}
A&=&\frac{\lambda}{4\sqrt{\lambda Q_0}}\pink{2}\ppin{k\p}
\frac{\left(V_{B-k_1k\p}\Delta_{k_2-k\p}+V_{B-k\p k_2}\Delta_{-k_1 k\p}\right)}{\ok 1}e^{-i\ok 2 t}\\
&&\times\int_{0}^t d\tau_1 (t-\tau_1)
e^{-i(\ok 0-\ok 2)\tau_1}
e^{-\sigma^2(k_1-k_0)^2-i(k_1-k_0)x_{\tau_1}}
|k_2\rangle_0 \nonumber
\eea
and
\bea
B&=&-i\frac{\lambda}{8}\pink{2}\ppin{k\p}
\frac{V_{B-k_1k\p}V_{B-k\p k_2}}{\ok 1\okp{}}e^{-i\ok 2 t} \label{f0d}\\
&&\times\int_{0}^t d\tau_1 \left[\frac{1}{(\ok 1+\okp{})^2}+\frac{1}{(\ok 2-\okp{})^2}
\right]
e^{-i(\ok 0-\ok 2)\tau_1}
e^{-\sigma^2(k_1-k_0)^2-i(k_1-k_0)x_{\tau_1}}
|k_2\rangle_0.\nonumber
\eea
While the term $A$ looks like that seen in the previous processes, the term $B$ is different, in that it does not contain a $t-\tau$ factor.  We will see that it is the only term in this section that contributes to elastic scattering.

\subsection{No Zero Modes from Four Zero Modes}

\begin{figure}[htbp]
\centering
\includegraphics[width = 0.9\textwidth]{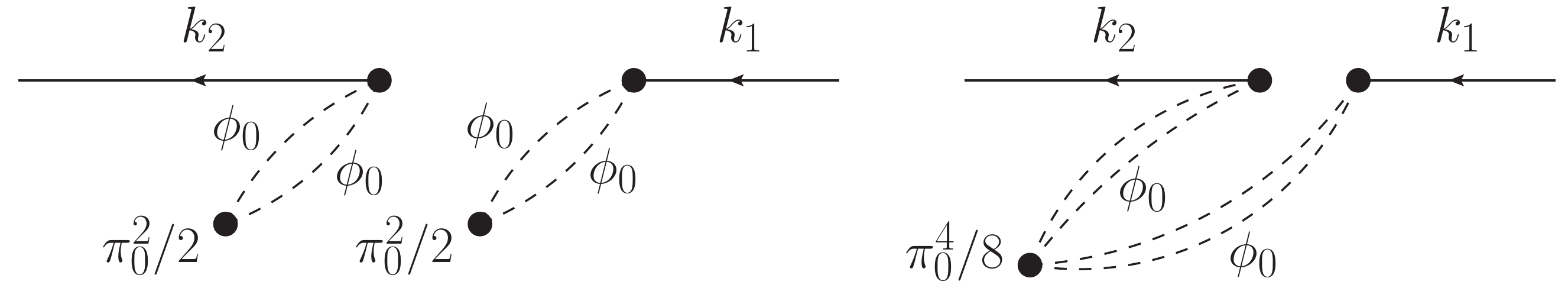}
\caption{This is as in Fig.~\ref{fig4} except that all four zero modes are annihilated by the kink center of mass kinetic term in the free evolution operator.} \label{fig40}
\end{figure}

The last process that leads to a single meson creates two zero modes in each of two interactions in Eq.~(\ref{h32}), and lets them both be destroyed by the $e^{-i\pi_0^2T/2}$ in the free evolution operator.  It is drawn in Fig.~\ref{fig40}.

Consider first $\tau_1<\tau_2$.  Now, as always, two zero modes are created at $\tau_1$ and two more at $\tau_2$.  There are two ways in which the zero modes may be destroyed.  First, the linear $-i\pi_0^2(\tau_2-\tau_1)/2$ term in the evolution operator may destroy two zero modes between times $\tau_1$ and $\tau_2$, and then the linear $-i\pi_0^2(t-\tau_2)/2$ term in the evolution operator may destroy two zero modes between times $\tau_2$ and $t$.  This contributes a factor of
\beq
[-i\pi_0^2(\tau_2-\tau_1)/2,\phi_0^2][-i\pi_0^2(t-\tau_2)/2,\phi_0^2]=(\tau_1-\tau_2)(t-\tau_2)
\eeq
to the $\phi_0^0$ term in the final state with respect to the $\phi_0^4$ term calculated in Sec.~\ref{foursez}.  

But it may also be that all four zero modes survive until $\tau_2$, and so are annihilated by the $-\pi_0^4(t-\tau_2)^2/8$ quadratic term in the free evolution operator between times $\tau_2$ and $t$.  This possibility contributes a factor of
\beq
[-\pi_0^4(t-\tau_2)^2/8,\phi_0^4]=-3(t-\tau_2)^2.
\eeq
Of course these processes, having the same final state, add coherently and so lead to a total weight which is the sum of these factors
\beq
 (t-\tau_2)(-3t+\tau_1+2\tau_2). \label{fac}
\eeq

The contribution (\ref{f4a}) to the $\phi_0^4$ sector of the final state then becomes
\bea
U_2^A(t)|t=0\rangle_0&=&-\frac{\lambda}{8}\pink{2}I_A e^{-i\ok 2 t} 
\frac{V_{BB-k_1}V_{BB k_2}}{\ok 1}
e^{-\sigma^2(k_1-k_0)^2-i(k_1-k_0)x_0}
|k_2\rangle_0\nonumber\\
I_A&=&\int_{0}^t d\tau_1\int_{\tau_1}^t d\tau_2 e^{i\ok 2\tau_2-i\ok 1\tau_1} (t-\tau_2)(-3t+\tau_1+2\tau_2)
.
\eea
Let us first integrate $\tau_2$, as usual dropping $\tau_2=t$ as its contribution vanishes after the other integrals have been performed
\bea
I_A&=&\int_{0}^t d\tau_1 e^{-i\ok 1\tau_1}\left(t+i\frac{\partial}{\partial\ok 2}\right)\left(-3t+\tau_1-2i\frac{\partial}{\partial\ok 2}\right)\int_{\tau_1}^t d\tau_2 e^{i\ok 2\tau_2}\\ 
&=&i\int_{0}^t d\tau_1 e^{-i(\ok 1 - \ok 2)\tau_1}\left(t-\tau_1+i\frac{\partial}{\partial\ok 2}\right)\left(-3(t-\tau_1)-2i\frac{\partial}{\partial\ok 2}\right)\frac{1}{\ok 2}\nonumber\\ 
&=&\frac{i}{\ok 2}\int_{0}^t d\tau_1 e^{-i(\ok 1 - \ok 2)\tau_1}\left[
-3(t-\tau_1)^2+\frac{5i(t-\tau_1)}{\ok 2} +\frac{4}{\ok{2}^2}
\right].\nonumber
\eea
On the other hand, performing the $\tau_1$ integration first leads to
\bea
I_A&=&\int_{0}^t d\tau_2e^{i\ok 2\tau_2}(t-\tau_2)\left(-3t+i\frac{\partial}{\partial\ok 1}+2\tau_2\right)\int_{0}^{\tau_2} d\tau_1 e^{-i\ok 1\tau_1} \\
&=&\frac{i}{\ok 1}\int_{0}^t d\tau_2e^{i(\ok 2-\ok 1)\tau_2}(t-\tau_2)\left(-3(t-\tau_2)-\frac{i}{\ok 1}\right).\nonumber
\eea

Consider now $\tau_1>\tau_2$.  The factor that one must now include is obtained by exchanging $\tau_1$ and $\tau_2$ in Eq.~(\ref{fac})
\beq
(t-\tau_1)(-3t+\tau_2+2\tau_1). 
\eeq
This modifies the contribution (\ref{f4b}) to
\bea
U_2^B(t)|t=0\rangle_0&=&-\frac{\lambda}{8}\pink{2} I_Be^{-i\ok 2 t}
\frac{V_{BB-k_1}V_{BB k_2}}{\ok 1}
e^{-\sigma^2(k_1-k_0)^2-i(k_1-k_0)x_0}
|k_2\rangle_0\nonumber\\
I_B&=&\int_{0}^t d\tau_1\int_{0}^{\tau_1} d\tau_2 e^{i\ok 2\tau_2-i\ok 1\tau_1}(t-\tau_1)(-3t+\tau_2+2\tau_1).
\eea
Now we first integrate $\tau_1$, dropping $\tau_1=t$
\bea
I_B&=&\int_{0}^t d\tau_2e^{i\ok 2\tau_2}\left(t-i\frac{\partial}{\partial\ok 1}\right)\left(-3t+\tau_2+2i\frac{\partial}{\partial\ok 1}\right)
\int_{\tau_2}^t d\tau_1e^{-i\ok 1\tau_1}\\
&=&-i\int_{0}^t d\tau_2e^{-i(\ok1-\ok 2)\tau_2}\left(t-\tau_2-i\frac{\partial}{\partial\ok 1}\right)\left(-3(t-\tau_2)+2i\frac{\partial}{\partial\ok 1}\right)\frac{1}{\ok 1}\nonumber\\
&=&-\frac{i}{\ok 1}\int_{0}^t d\tau_2e^{-i(\ok1-\ok 2)\tau_2}\left[-3(t-\tau_2)^2-\frac{5i(t-\tau_2)}{\ok 1}+\frac{4}{\ok{1}^2} 
\right].\nonumber
\eea
On the other hand, integrating $\tau_2$ first
\bea
I_B&=&\int_{0}^t d\tau_1 e^{-i\ok 1\tau_1}
(t-\tau_1)\left(-3t-i\frac{\partial}{\partial\ok 2}+2\tau_1\right)
\int_{0}^{\tau_1} d\tau_2 e^{i\ok 2\tau_2}\\
&=&-\frac{i}{\ok 2}\int_{0}^t d\tau_1 e^{-i(\ok 1-\ok 2)\tau_1}
(t-\tau_1)\left(-3(t-\tau_1)+\frac{i}{\ok 2}\right).
\nonumber
\eea

Again we replace the dummy variables $\tau_2$ with $\tau_1$ and average over integration orders to obtain
\beq
I_A+I_B=\int_{0}^t d\tau_1 e^{-i(\ok 1-\ok 2)\tau_1} \left[ 
-2(t-\tau_1)\left(\frac{1}{\ok{2}^2}+\frac{1}{\ok{1}^2}\right)
+\frac{2i}{\ok{2}^3}-\frac{2i}{\ok{1}^3}
\right].
\eeq
We note that at $k_1=\pm k_2$, corresponding to the average value in elastic scattering, the terms that are linearly divergent in $t-\tau_1$ are nonzero but the constant piece vanishes.  As a result, these terms will not contribute to our final amplitude.  The contribution to the final state is
\bea
\left( U_2^A(t)+U_2^B(t)
\right)|t=0\rangle_0&=&\frac{\lambda}{4}\pink{2}
\frac{V_{BB-k_1}V_{BBk_2}}{\ok 1}e^{-i\ok 2 t}\nonumber\\
&&\hspace{-4cm}\ \ \times\int_{0}^t d\tau_1 \left[ 
(t-\tau_1)\left(\frac{1}{\ok{2}^2}+\frac{1}{\ok{1}^2}\right)
-\frac{i}{\ok{2}^3}+\frac{i}{\ok{1}^3}
\right]\nonumber\\
&&\hspace{-4cm}\ \ \times
e^{-i(\ok 0-\ok 2)\tau_1}
e^{-\sigma^2(k_1-k_0)^2-i(k_1-k_0)x_{\tau_1}}
|k_2\rangle_0=C+D\nonumber
\eea
where 
\bea\label{f0e}
C&=&\frac{\lambda}{4\sqrt{\lambda Q_0}}\pink{2}
\frac{V_{BB-k_1}\Delta_{k_2 B}+V_{BB k_2}\Delta_{-k_1 B}}{\ok 1}e^{-i\ok 2 t}\\
&&\times\int_{0}^t d\tau_1 (t-\tau_1)
e^{-i(\ok 0-\ok 2)\tau_1}
e^{-\sigma^2(k_1-k_0)^2-i(k_1-k_0)x_{\tau_1}}
|k_2\rangle_0 \nonumber
\eea
and
\bea
D&=&\frac{i\lambda}{4}\pink{2}
\frac{V_{BB-k_1}V_{BBk_2}}{\ok 1}e^{-i\ok 2 t}
\\&&\times
\int_{0}^t d\tau_1 \left[\frac{1}{\ok{1}^3}-\frac{1}{\ok 2^3}
\right]
e^{-i(\ok 0-\ok 2)\tau_1}
e^{-\sigma^2(k_1-k_0)^2-i(k_1-k_0)x_{\tau_1}}
|k_2\rangle_0.\nonumber
\eea

In this section, like the two before it, we have been careful to distinguish $k_0$, $k_1$ and $k_2$, even though they only differ by of order $O(1/\sigma)$.  Our care has paid off, because these differences were multiplied by factors of $t-\tau$ and even $(t-\tau)^2$ in terms where zero modes were canceled.  These factors resulted from the fact that the free evolution leads to a constant rate of demotion from $\phi_0^m$ to $\phi_0^{m-2}$.

However, now we have already calculated these factors, and they are not present in $D$.  Therefore, in $D$, one can safely take our limit $m\sigma\rightarrow\infty$, which implies that, in the support of our $e^{-\sigma^2(k_1-k_0)^2}$ weight, $k_1$ may be replaced with $k_0$.  Thus, with the usual argument that $x/\sigma$ is negligible when multiplied by $\g_B(x)$, we may write
\bea
D&=&\frac{i\lambda}{4}\pink{2}
\frac{V_{BBk_2}}{\ok 0}e^{-i\ok 2 t}\int_{0}^t d\tau_1 \left[\frac{1}{\ok{0}^3}-\frac{1}{\ok 2^3}
\right]
e^{-i(\ok 0-\ok 2)\tau_1}
\\&&\times
\int dx \V3\g_B^2(x) \g_{-k_0}(x)
e^{-\sigma^2(k_1-k_0)^2-i(k_1-k_0)(x_{\tau_1}-x)}
|k_2\rangle_0\nonumber\\
&=&\frac{i\lambda}{4}\frac{\sqrt{\pi}}{2\pi\sigma}\pin{k_2}
\frac{V_{BB-k_0}V_{BBk_2}}{\ok 0}e^{-i\ok 2 t}\left[\frac{1}{\ok{0}^3}-\frac{1}{\ok 2^3}
\right]
\nonumber\\&&\times
\int_{0}^t d\tau_1 e^{-x_{\tau_1}^2/(4\sigma^2)-i(\ok 0-\ok 2)\tau_1} |k_2\rangle_0\nonumber\\
&=&\frac{i\lambda}{4}\pin{k_2}
\frac{V_{BB-k_0}V_{BBk_2}}{k_0}e^{-i\ok 2 t}e^{-i(\ok 0 - \ok 2) t_c}\left[\frac{1}{\ok{0}^3}-\frac{1}{\ok 2^3}
\right]e^{-\sigma^2(k_0+k_2)^2}
|k_2\rangle_0.\nonumber
\eea
In the support of the $e^{-\sigma^2(k_0+k_2)^2}$, so that $k_2=-k_0+O(1/\sigma)$, we can see that the term in square brackets is $1/\ok{0}^3$ times a factor of order $O(1/(m\sigma))$  and so vanishes as $m\sigma\rightarrow\infty$.  Thus $D$ will not contribute to the amplitude and we will not consider it further.

\subsection{The Total}

Finally we are ready to add the contributions in Eqs.~(\ref{f0a}), (\ref{f0b}), (\ref{f0c}), (\ref{f0d}) and (\ref{f0e}) to the one-meson, no zero-mode part of the final state.  Recall that these are the contributions arising from interactions that created zero modes, that were later annihilated.  The sum is
\bea
U_2(t)|t=0\rangle_0&=&B-\frac{\lambda}{4}\pink{2}
\frac{S_2}{\ok 1}e^{-i\ok 2 t}
\\&&\times
\int_{0}^t d\tau_1 (t-\tau_1)
e^{-i(\ok 0-\ok 2)\tau_1}
e^{-\sigma^2(k_1-k_0)^2-i(k_1-k_0)x_{\tau_1}}
|k_2\rangle_0. \nonumber
\eea
The quantity $S_2$ was defined in Eq.~(\ref{s2}) where it was noted that $S_2=0$ as a result of the Ward Identity (\ref{ward}).  This leaves $B$.

The quantity $B$ was defined in Eq.~(\ref{f0d}).  It is
\bea
B&=&-i\frac{\lambda}{8}\frac{\sqrt{\pi}}{2\pi\sigma}\pin{k_2}\ppin{k\p}
\frac{V_{B-k_0k\p}V_{B-k\p k_2}}{\ok 0\okp{}}e^{-i\ok 2 t} \label{b}\\
&&\times\left[\frac{1}{(\ok 0+\okp{})^2}+\frac{1}{(\ok 2-\okp{})^2}
\right]
\int_{0}^t d\tau_1 
e^{-x_{\tau_1}^2/(4\sigma^2)-i(\ok 0-\ok 2)\tau_1}
|k_2\rangle_0\nonumber\\
&=&-i\frac{\lambda}{8}\pin{k_2}\ppin{k\p}
\frac{V_{B-k_0k\p}V_{B-k\p k_2}}{k_0\okp{}}e^{-i\ok 2 t} \nonumber\\
&&\times\left[\frac{1}{(\ok 0+\okp{})^2}+\frac{1}{(\ok 2-\okp{})^2}
\right]
e^{-\sigma^2(k_0+k_2)^2-i(\ok 0-\ok 2)t_c}
|k_2\rangle_0.\nonumber
\eea

In the support of $e^{-\sigma^2(k_0+k_2)^2}$ we may set $\ok 0=\ok 2$ and so manipulate
\bea
V_{B-k_0k\p}V_{B-k\p k_2}\left[\frac{1}{(\ok 0+\okp{})^2}+\frac{1}{(\ok 2-\okp{})^2}\right]\,\,&=&\,\,\frac{\Delta_{-k_0k\p}\Delta_{-k_0-k\p}}{\lambda Q_0}\left[(\ok 0-\okp{})^2+(\ok 0+\okp{})^2
\right]\nonumber\\
&=&\frac{2\Delta_{-k_0k\p}\Delta_{-k_0-k\p}}{\lambda Q_0}\left(\ok{0}^2+\okp{}^2\right).
\eea
Here we replaced $V_{B-k\p k_2}$ with $V_{B-k\p -k_0}$, which yields a phase $e^{-i(k_2+k_0)x}$.  However, the $\g_B(x)$ is supported at $x\sim O(1/m)$ and $k_2+k_0$ is of order $O(1/\sigma)$ so the argument of the phase is of order $1/(\sigma m)$ which tends to zero, so the phase factor tends to unity.

Therefore we conclude
\bea
U_2(t)|t=0\rangle_0&=&-i\frac{1}{4Q_0k_0}\pin{k_2}e^{-i\ok 2 t}
e^{-\sigma^2(k_0+k_2)^2+i(k_0+k_2)x_0} \label{a4} \\
&&\times \ppin{k\p}\left(\ok{0}^2+\okp{}^2\right)
\frac{\Delta_{-k_0k\p}\Delta_{-k_0-k\p}}{\okp{}}
|k_2\rangle_0.\nonumber
\eea
This, together with the terms found in Sec.~\ref{calcsez}, is the part of the final state corresponding to meson with no zero modes that is not forward scattered.

\section{The Elastic Scattering Probability}
Adding together the contributions to the final state from Eqs.~(\ref{a1}), (\ref{a2}), (\ref{a3}) and (\ref{b})  finally we find
\bea
U_2(t)|t=0\rangle&=&-i\pin{k_2} e^{-i\ok 2 t}R(k_2)e^{-\sigma^2(k_0+k_2)^2+i(k_0+k_2)x_0}|k_2\rangle_0  \label{fin}
\eea
where the reflection coefficient is
\bea
R(k_2)&=&{\lambda}(A(k_2)+B(k_2)+C(k_2)+D(k_2))
\eea
and
\bea
A(k_2)&=& \frac{1}{{8k_0}}\ppin{k\p}\left(\frac{1}{\left(\ok{0}+\okp{}\right)^2}+\frac{1}{\left(\ok{2}-\okp{}\right)^2}\right)
\frac{V_{B-k_0k\p}V_{B-k\p k_2}}{\okp{}}
\\
B(k_2)&=&\frac{V_{\I k_0-k_2}}{4k_0}
\nonumber\\
C(k_2)&=&-\frac{1}{{4k_0}}\ppin{k\p}\frac{V_{-k_0k_2k\p}V_{\I-k\p}}{ \okp{}^2}
\nonumber\\
D(k_2)&=& \frac{1}{8k_0}\ppinkp{2}\frac{(\okp 1+\okp 2) V_{k_0-k\p_1-k\p_2}V_{-k_0k\p_1k\p_2}}{ \okp 1\okp 2\left(\ok {0}^2-\left(\okp 1+\okp 2\right)^2+i\epsilon\right)}
.\nonumber
\eea
For example, $A(k_2)$ is just the coefficient in $B$ in Eq.~(\ref{b}) divided by $\lambda$.   We remind the reader that $U_2$ is not unitary, as we have defined it to be just to be the part of the evolution operator that leads to one nonforward meson and no zero modes. Note that at $k_2=-k_0$ one may simplify
\beq
A(-k_0)= \frac{1}{{4k_0}\lambda Q_0}\ppin{k\p}\left(\frac{\ok{0}^2+\okp{}^2}{\okp{}}\right)
\Delta_{-k_0k\p}\Delta_{-k_0-k\p}.
\eeq

Following Ref.~\cite{elas1}, it is easy to see that the probability of elastic scattering is $|R(-k_0)|^2$.  This calculation is done using the reduced inner product of \cite{menorm}, which carefully removes the divergences arising from the infinite moduli space.  Using
\beq
|k_1\rangle_0=\Bd 1\vac_0\hsp B^{\ddag\dag}_{k_1}=\frac{B_{k_1}}{2\ok 1}\hsp \langle 0|0\rangle_{\rm{red}}=\sqrt{Q_0}
\eeq
one finds that at leading order the reduced inner product of $|k_1\rangle$ and $|k_2\rangle$ is $\sqrt{Q_0}2\pi \delta(k_1-k_2)/(2\ok 1)$.  Subleading corrections are computed in Ref.~\cite{menorm} and it is argued that they vanish in the present case in Ref.~\cite{elas1}.

The reduced norm squared of the elastic scattered part of the final state (\ref{fin}) is then
\beq
\langle t=0|U_2^\dag(t)U_2(t)|t=0\rangle_{\rm{red}}=\sqrt{Q_0}|R(-k_0)|^2\frac{\sqrt{\pi}}{4\sqrt{2}\pi\sigma\ok 0}.
\eeq
Here we have used the fact that $\sigma m\rightarrow\infty$ to approximate $R$ to be independent of $k_2$ over the support of the Gaussian, so that it could be pulled out of the integral, evaluated at $-k_0$.

On the other hand, the reduced norm squared of the total final state is equal to the reduced norm squared of the initial state $|t=0\rangle$, as a result of the unitarity of the evolution, which is
\beq
\langle t=0|t=0\rangle_{\rm{red}}=\sqrt{Q_0}\frac{\sqrt{\pi}}{4\sqrt{2}\pi\sigma\ok 0}.
\eeq
The probability of elastic scattering is just the ratio of these two reduced norms
\beq
P=\frac{\langle t=0|U_2^\dag(t)U_2(t)|t=0\rangle_{\rm{red}}}{\langle t=0|t=0\rangle_{\rm{red}}}=|R(-k_0)|^2.
\eeq
Indeed, the reduced norm was developed just to solve this problem.

\section{Applications}

After a long calculation, we have recovered the results of Ref.~\cite{elas1}.  What have we gained?

We have drawn diagrams corresponding to each process.  Yet no Feynman rules have been given that would derive the corresponding contribution to the amplitude from the diagrams.  We intend to use this collection of examples to guide the derivation of such Feynman rules for kink sector perturbation theory.  With this, we hope that such calculations in the future may be much faster.  Indeed, the fact that the derivation of the elastic scattering amplitude in Ref.~\cite{elas1} was so short, gives us hope that such a simplification is possible.  

A more streamlined framework will allow for higher order computations.  These have several potential applications.  First, by summing bubble diagrams, one may see a complex shift in the location of the pole corresponding to the twice-excited shape mode resonance.  The width of this resonance should correspond to the lifetime of this unstable state calculated in Ref.~\cite{alberto}, which agrees with the classical field theory calculation of Ref.~\cite{mm}.  One can test to see whether, like in the vacuum sector, also in the kink sector the lifetimes of unstable states may be read off of the imaginary parts of the self-energies.

The situation potentially differs qualitatively from the familiar vacuum sector case when one goes beyond leading order.  Here zero modes created at one bubble may annihilate those created at another.  It remains to be seen whether this simply leads to subleading corrections corresponding to larger bubbles, or else a qualitative change in the structures of these resonances.  Either way, we hope to calculate these subleading corrections as they may yield, for the first time, the correction to the lifetime of an unstable excited soliton state.

Finally, we would like to study higher order diagrams to search for a kink sector LSZ reduction theorem.  In this paper, and in Ref.~\cite{memult}, we have observed that initial and final state corrections always seem to cancel, by a number of different mechanisms.  This leads one to wonder just how generic this result is, and whether bubbles on external legs can be easily summed.

If one considers initial conditions with multiple mesons, one may also study meson fusion.  Using coherent states to create a classical limit as in \cite{alberto}, this should allow a study of the negative radiation pressure observed in Refs.~\cite{tomrad1,tomrad2,tomrad3}.

Of course, kinks themselves have limited phenomenological interest.  In general, 1+1d scalar models with kinks are instead used as toy models either for QCD \cite{dhn2} or for quantum gravity \cite{stotzel95,zhong23}.  In the near future we hope to generalize linearized soliton perturbation theory to solitons in more dimensions, and so the answers to the above questions may have more relevant applications.

\appendix
\section{Ward Identities}


Consider the $n$-point functions
\beq
V_{A_1\cdots A_n}=\int dx \V{n} \g_{A_1}(x)\cdots \g_{A_n}(x)
\eeq
where $A_i$ runs over continuum modes $k$, shape modes $S$ and the zero mode $B$. These correspond to $n$-point functions with external legs $A_i$ corresponding to various zero and normal modes. The generalization containing factors of $\I(x)$ is obvious.  

Now consider an $n$-point function containing at least one zero mode.  The $n$-point function is symmetric, so let us put the zero mode in the last index $A_n=B$. Then $V_{A_1\cdots A_{n-1}B}$ satisfies a Ward Identity corresponding to translation invariance.  Schematically the Ward Identity is
\beq
V_{A_1\cdots A_{n-1}B}=\frac{1}{\slq}\sum_{i=1}^{n-1}\ppin{A\p}\Delta_{A_i-A\p}V_{A_1\cdots A_{i-1}A\p A_{i+1}\cdots A_{n-1}}.
\eeq
The matrix $\Delta$, defined in Eq.~(\ref{ddef}), plays the role of the momentum operator.  Here, breaking from our usual notation, the symbol $\dint$ includes not only shape modes but also the zero mode.  The constant factor of $\slq$ is the result of various conventions.

These are derived by noting that
\beq
\V{n}\g_B(x)=-\frac{\partial_x{\V{n-1}}}{\sqrt{\lambda Q_0}}
\eeq
and integrating by parts to move the derivative onto the other factors $\g_A(x)$.  The identity, in the form of the normal mode completeness relation (\ref{crapp}), which is a standard result in Sturm-Liouville theory, is inserted to turn $\g_A\p(x)$ into a $\Delta_{A_iA_j}$ matrix, which represents translations on the normal modes.  In this appendix, some such Ward Identities will be derived.

Note that the only continuous global symmetry in our model is translation invariance.  However, in more general models in which global symmetries are explicitly broken by classical solutions, we expect the same results to hold for the corresponding zero modes.  Note that this is true even if, as in the present case, the ground state in the soliton sector preserves the classically-broken symmetry as a result of the Coleman-Mermin-Wagner theorem.

\subsection{Warm Up}

In the special case $N=3$ we can use the fact that $\V2$ satisfies the Sturm-Liouville equations of motion of the normal modes to simplify the Ward identities further.  This will be the first approach below.

Using
\beq
\g_B(x)=-\frac{f\p(x)}{\sqrt{Q_0}}
\eeq
one can expand
\bea
V_{B k_2 k_1}&=&\int dx \V3 \g_B(x) \g_{k_2}(x)\g_{k_1}(x)\\
&=&-\frac{1}{\sqrt{Q_0}}\int dx \left( \V3 f\p(x) \right) \g_{k_2}(x)\g_{k_1}(x)\nonumber\\
&=&-\frac{1}{\slq}\int dx \partial_x\left( \V2 + C_{k_1k_2}\right) \g_{k_2}(x)\g_{k_1}(x)\nonumber
\eea
where $C_{k_1k_2}$ is independent of $x$ but otherwise arbitrary.

\subsubsection{Approach One}
 
As $C_{k_1k_2}$ is $x$-independent, its derivative vanishes and we may drop it.  Now, cut off the integration at $\pm\hat{x}$, such that $|\hat{x}|\gg 1/m$ and integrate by parts
\bea
V_{B k_2 k_1}&=&-\frac{m^2}{\slq}\left( \g_{k_2}(\hat x)\g_{k_1}(\hat x)-\g_{k_2}(-\hat x)\g_{k_1}(-\hat x)
\right)\\
&&
+\frac{1}{\slq}\int^{\hat x}_{-\hat x} dx  \V2\left(\g_{k_2}(x)\g\p_{k_1}(x)+\g\p_{k_2}(x)\g_{k_1}(x)\right)\nonumber\\
&=&-\frac{m^2}{\slq}\left( \g_{k_2}(\hat x)\g_{k_1}(\hat x)-\g_{k_2}(-\hat x)\g_{k_1}(-\hat x)
\right)\nonumber\\
&&
+\frac{1}{\slq}\int^{\hat x}_{-\hat x} dx  \left[\g\p_{k_1}(x)\left(\ok{2}^2+\partial_x^2\right)\g_{k_2}(x)+\g\p_{k_2}(x)\left(\ok{1}^2+\partial_x^2\right)\g_{k_1}(x)
\right]\nonumber\\
&=&-\frac{m^2}{\slq}\left( \g_{k_2}(\hat x)\g_{k_1}(\hat x)-\g_{k_2}(-\hat x)\g_{k_1}(-\hat x)
\right)+\frac{1}{\slq}\left( \g\p_{k_2}(\hat x)\g\p_{k_1}(\hat x)-\g\p_{k_2}(-\hat x)\g\p_{k_1}(-\hat x)
\right)\nonumber\\
&&
+\frac{\left(\ok{1}^2-\ok{2}^2\right)\Delta_{k_1k_2}}{\slq}\nonumber\\
&=&-\frac{m^2+k_1k_2}{\slq}\left( \g_{k_2}(\hat x)\g_{k_1}(\hat x)-\g_{k_2}(-\hat x)\g_{k_1}(-\hat x)
\right)+\frac{\left(\ok{1}^2-\ok{2}^2\right)\Delta_{k_1k_2}}{\slq}.\nonumber
\eea
Note that the term in the first parenthesis is proportional to $e^{\pm i\hat x (k_1+k_2)}$ which, by the Riemann-Lebesque Lemma, will vanish when folded into any integrable function of $k_1+k_2$, such as our normalizable wave functions.  One might have expected it to be proportional to $\delta(k_1+k_2)$, however it remains finite when $k_1+k_2=0$ and so the constant of proportionality is zero.  The left side also remains finite at $k_1+k_2=0$ because $\g_B(x)$ has compact support.  Thus, taking the limit $\hat x\rightarrow \infty$, one arrives at
\beq
V_{B k_2 k_1}=\frac{\left(\ok{1}^2-\ok{2}^2\right)\Delta_{k_1k_2}}{\slq}. \label{warda}
\eeq

In the above derivation, $k_1$ and $k_2$ could be continuum modes or shape modes.  However, the derivation also applies to the case in which $k_1$ or $k_2$ is the zero mode.  In this case, the corresponding frequency in the Sturm-Liouville equation satisfied by $\g(x)$ vanishes and so one obtains
\beq
V_{B B k}=\frac{\ok{}^2\Delta_{kB}}{\slq}\hsp V_{BBB}=0. \label{warda2}
\eeq

\subsubsection{Approach Two}

If we keep the $C_{k_1k_2}$ when integrating by parts, one arrives at
\bea
V_{B k_2 k_1}&=&-\frac{(m^2+C_{k_1k_2})}{\slq}\left( \g_{k_2}(\hat x)\g_{k_1}(\hat x)-\g_{k_2}(-\hat x)\g_{k_1}(-\hat x)
\right)\label{ve}\\
&&
+\frac{1}{\slq}\int dx \left( \V2 + C_{k_1k_2}\right) \left(\g_{k_2}(x)\g\p_{k_1}(x)+\g\p_{k_2}(x)\g_{k_1}(x)\right)\nonumber
\eea
 where $\hat x$ is a spatial cutoff, which should be taken to infinity.  The boundary term on the first line oscillates rapidly and so vanishes as a distribution unless $k_1+k_2=0$.  It therefore may only contribute a divergent term at $k_1+k_2=0$, but $V_{Bk_1k_2}$ has no such large $x$ divergence as $\g_B(x)$ has compact support.   Therefore, the boundary term always vanishes and we will drop it.  Now insert the completeness relation
\beq
\delta(x-y)=\g_B(x)\g_B(y)+\ppin{k\p}\g_{k\p}(x)\g_{-k\p}(y) \label{crapp}
\eeq
to change the $\g\p(x)$ terms to $\g\p(y)$,  leaving the boundary terms implicit
\bea
V_{B k_2 k_1}&=&
\frac{1}{\slq}\int dx\int dy \delta(x-y)  \left(\V2 + C_{k_1k_2}\right)\left(\g_{k_2}(x)\g\p_{k_1}(y)+\g_{k_1}(x)\g\p_{k_2}(y)\right)\nonumber\\
&=&
\frac{1}{\slq}\int dx  \left(\V2 + C_{k_1k_2}\right)\Bigg[\g_B(x) \g_{k_2}(x)\Delta_{Bk_1}+\g_B(x) \g_{k_1}(x)\Delta_{Bk_2} \nonumber\\
&& +\ppin{k\p}\left(
\g_{k_2}(x)\g_{-k\p}(x)\Delta_{k\p k_1}+
\g_{k_1}(x)\g_{-k\p}(x)\Delta_{k\p k_2}
\right) \Bigg]\nonumber\\
&=&
\frac{1}{\slq}\left[V_{Bk_2}\Delta_{Bk_1}+ V_{Bk_1}\Delta_{Bk_2}
+\ppin{k\p}\left(V_{k_2-k\p}\Delta_{k\p k_1}+V_{k_1-k\p}\Delta_{k\p k_2}
\right)\right] \label{cald}.
\eea
Here the $C$ terms each vanish as a result of the orthonormality of the normal modes $\g$ as well the antisymmetry of $\Delta$. 

This Ward Identity relates 3-point functions with contractions of 2-point functions with $\Delta$.  We will see below that it can be generalized to an expression relating any $n$-point function with a contraction of $(n-1)$-point functions with $\Delta$.  

\subsection{Infrared Divergences}

One needs to be aware of the infrared divergences that arise when some subset of the $k_i$ sum to zero.  These result from the fact that the $e^{-i k_i x}$ factors in the corresponding $\g_{k_i}(x)$ at large $|x|$ have a product that does not oscillate, and so some integrals diverge.  For example, consider the manipulation
\bea
\Delta_{k_1 k_2}&=&\int dx \g_{k_1}(x) \partial_x \g_{k_2}(x)=\g_{k_1}(x)\g_{k_2}(x)|^\infty_{-\infty}-\int dx \g_{k_2}(x)\partial_x \g_{k_1}(x)\\
&&=\g_{k_1}(\hat x)\g_{k_2}(\hat x)|^\infty_{-\infty}-\Delta_{k_2 k_1}.\nonumber
\eea
Generally we drop the boundary term and summarize the result by stating that $\Delta_{k_1 k_2}$ is antisymmetric.  This is justified because, if we take the limit $|\hat x|\rightarrow\infty$ of the boundary term at the end, it oscillates rapidly in $|\hat x|$
and so vanishes as a distribution.  However this argument fails if $k_1=-k_2$.  Thus the antisymmetry is only up to a correction with support at $k_1=-k_2$, such as a Dirac $\delta$ function.  In practice, in the case of kinks in gapped theories considered here, this term is proportional to $k_1\delta(k_1+k_2)$ which in fact is antisymmetric.  In principle, such contributions may lead to finite effects in quantities of interest, and one must always be aware of them, and must determine when they may contribute.  For example, in the case of the one-loop mass correction, the general formula of Ref.~\cite{cahill76} is proportional to $(\ok{}-\omega_p)^2$ and so it vanishes even when the coefficient contains a $\delta(k-p)$.  

In general we expect such divergences in our $(n-1)$-point functions on the right hand side of the Ward Identities, but we do not expect them in the $n$-point functions on the left hand side because these include a $\g_B(x)$ which has compact support.  Let us now show that this expectation is fulfilled in the case at hand, and a divergence on the right hand side of the Ward Identity does not lead to one on the left hand side.

The $3$-point function $V_{Bk_1k_2}$ plays an important role, as the vertex factor connecting a zero mode to two mesons.  We now ask whether it is sensitive to $\delta$ function terms in $V_{k_1 k_2}$.  In the last line of (\ref{cald}), one can see that the shift $V_{k_1k_2}\rightarrow V_{k_1k_2}+\delta(k_1+k_2)$ leads to the shift $V_{Bk_1k_2}\rightarrow V_{Bk_1k_2}+(\Delta_{k_2k_1}+\Delta_{k_1k_2})/\sqrt{\lambda Q_0}$. This of course would vanish were $\Delta_{k_1k_2}$ truly antisymmetric, but as we reviewed above this argument fails at $k_1=-k_2$.  However, a finite contribution on a codimension one surface like $k_1=-k_2$ also vanishes in the sense of a distribution, while we recall that an infinite distribution is excluded by the fact that $\g_B(x)$ has compact support.  Therefore again we are not interested in such contributions, and so we conclude that a $\delta(k_1+k_2)$ contribution to the 2-point function does not affect the 3-point function. 



\subsection{Computation}

This time, expand
\bea
V_{BB k_2 k_1}&=&\int dx \V4 \g_B^2(x) \g_{k_2}(x)\g_{k_1}(x)\\
&=&-\frac{1}{\sqrt{Q_0}}\int dx \left( \V4 f\p(x) \right)\g_B(x) \g_{k_2}(x)\g_{k_1}(x)\nonumber\\
&=&-\frac{1}{\slq}\int dx \partial_x\left( \V3 + C_{k_1k_2}\right)\g_B(x) \g_{k_2}(x)\g_{k_1}(x)\nonumber
\eea
where $C_{k_1k_2}$ is independent of $x$ but otherwise arbitrary.

As $\g_B(x)$ vanishes asymptotically, no boundary term is introduced when we integrate by parts
\bea
V_{BB k_2 k_1}&=&\frac{1}{\slq}\int dx  \left(\V3 + C_{k_1k_2}\right)\partial_x\left(\g_B(x) \g_{k_2}(x)\g_{k_1}(x) \right)\\
&=&\frac{1}{\slq}\int dx  \left(\V3 + C_{k_1k_2}\right)\left(\g_B(x) \g_{k_2}(x)\g\p_{k_1}(x) \right.\nonumber\\
&&\left.+\g_B(x) \g_{k_1}(x)\g\p_{k_2}(x) +\g_{k_1}(x) \g_{k_2}(x)\g\p_{B}(x) \right).\nonumber
\eea
Note that the $C_{k_1k_2}$ terms vanish as they are the integral of a total derivative of a bounded function.  Of course this is obvious because $C_{k_1k_2}$ is arbitrary.

Now insert the completeness relation
\beq
\delta(x-y)=\g_B(x)\g_B(y)+\ppin{k\p}\g_{k\p}(x)\g_{-k\p}(y)
\eeq
to change the $\g\p(x)$ terms to $\g\p(y)$
\bea
V_{BB k_2 k_1}&=&
\frac{1}{\slq}\int dx\int dy \delta(x-y)  \left(\V3 + C_{k_1k_2}\right)\left(\g_B(x) \g_{k_2}(x)\g\p_{k_1}(y) \right. \label{ward}\\
&&\left.+\g_B(x) \g_{k_1}(x)\g\p_{k_2}(y) +\g_{k_1}(x) \g_{k_2}(x)\g\p_{B}(y) \right)\nonumber\\
&=&
\frac{1}{\slq}\int dx  \left(\V3 + C_{k_1k_2}\right)\left[\g^2_B(x) \g_{k_2}(x)\Delta_{Bk_1}+\g^2_B(x) \g_{k_1}(x)\Delta_{Bk_2} \right.\nonumber\\
&&\left.\hspace{-2cm} +\ppin{k\p}\left(
\g_B(x) \g_{k_2}(x)\g_{-k\p}(x)\Delta_{k\p k_1}+
\g_B(x) \g_{k_1}(x)\g_{-k\p}(x)\Delta_{k\p k_2}
+\g_{k_1}(x) \g_{k_2}(x)\g_{-k\p}(x)\Delta_{ k\p B}\right) \right]\nonumber\\
&&\hspace{-2cm}=\frac{1}{\slq}\left[ V_{BBk_2}\Delta_{Bk_1}+V_{BBk_1}\Delta_{Bk_2}+\ppin{k\p}\left(V_{Bk_2-k\p}\Delta_{k\p k_1}+V_{Bk_1-k\p}\Delta_{k\p k_2}+V_{k_1 k_2-k\p}\Delta_{k\p B}\right)
\right].\nonumber
\eea
Now shifting $V_{k_1k_2-k\p}$ by $\delta(k_1+k_2-k\p)$ changes the answer, with a shift proportional to $\Delta_{B,k_1+k_2}$.

\section{Subleading Corrections to Stokes Scattering} \label{stokesez}

\subsection{Shape Modes}

In some models, the kink possesses shape modes.  In that case, the virtual meson above could be a shape mode.  That invalidates two of the arguments made above.

First of all, several times about we stated that the $k\p$ integrand oscillates so rapidly that once $k\p$ is integrated out, the contribution to the amplitude will be exponentially suppressed by interference.  If $k\p$ is discrete, this argument does not work.

Second, we used the reduced inner product.  The kink has an infinite moduli space of classical solutions, related by translation invariance.  By choosing one kink solution, we have fixed the translation symmetry.  This can be done consistently in the ratio of matrix elements, like in our formula for the probability.  However, when one fixes a symmetry, a determinant term must be included.

This determinant was calculated in Ref.~\cite{menorm}.  Including it in the inner product, we found that the inner product is nonvanishing not only when all mesons have the same momenta, but also the inner product is nonvanishing between two states which differ by one meson with momentum $k\p$.  However, in this case, there is a suppression factor that is schematically $\sl \Delta_{k\p B}$.

As the zero mode $B$ is localized close to the kink, if the virtual meson has traveled far, then it will be disjoint from $\g_B(x)$ and this term will cancel.  However, a shape mode is bound to the kink and so cannot travel far.  Therefore, one can expect a contribution to the inner product arising from virtual shape modes.  

However this contribution is suppressed by a factor of $\sl$, and so one must consider evolution $U_1(t)$ at order $O(\sl)$.  This evolution has been comprehensively studied in Refs.~\cite{memult,stokes}.  The conclusion is that, if the kink starts in its ground state, the only allowed process is the creation of two quanta.  If both are continuum mesons, this process is called meson multiplication.  If one is a shape mode, this is called Stokes scattering.

We thus conclude that Stokes scattering, included in $U_1(t)|t=0\rangle_0$, may in principal lead to a nonvanishing inner product with respect to a nonforward meson, and so contribute to our process.

Of course this cannot really happen, as the conservation of energy would imply $\ok 2=\ok 1-\os$, which is the wrong energy for the recoil meson.  But in this Appendix we will try to show how this contribution vanishes.

\subsection{Stokes Scattering}

Consider the interaction $H^{(1)\prime}$ from Eq.~(\ref{vkkk}).  It acts on meson 1 as in Eq.~(\ref{vkkka}).  At leading order, this leads to the final state
\bea
U_1(t)|t=0\rangle_0&=&-i\int_0^t d\tau_1 e^{-iH\p_2(t-\tau_1)}H^{(1)\prime}_3e^{-iH\p_2\tau_1}\pin{k_1}e^{-\sigma^2(k_1-k_0)^2-i(k_1-k_0)x_0}|k_1\rangle_0\nonumber\\
&=&-i\frac{\sl}{4}\pin{k_1}\ppin{k_2}\ppin{k\p}e^{-i(\ok 2+\okp{})t}\int_0^t d\tau_1 e^{-i(\ok 0-\ok 2-\okp{})\tau_1}\nonumber\\
&&\times\frac{V_{-k_1k_2k\p}}{\ok 1}e^{-\sigma^2(k_1-k_0)^2-i(k_1-k_0)x_{\tau_1}}|k_2k\p\rangle_0.
\eea
Stokes scattering corresponds to the case $k\p=S$ and $k_2$ is a continuum mode.  Of course, this expression is symmetric in $k\p$ and $k_2$ and so if $k_2=S$ then one can rename it $k\p$.  This freedom leads to a factor of 2.

Abusing our notation again, we will define $U_1(t)$ be the Stokes scattering part, which is equivalent to considering only Stokes terms in the definition of $H^{(1)\prime}_3$
\bea
U_1(t)|t=0\rangle_0&=&-i\frac{\sl}{2}\pink{2}e^{-i(\ok 2+\os)t}\int_0^t d\tau_1 e^{-i(\ok 0-\ok 2-\os)\tau_1}\\
&&\times\sum_S\frac{V_{-k_1k_2S}}{\ok 1}e^{-\sigma^2(k_1-k_0)^2-i(k_1-k_0)x_{\tau_1}}|k_2S\rangle_0.\nonumber
\eea
Now, we make the usual approximation that $\ok 1$ in the denominator is $\ok 0$, and again using the fact that $\g_S(x)$ has compact support we find
\bea
U_1(t)|t=0\rangle_0&=&-i\frac{\sl}{2}\frac{\sqrt{\pi}}{2\pi\sigma}\pin{k_2}\frac{V_{-k_0k_2S}}{\ok 0}e^{-i(\ok 2+\os)t}\\
&&\times \int_0^td\tau_1 e^{-i(\ok 0-\ok 2-\os)\tau_1}e^{-x^2_{\tau_1}/(4\sigma^2)}|k_2S\rangle_0\nonumber\\
&=&-i\frac{\sl}{2}e^{-i\ok 0t_c}\pin{k_2}\sum_S\frac{V_{-k_0k_2S}}{k_0} e^{-\sigma^2(\ok 0/k_0)^2(\ok 0-\ok 2-\os)^2-i(\ok 2+\os)(t-t_c)}|k_2S\rangle_0.\nonumber
\eea

\subsection{Reduced Inner Product}

To obtain the corresponding contribution to the probability, one needs to project this final state onto one meson final states, using the projection
\beq
\mathcal{P}=\frac{1}{\sqrt{Q_0}}\pin{k_2}2\ok 2|k_2\rangle\langle k_2|.
\eeq
What is the reduced inner product of this final state with a single meson state $|k_2\rangle$?

There are two contributions.  The first comes from the leading part $|k_2\rangle_0$ of $|k_2\rangle$.  Using the master formula (4.14) of Ref.~\cite{menorm} with $\gamma^{02}(k_2 S)=1/2$, one finds that the term contracting the shape mode and the zero mode is
\beq
{}_0\langle k_2|U_1(t)|t=0\rangle_{0 \ \rm{red}}=-i\frac{\sl}{8}e^{-i\ok 0t_c}\sum_S\frac{V_{-k_0k_2S}\Delta_{SB}}{k_0\os\ok 2} e^{-\sigma^2(\ok 0/k_0)^2(\ok 0-\ok 2-\os)^2-i(\ok 2+\os)(t-t_c)}. \label{reda}
\eeq
Here we have remembered the factor of $1/(2\ok 2)$ from $B^{\ddag\dag}_{k_2}$ which is built into the normalization of our states $|k_2\rangle_0$.

However there are also contributions, at the same order, from the two-meson quantum corrections $|k_2\rangle_1$
\bea
\gamma_{1 k_2}^{02}(k\p_1,k\p_2)&=& -\frac{2\pi\delta(k\p_2-k_2)}{4}\left(\Delta_{k\p_1 B}+\sqrt{\lambda Q_0}\frac{V_{\I  k\p_1}}{\okp1}\right)+\frac{\sqrt{\lambda Q_0}V_{-k_2 k\p_1 k\p_2}}{4\ok2\left(\ok2-\okp1-\okp2\right)}\nonumber\\
&&-\frac{2\pi\delta(k\p_1-k_2)}{4}\left(\Delta_{k\p_2 B}+\sqrt{\lambda Q_0}\frac{V_{\I  k\p_2}}{\okp 2}\right).
\eea
Here $\gamma^{mn}_{i\psi}(k_1\cdots k_n)$ is the coefficient that arises when decomposing the state $|\psi\rangle$ into the basis $\phi_0^m|k_1\cdots k_n\rangle_0$ at order $O(\lambda^{i/2})$.  It is defined to include a factor of $Q_0^{i/2}$ so that it contains no powers of the coupling $\lambda$.

We are interested in the case where $k\p_1$ or $k\p_2$ is a shape mode.  At late times, the meson $k_2$ is far from the kink, and so cannot interact with a shape mode.  As a result, after all of the usual integrations, the $V_{-k_2 k\p_1 k\p_2}$ term will vanish.  

Now $k_2$ is the momentum of the asymptotic meson, so it is by assumption not a shape mode, as we are calculating the amplitude to produce an asymptotic meson.  Therefore, in the $\delta(k\p_2-k_2)$ term, it must be that $k\p_1$ is the shape mode, and similarly for the other $\delta$ term.

We need to sum over whether $k\p_1$ or $k\p_2$ is the shape mode.  Now, remembering the factors of $1/(2\os)$ and $1/(2\ok 2)$ from the contractions of the two mesons, one can see that the $\Delta$ terms cancel half of (\ref{reda}).  The other half is canceled by the contribution from Eq.~(\ref{finale})
\bea
{}_0\langle k_2|U_1(t)|t=0\rangle_{0\ \rm{red}}&=&
i\frac{\sl}{16}e^{-i\ok 0t_c}\ppin{k\p}
\frac{V_{-k_0k_2k\p}\Delta_{-k\p B}}{k_0 \okp{}\ok 2 }
\\&&\times
e^{-\sigma^2(\ok 0/k_0)^2(\ok 0-\ok 2-\okp{})^2-i(\ok 2+\okp{}) (t-t_c)}.\nonumber
\eea

This leaves the two tadpole terms, which are proportional $V_{\I S}$.  They yield an inner product of 
\beq
{}_0\langle k_2|U_1(t)|t=0\rangle_{0\ \rm{red}}=i\frac{\lambda\sqrt{Q_0}}{8}e^{-i\ok 0t_c}\frac{V_{-k_0k_2S}V_{
\I S}}{k_0\os^2\ok 2} e^{-\sigma^2(\ok 0/k_0)^2(\ok 0-\ok 2-\os)^2-i(\ok 2+\os)(t-t_c)}. 
\eeq
This term appears to be a disaster, as it contributes to the final probability with a final energy $\ok 2$ that is not close to the initial energy $\ok 0$.  

However the term looks reminiscent of the tadpole interactions studied in Subsec.~\ref{tadsez}.  Indeed, the inner product of the first term in the parentheses in Eq.~(\ref{guscio}) with ${}_0\langle k_2|$ exactly cancels this term.

\section* {Acknowledgement}

\noindent
JE is supported by NSFC MianShang grants 11875296 and 11675223. HL thanks BY Zhang for the helpful discussion.

\end{document}